\documentclass[reprint,superscriptaddress,amsmath,amssymb,aip,jap,floatfix]{revtex4-1}
\usepackage[version=3]{mhchem}
\usepackage{siunitx,natmove,graphicx}

\DeclareSIUnit{\gcc}{\g\per\cubic\cm}
\DeclareSIUnit{\mgcc}{\milli\g\per\cubic\cm}

\begin{document}

\title{Vibration Damping of Carbon Nanotube Assembly Materials}

\author{Jingna Zhao}
\author{Fulin Wang}
\author{Xin Zhang}
\author{Linjie Liang}
\author{Xueqin Yang}
\author{Qingwen Li}
\author{Xiaohua Zhang}
\email{xhzhang2009@sinano.ac.cn}
\affiliation{Division of Advanced Nano-Materials, Suzhou Institute of Nano-Tech
and Nano-Bionics, Chinese Academy of Sciences, Suzhou 215123, China}

\begin{abstract}
Vibration reduction is of great importance in various engineering applications, and a material that exhibits good
vibration damping along with high strength and modulus has become more and more vital. Owing to the superior mechanical
property of carbon nanotube (CNT), new types of vibration damping material can be developed. This paper presents recent
advancements, including our progresses, in the development of high-damping macroscopic CNT assembly materials, such as
forests, gels, films, and fibers. In these assemblies, structural deformation of CNTs, zipping and unzipping at CNT
connection nodes, strengthening and welding of the nodes, and sliding between CNTs or CNT bundles are playing important
roles in determining the viscoelasticity, and elasticity as well. Towards the damping enhancement, strategies for
micro-structure and interface design are also discussed.
\end{abstract}

\maketitle

\section{Introduction}

Vibrations are of concern to structural materials for the safety and comfort of civil infrastructures. For example,
passenger comfort or the ride quality in aircraft and automobile is greatly affected by the vibrations caused by outside
disturbances, such as aeroelastic effects and rough road surfaces. In other cases, civil engineering structures located
in environments where earthquakes or large wind forces are common will be subjected to serious vibrations. Furthermore,
vibration also affects the tracking and pointing characteristics and accuracy of weapon systems mounted on aircraft and
land systems. Unfortunately, most of the materials are inherently poor in vibration reduction \cite{chung.ddl_2001,
lakes.rs_2002, luo.jl_2015, zhou.xq_2016}. A material that exhibits good vibration reduction along with high specific
stiffness/modulus and strength becomes very vital for everyday life. Vibration reduction is usually reflected by the
damping capacity, which describes how much mechanical energy can be dissipated during the vibration. Damping of a
structure can be attained by passive or active methods \cite{chung.ddl_2001}. The passive methods depend on the inherent
ability of certain materials to absorb the vibrational energy, while the active methods make use of sensors and
actuators to attain vibration sensing and activation to suppress the vibration in real time.

Metals and polymers are dominantly good materials for vibration damping, owing to their inherent viscoelasticity. Other
mechanisms, especially structural defects such as dislocations, phase boundaries, grain boundaries, and various
interfaces can also contribute greatly to damping. For composite structures, the overall damping capacity greatly
depends on the reinforcement used and is proportional to the individual damping capacities of the reinforcement
\cite{chung.ddl_2003}. Therefore, the microstructure design and interface control have become the most efficient ways to
improve the damping performance \cite{chandra.r_1999, khan.su_2011, kumar.ks_2014, zhang.p_2015}.

Carbon nanotubes (CNTs), with their exceptional mechanical, electrical, and thermal properties, could be building blocks
for various macroscopic structural materials (architectures), such as one-dimensional (1D) fibers, 2D films and
buckypapers, and 3D forests, gels, and sponges \cite{liu.lq_2011, lu.wb_2012, liu.yd_2014, di.jt_20161, zhang.xh_2016,
hashim.dp_2012, lin.zq_2016}. These CNT-based materials have shown great potentials for civil, military, and aerospace
applications. Different from metals and polymers, the CNT assembly materials possess rich contacts, connections,
cross-links, and thus different types of interfaces between the nano components. For example, zipping/unzipping of
adjacent CNTs, sliding between CNTs or CNT bundles, buckling of CNTs, and self-organization upon loading have become the
main sources of viscoelasticity for the assembly materials \cite{cao.ay_20051, xu.m_2010, wang.c_2012, won.y_2013,
zhao.jn_2015, li.yp_2015, shen.zq_2017}.

Here, we review recent progresses on the damping performance of various CNT assembly materials, and report our studies
on the dynamic properties of CNT fibers and films along with their related strengthening and toughening strategies.
First, the basic concepts of viscoelasticity are provided in Sec.\ \ref{sec:concept}. Then, the major progress report is
organized in Sec.\ \ref{sec:CNT} according to the assembly way of the macroscopic CNT materials, namely CNT forests,
gels, films and buckypapers, and fibers. Finally, a brief discussion on the strategies to build high-damping assemblies
is given in Sec.\ \ref{sec:discuss}.

\section{Basic Concepts of Viscoelasticity}
\label{sec:concept}

\begin{figure*}[t!]
\centering
\includegraphics[width=0.75\textwidth]{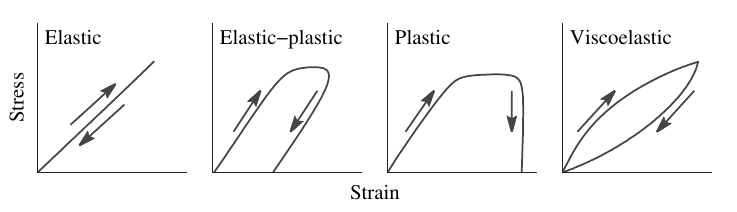}
\caption{Stress-strain curves illustrating different types of tensile behavior, of the elastic, elastic-plastic,
plastic, and viscoelastic materials, respectively.}
\label{fig:tensilebehaviors}
\end{figure*}

Elasticity deals with elastic stresses and strains, their relationship, and the external forces that cause them
\cite{ferry.jd_1980, meyers.ma_2009}. For all solid materials there is a domain in stress space in which strains are
reversible due to small relative movements of atoms. Thus, the classical Hooke's law of elasticity applies. In reality,
all materials deviate from Hooke's law in various ways, usually by exhibiting both viscous-like and elastic
characteristics when undergoing deformation \cite{lakes.r_2009, meyers.ma_2009}. Therefore, viscoelasticity considers in
addition a dissipative phenomenon due to ``internal friction'', such as between molecules in polymers, between cells in
wood, or between nanoparticles and matrix in nanocomposites \cite{lemaitre.j_2001}. It describes the ability of a
material to both dissipate energy (viscous) and reversibly deform (elastic). The capacity of a system to dissipate
dynamic (vibration) energy is primarily described or measured by ``damping'', either by energy dissipation or by a force
that works to dissipate energy.

It is necessary to point out that viscoelasticity is not plasticity, with which it is often confused. A viscoelastic
material will return to the original shape when the deforming force is removed, despite the fact that it might take a
long time for the recovery. On the contrary, a plastic material can not recover after the load is removed. {\bf Figure
\ref{fig:tensilebehaviors}} schematically compares the stress-strain curves for different types of tensile behavior. For
viscoelastic materials, although the original shape is recovered on removal of the load, some permanent deformation
remains owing to plastic deformation or molecular slippage of an irreversible nature.

Two major types of experiment are performed on viscoelastic materials: transient and dynamic. Transient experiments
involve deforming the material and studying the response of the material with time. Creep and stress-relaxation
experiments are two transient experiments. In the dynamic tests, either stress or strain is varied cyclically with time,
and the response is measured at various different frequencies of deformation.

\subsection{Creep-Recovery and Stress Relaxation}

\begin{figure}[t!]
\centering
\includegraphics[width=0.48\textwidth]{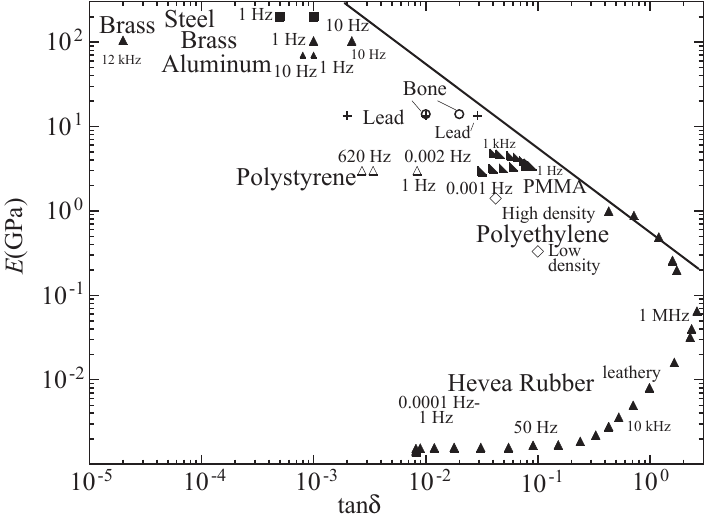}
\caption{Stiffness-loss map for some materials at near room temperature \cite{lakes.r_2009}. The diagonal line
represents $E'' = E' \tan \delta = 0.6$ GPa. Most materials occupy the region to the left of that line.}
\label{fig:lossmap}
\end{figure}

\begin{figure*}[t!]
\centering
\includegraphics[width=0.80\textwidth]{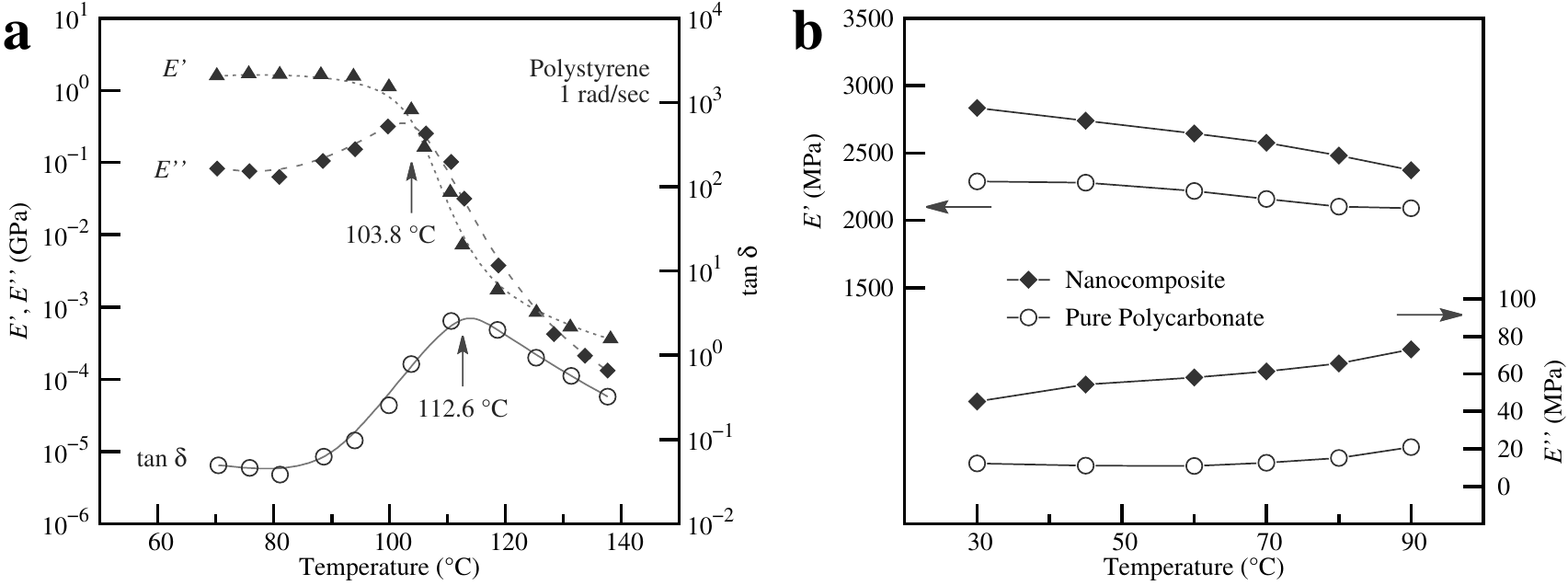}
\caption{(a) DMA thermal curves for PS film, step heating with isothermal frequency sweep every 5 \si{\celsius} in
nitrogen \cite{achorn.pj_1994}. (b) Storage modulus and loss modulus as functions of test temperature for constant
strain amplitude (0.35\%), at 1 Hz, for the pure polycarbonate and the nanocomposite samples containing 1.5 wt\% of
SWCNTs \cite{suhr.j_2006}.}
\label{fig:DMA-polymer}
\end{figure*}

Viscoelastic behavior manifests itself in creep and in stress relaxation. Creep is a progressive deformation of a
material under constant stress, and, vice versa, stress relaxation is the gradual decrease of stress when the material
is held at constant strain. Therefore, creep testing involves loading a sample with a set weight and watching the strain
change over time, and recovery tests look at how the material relaxes once the load is removed. For stress relaxation, a
sample is held at a set length and the force it generates is measured. Creep and creep-recovery tests are very useful
for studying materials under very low shear rates or frequencies, under long test times, or under real use conditions.

In order to model the creep-recovery behavior, different combinations of springs (elastic sections) and dashpots
(viscous sections) are suggested. For example, the Maxwell model with the spring and dashpot in series, and the
Kelvin-Voigt model with the spring and dashpot in parallel describe totally different creep-recovery behaviors.

\subsection{Dynamic Mechanical Analysis}

Viscoelasticity is widely studied using dynamic mechanical analysis (DMA) where an oscillatory force (stress) is applied
to a material and the resulting displacement (strain) is measured. Under a sinusoidally varying stress $\sigma =
\sigma_0 \sin \omega t$ where $\omega$ is the angular frequency, the strain $\varepsilon$ is given by $\varepsilon =
\varepsilon_0 \sin \omega t$ for elastic solids. For materials that exhibit damping, the stress and strain are not in
phase; the strain lags behind the stress by a phase angle $\delta$, which defines an in-phase and out-of-phase component
of the stress, $\sigma' = \sigma_0 \cos \delta$ and $\sigma'' = \sigma_0 \sin \delta$. Therefore, a complex modulus,
$E^* = E' + i E''$, is used to describe the stress-to-strain ratio, where $E' = \sigma' / \varepsilon_0 = |E^*| \cos
\delta$ and $E'' = \sigma'' / \varepsilon_0 = |E^*| \sin \delta$ are the storage modulus and loss modulus. A further
quantity of viscoelastic materials is the loss tangent, $\tan \delta = E''/E'$, where the ratio of the two modulus
quantities represents an extremely useful damping quantity.

\subsection{Common Viscoelastic Materials}

{\bf Figure \ref{fig:lossmap}} provides a comparison of different viscoelastic materials in their stiffness-loss map
\cite{lakes.r_2009}. Some single crystal materials exhibit the lowest losses, such as fused silica and sapphire,
followed by common metals. Polymers in the transition region exhibit the highest losses commonly found in solids; the
loss tangent can exceed unity. For common materials, high stiffness is associated with low loss, as reflected by the
diagonal line as the upper limit for the loss modulus (Figure \ref{fig:lossmap}). (For more and detailed information,
see Chapter 7 in R.\ Lakes, Viscoelastic Materials, 2009 \cite{lakes.r_2009}.)

{\bf Figure \ref{fig:DMA-polymer}}a shows a typical DMA result in temperature sweep mode for polystyrene (PS)
\cite{achorn.pj_1994}. Generally, at the glass transition temperature, $T_g$, $E'$ begins to decrease, while $E''$ and
$\tan \delta$ go through maxima, as the temperature is increased. Notice that, different values of $T_g$ can be defined,
by using either the temperature of the loss modulus peak or the temperature of the $\tan \delta$ peak.

In order to improve the resilience and strength of polymers, extensive use of different types of filler materials have
been considered. Nanometer-size fillers having large surface areas, such as nanotubes, nanorods, and nanofibers, have
added advantages with greater interactions at the interfaces \cite{zhou.x_2004, suhr.j_2005, ajayan.pm_2006,
suhr.j_2006, wang.zw_2011, agrawal.r_2013, carponcin.d_2015, luo.jl_2015, chu.yc_2016, eftekhari.m_2016}. For example,
the introduction of single-walled CNTs (SWCNTs) into polycarbonate can activate interfacial slip at the tube-polymer
interfaces at relatively low dynamic strain levels, by raising temperature to $\sim$90 \si{\celsius}, and thus enhance
the damping performance for vibration and acoustic suppression \cite{suhr.j_2006}. As shown in Figure
\ref{fig:DMA-polymer}b, after loading CNTs, the loss modulus increases much greater than the storage modulus.

\section{Vibration Damping in Carbon Nanotube Assemblies}
\label{sec:CNT}

Recent studies have shown that CNTs are ideal scaffolds to design and architect high-performance composites, at high CNT
volume fractions or even fully composed by CNTs themselves \cite{zhang.xh_2016}. For example, the aligned pure or neat
CNT films can exhibit a tensile strength and Young's modulus of 3.2 GPa and 172 GPa \cite{wang.yj_2015, zhang.lw_2015,
yu.xp_2016}, and the high-volume-fraction CNT/bismaleimide (BMI) composites can be even as strong as 3.8--6.94 GPa in
strength \cite{cheng.qf_2009, cheng.qf_2010, wang.x_20131, han.y_2015}. In these high-strength materials, the CNTs
should be highly packed and super aligned. Generally, these high-strength or high-stiffness CNT assembly materials show
low loss tangent in dynamic tests. On the contrary, the loosely assembled macroscopic CNT materials, especially CNT
forests, aerogels, and sponges, are naturally high-performance damping materials \cite{koratkar.na_2003, pathak.s_2009,
xu.m_2010, zhang.q_20102, bradford.pd_2011, won.y_2013, liu.ql_2015, shen.zq_2017}, owing to the sliding,
zipping/unzipping, buckling, and cross-linking between/of the CNTs \cite{cao.ay_20051, xu.m_2010, wang.c_2012,
li.y_2012, won.y_2013, zhao.jn_2015, li.yp_2015, shen.zq_2017}. Here, to provide a detailed and systematic overview of
the various CNT assembly materials, recent progresses are discussed according to the assembly way for CNTs.

\subsection{CNT Forests}

\begin{figure*}[t!]
\centering
\includegraphics[width=0.85\textwidth]{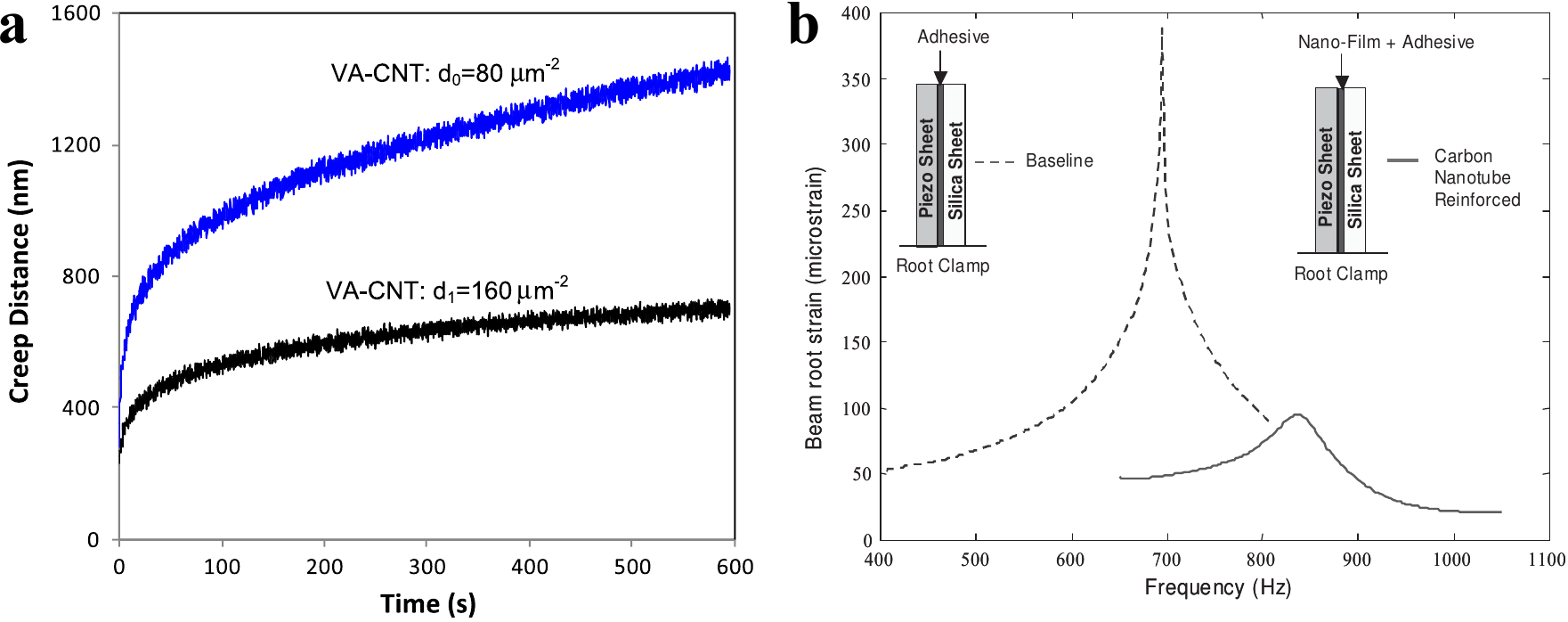}
\caption{(a) The indentation displacement-time curves of CNT forests with a low and high number density
\cite{zhang.q_20102}. (b) Comparison of the dynamic response of the baseline beam and the CNT-reinforced sandwich beam
for a frequency-sweep test at 50 V \cite{koratkar.n_2002}.}
\label{fig:forest}
\end{figure*}

CNT forests, also called vertically aligned CNTs or CNT arrays, have drawn considerable interest for a large range of
potential applications, such as field emission electron sources, thermal interface materials, black body, composites,
and CNT fibers \cite{fan.ss_1999, zhang.m_2004, wang.d_2008, mizuno.k_2009}. Notice that, for most cases, the forests
are composed of multi-walled CNTs (MWCNTs), and for better mechanical performances, few-walled CNTs (wall number 2--4)
are preferred \cite{fang.c_2010, jia.jj_2011}. Due to the empty spaces between the aligned CNTs, as reflected by the low
number density of 80--160 \si{\per\square\um} or mass density of 1--90 \si{\micro\g\per\cubic\mm} \cite{cao.ay_20051,
zhang.q_20102, polsen.es_2013}, the CNTs can easily bend upon compression or indentation. Thus, such assembly exhibits
typical viscoelastic creep behavior, that is, when the indentation stress is increased and then held at a constant
value, the indenter continues to penetrate into the specimen \cite{zhang.q_20102}. As shown in {\bf Figure
\ref{fig:forest}}a, a denser CNT forest has fewer geometric freedoms for the tube movement, and thus less creep
deformation. However, due to the lack of cross-linking and interconnection between CNTs, such CNT forests still show
strong compressive plasticity \cite{pathak.s_2009, zhang.q_20102}.

A more detailed study showed that the energy dissipation ability of CNT forests can be remarkably reduced by introducing
solvents between the CNTs \cite{eom.k_2013}. For example, water can induce a transition from the viscoelastic behavior
into an elastic one, as hydration mediates the interaction between CNTs and thus reduces the frictional dissipation
during the loading-unloading process. Further removal of water molecules allows the CNT forest to recover its
viscoelastic behavior. Decane shows stronger ability to reduce the dissipation, indicating that the choice of solution
is important to tune the viscoelasticity.

However, as CNT forests are loosely assembled and thus have weak mechanical performances in stretching, compressing, and
bending, the most possible application of CNT forest is the structural damping enhancement. To do so, a piezo-silica
sandwich beam has been embedded with a CNT forest, by stacking the CNTs in between a top piezoelectric sheet and a
bottom silica substrate \cite{koratkar.n_2002}. Figure \ref{fig:forest}b compares the dynamic response of the sandwich
beam with the baseline beam (without CNT reinforcement) for a frequency sweep test, at the same beam size. Clearly the
CNT reinforcement has resulted in a very significant increase in the damping as well as the stiffness; the increased
damping is reflected by the reduced dynamic amplification at resonance while the increased stiffness is manifested as a
shift in the first bending frequency from $\sim$700 Hz to $\sim$840 Hz.

Another similar sandwich composites with improved energy dissipation and damping properties were fabricated by stacking
the freestanding CNT forests and carbon fiber fabrics, with a following epoxy infiltration using a vacuum-assisted resin
transfer mold \cite{zeng.y_2010}. As compared to the conventional carbon fiber/epoxy laminate composites, the sandwich
composites exhibit higher flexural rigidity and damping, which is achieved due to the rich interfacial areas inside the
nanocomposites and the high thermal conductivity of CNTs.

In a following study where the silicon wafer covered with a CNT forest was tested, the mechanism for the enhanced
damping was discussed. The primary energy dissipation is also believed to be a general result of the interfacial
friction between individual CNTs, caused by their entanglements under cyclic deformation \cite{mantena.pr_2013}.

Nowadays, CNT forests have been widely and successfully applied to enhance the damping performance of SiC fiber cloth
\cite{veedu.vp_2006}, woven fiber-glass composites \cite{kim.k_2015}, metal layers \cite{boddu.vm_2016}, amorphous
diamond film \cite{teo.eht_2007}, and polymer matrix \cite{silva.gg_2011, urk.d_2016}.

\subsection{CNT Gels}

Besides such effect of intertube friction, the CNT entanglement plays more important roles in determining the
viscoelasticity. (Even in CNT forests, the effect of entanglement is still non-negligible \cite{mantena.pr_2013}.) This
is because the entanglement can cause stronger intertube interaction than the alignment upon the structural deformation
on CNT assemblies.

\begin{figure*}[t!]
\centering
\includegraphics[width=0.80\textwidth]{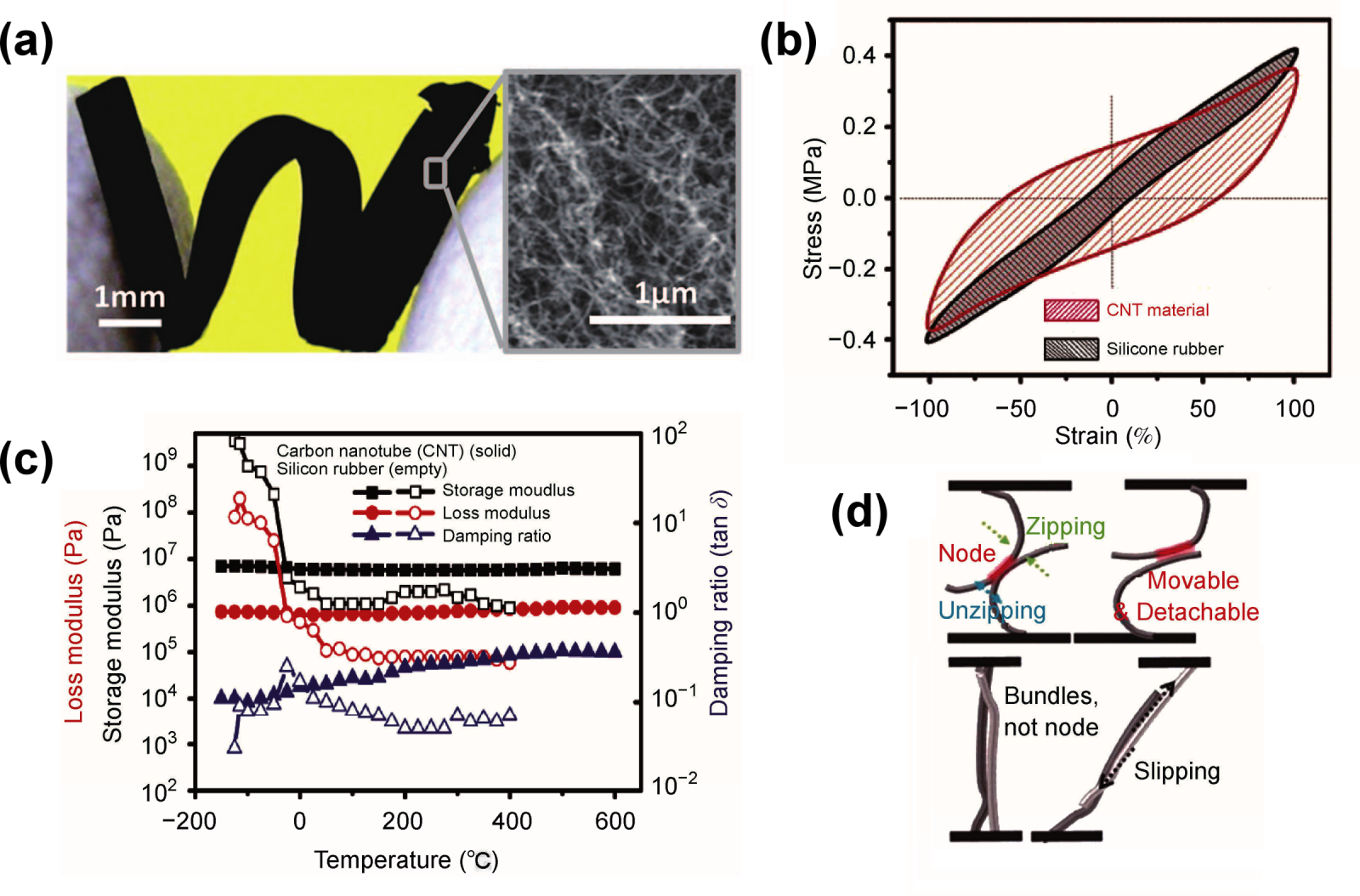}
\caption{Viscoelasticity of densified CNT gels \cite{xu.m_2010}. (a) Photograph and SEM image of the CNT gel. (b)
Stress-strain curves of CNT gel and silicone rubber. (c) Temperature dependence of the storage modulus, loss modulus,
and damping ratio of the CNT material, compared with silicone rubber. (d) Schematic of the zipping and unzipping of CNT
connections nodes, as compared to the sliding between CNTs.}
\label{fig:CNTgel}
\end{figure*}

In one such assembly, as a result of water-assisted chemical vapor deposition (CVD) and a post-treatment of compression,
each CNT makes contacts with numerous other CNTs, see the photograph and scanning electron microscopy (SEM) image in
{\bf Figure \ref{fig:CNTgel}}a \cite{xu.m_2010}. (Such material is also called a gel, aerogel, or sponge, due to the
feature of entanglement and the small mass density of 36 \si{\mgcc}.) The stress-strain behavior from shear-mode DMA
showed a strain up to 100\%, high nonlinearity, and a closed hysteresis without abrupt changes, which are typical of
viscoelastic, energy-dissipative, and highly deformable materials (Figure \ref{fig:CNTgel}b). As compared to silicone
rubber, the CNT gel had a larger enclosed area of the hysteresis loop, corresponding to a higher damping ratio
($E''/E'\approx 0.3$) at room temperature. (Notice that, such damping ratio was nearly constant in the frequency range
of 0.1--100 Hz.) Interestingly, the viscoelastic properties measured in ambient \ce{N2} were nearly constant over an
exceptionally wide temperature range (from $-140$ to 600 \si{\celsius}) in contrast to the rubber, which showed large
variation due to hardening (at $-55$ \si{\celsius}) and degradation (at 300 \si{\celsius}), see Figure
\ref{fig:CNTgel}c. Actually, due to the high thermal stability of CNT, the viscoelasticity could be unvaried beyond the
limitation at $-190$ and $>$900 \si{\celsius}. Besides the damping ratio, the creep and creep recovery of CNT gels are
also found to be at the same level as silicone rubber, still with an extreme temperature stability \cite{xu.m_20111}.
Furthermore, with increasing the gel's mass density, from 3.3 to 54 \si{\mgcc}, both the storage modulus and damping
ratio concurrently increased \cite{xu.m_2011}. 

\begin{figure}[t!]
\centering
\includegraphics[width=0.45\textwidth]{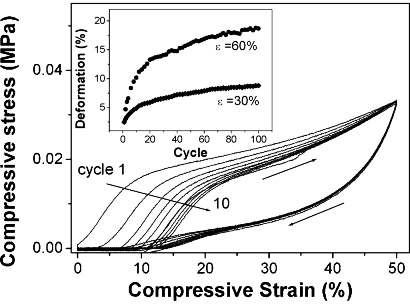}
\caption{Cyclic stress-strain curves at a maximum of 50\% in air \cite{gui.xc_2010}. Inset, recorded deformations
developed by compression for 100 cycles at different set strains of 30\% and 60\%, respectively.}
\label{fig:collapse}
\end{figure}

\begin{figure*}[t!]
\centering
\includegraphics[width=0.80\textwidth]{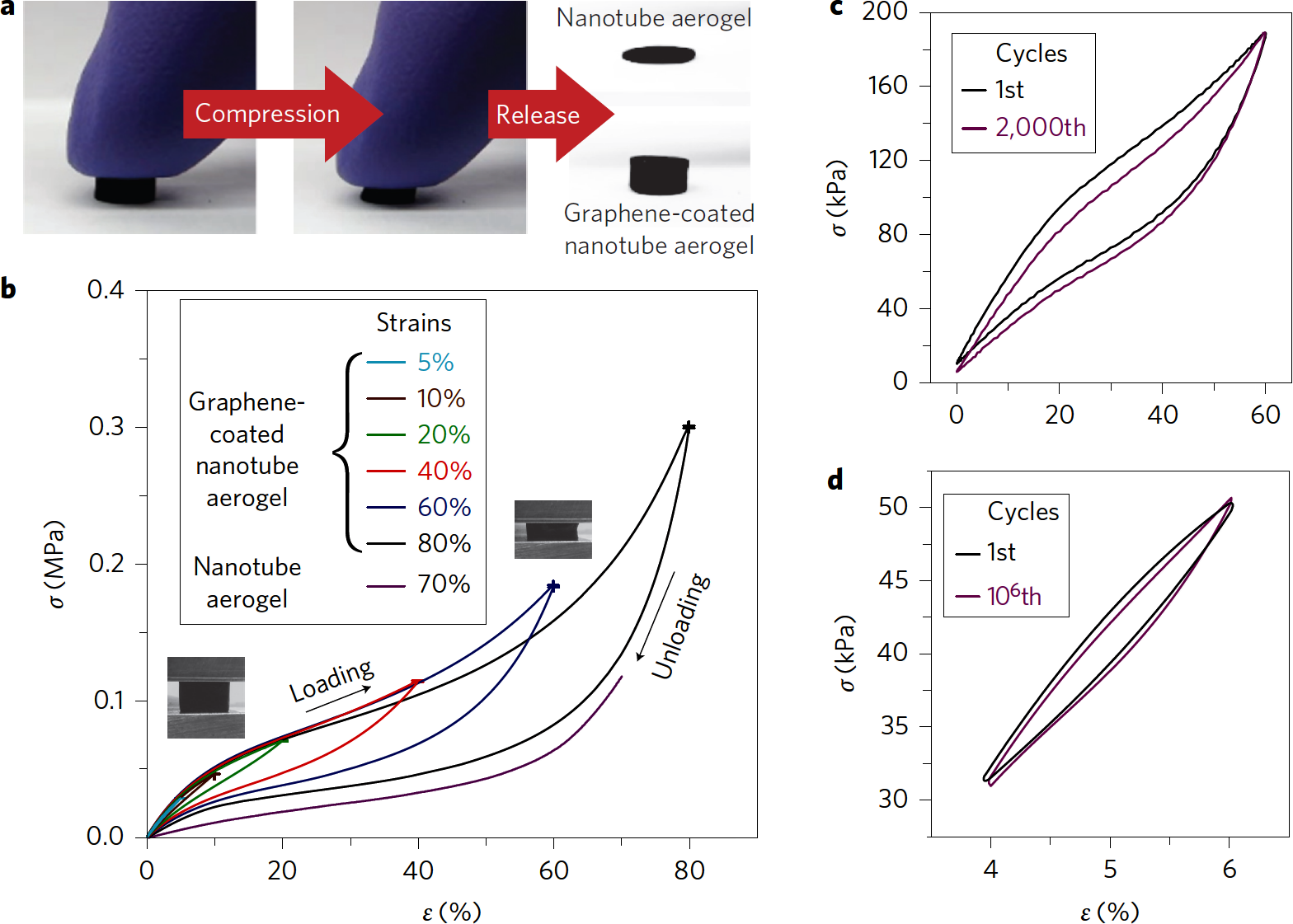}
\caption{Mechanical properties of CNT gels \cite{kim.kh_2012}. (a) CNT gels collapse and graphene-coated CNT gels
recover their original shape after compression by $>$90\%. (b) Stress-strain curves for CNT gels along the loading
direction and for graphene-coated CNT gels during loading-unloading cycles. (c,d) Fatigue resistance of graphene-coated
CNT aerogel at 60\% strain, 1Hz, for the 1st and 2,000th cycles (c) and at 2\% strain, 100 Hz, for the 1st and $10^6$th
cycles (d).}
\label{fig:cnt-graphene}
\end{figure*}

In the CNT gel, each CNT makes contact with numerous other tubes, resulting in a high density of connections (nodes).
These nodes are separated mainly by isolated CNTs or CNT bundles (struts). Such intertube assembly of struts and nodes
should be the key for structural cohesiveness that allows for large deformations \cite{xu.m_2010, gui.xc_2010}. It is
also believed that the CNTs in a node can reversibly attach and detach through zipping and unzipping (Figure
\ref{fig:CNTgel}d). This process can dissipate energy because it overcomes the large van der Waals (vdW) attraction
between CNTs when unzipping, yet no energy is required for zipping. On the contrary, when there is a lack of such
connection nodes, sliding between CNTs dominates the structural deformation for the gel. However, the sliding can not
contribute much to the dissipation due to the small friction coefficient, and such sliding is rather a plastic and
irreversible process.

Besides the CVD-based manufacture \cite{xu.m_2010, gui.xc_2010, dai.zh_2016, wang.h_2017}, CNT gels are also created by
critical-point drying of solution-processed hydrogels \cite{hough.la_2006, bryning.mb_2007, zou.jh_2010, kim.kh_2011},
dry stacking of aligned CNT sheets drawn out from CNT forests \cite{faraji.s_2015}, and a frit compression method where
surfactants or surface modification of CNT are avoided \cite{whitby.rld_2008, whitby.rld_2010}. However, these pristine
CNT gels could collapse (irreversible thickness reduction for the CNT film) upon compressing, as shown in {\bf Figure
\ref{fig:collapse}}, corresponding to a plastic deformation \cite{gui.xc_2010}. Therefore, various post-treatments have
been proposed to improve the elasticity.

For example, by infiltrating pre-formed CNT gels with a low molecular-weight polyacrylonitrile (PAN) polymer and then
critical-point drying, a PAN/CNT composite gel can be obtained. After degassing at 140 \si{\celsius}, stabilizing at 210
\si{\celsius}, and pyrolyzing at 1010 \si{\celsius}, these gels can be converted into graphene-coated gels, with the
mass density increased from 8.8 to 14.0 \si{\mgcc} \cite{kim.kh_2012, kim.kh_2017}. The pure CNT gels usually collapse
upon high compression ratio, e.g., $>$90\%, while the graphene-coated gels can well recover their original shape ({\bf
Figure \ref{fig:cnt-graphene}}a). Typical viscoelastic behaviors are observed in the plots of compressive stress
$\sigma$ versus compressive strain $\varepsilon$ (Figure \ref{fig:cnt-graphene}b). The energy dissipation, the
hysteresis loop in the loading-unloading cycle, was negligible at a small strain of $\varepsilon=5$\%, but steadily
increased with $\varepsilon$.  Of great importance, after 2000 loading-unloading cycles at $\varepsilon=60$\% and 1 Hz,
and $1\times 10^6$ cycles at $\varepsilon=2$\% and 100 Hz, no significant plastic deformation or degradation in
compressive strength is observed, corresponding to a high structural robustness (Figure \ref{fig:cnt-graphene}c and d).
Obviously, the graphene coating can modify the intertube interactions at the CNT connections, making them slippery and
deformable into robust and elastic.

\begin{figure}[h]
\centering
\includegraphics[width=0.48\textwidth]{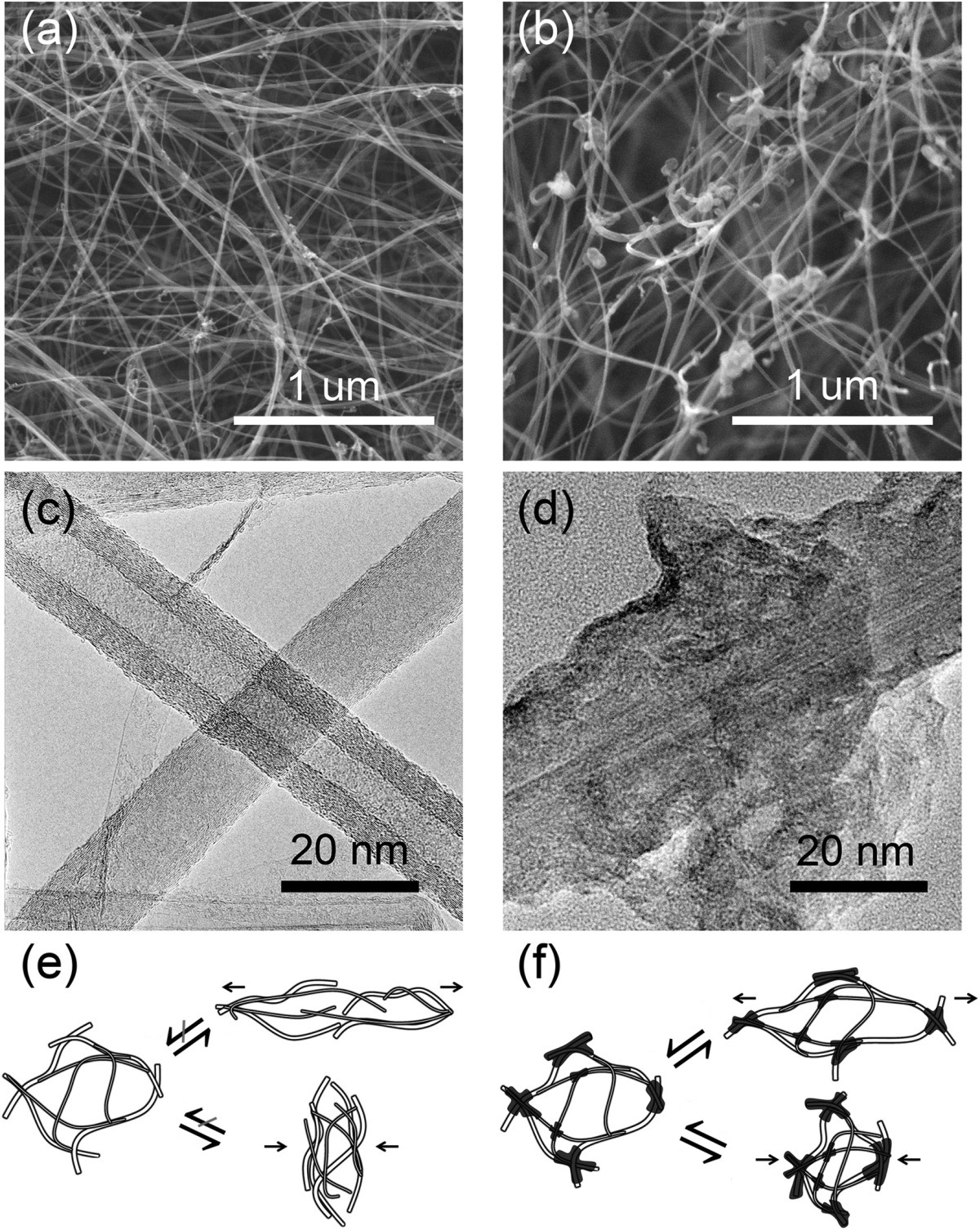}
\caption{Structural comparison between the CNT gels before and after joint welding: (a,b) SEM images, (c,d) TEM images,
and (e,f) schematics of network deformation under stretching and compressing \cite{wang.h_2017}.}
\label{fig:postCVD}
\end{figure}

Similarly, the weak connections can be strengthened by joint-welding with a post CVD treatment \cite{wang.h_2017}. The
raw CNT gel was obtained by stacking up the continuous CNT networks grown via an injection CVD (iCVD), also called a
floating catalyst CVD \cite{li.yl_2004, motta.m_2005, han.y_2015, zou.jy_2016, zou.jy_2017}. The raw gel showed a low
density of 1.78 \si{\mgcc}, and could not recover the original shape even upon slight compressing. To improve the
elasticity, a second CVD process was performed to fix the connection nodes in the 3D CNT networks. By tuning the
treatment time for the post CVD, the gel's density increased up to 4 and 12.75 \si{\mgcc} after 2- and 4-h treatments,
respectively. The CNTs in the untreated gel were free to slide with each other upon stretching and compressing,
resulting in an irreversible plastic deformation. After the treatment, amorphous carbon was deposited at the nodes,
effectively welding the CNT connections, as shown in {\bf Figure \ref{fig:postCVD}}a--d. As a result, the post-treated
gels became more and more elastic and could fully recover upon release from a compressive strain of 98\%. Due to the
increased elasticity, the compressive loss tangent at 1 Hz gradually decreased from $\sim$0.09 to $\sim$0.04 after a
treatment time of 3 h.

For these CNT gels, temperature-invariant viscoelasticity has become a common feature \cite{xu.m_2010, xu.m_20111,
kim.kh_2017, wang.h_2017}, different from traditional viscoelastic polymers which degrade in performance at elevated
temperatures. This is because CNTs can withstand high temperatures without any significant degradation
\cite{koratkar.n_2002}. Typically, oxidation of MWCNTs in air begins at 600 \si{\celsius}. The porous nature of CNT gel
also allows for rapid and efficient heat dissipation, which prevents significant head accumulation. Further, the thermal
stability of CNT gel depends on the contents of catalyst particles. The viscoelasticity can be unvaried up to 1200
\si{\celsius} for pure CNT gels while the existence of catalyst materials significantly hinders the up limit to just
$\sim$400 \si{\celsius} \cite{xu.m_2010}.

Besides the experimental studies described above, several molecular dynamics (MD) simulations have been reported to
understand the dissipation mechanism in CNT gels \cite{cranford.sw_2010, xie.b_2011, yang.xd_2011, li.y_2012,
li.y_20121, chen.h_2016}. These studies can provide us an in-depth understanding of the viscoelasticity.

\begin{figure}[t!]
\centering
\includegraphics[width=0.40\textwidth]{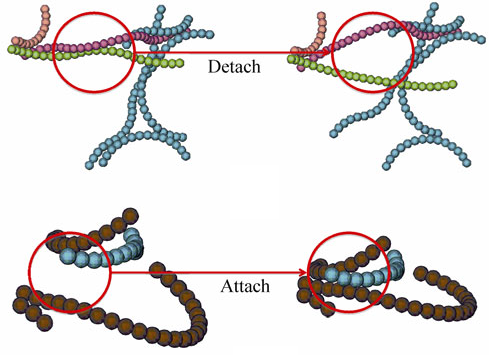}
\caption{Sudden detachment and attachment of individual CNTs during the loading process, which transforms elastic energy
into heat \cite{yang.xd_2011}.}
\label{fig:MDziping}
\end{figure}

Zipping and unzipping have been confirmed by a coarse-grained MD study \cite{li.y_2012}. In the coarse-grained model,
each CNT is mapped as a multi-bead chain. The intra-chain elasticity is described by a bond stretching and a bond angle
bending term, reproducing the tube's modulus and bending rigidity. The inter-chain binding is captured by a
Lennard-Jones type pair interaction between the coarse-grained beads. Simulations revealed that the temperature- and
frequency-invariant hysteresis of CNT gels \cite{xu.m_2010} is due to unstable detachments/attachments of individual
CNTs in the system induced by the vdW interactions \cite{yang.xd_2011}, as shown in {\bf Figure \ref{fig:MDziping}}.
Simulations also showed that the mechanical properties of such an entangled CNT network depend on the mass density of
the system. However, such single zipping/unzipping mechanism is not enough to describe the frequency and density
dependence on the viscoelasticity, as discussed below. More delicate simulation model is necessary, where other physical
processes like sliding between CNTs, rotation and bending of CNTs, and bundling of CNTs should be included
\cite{xie.b_2011}.

\subsection{Densified CNT Assemblies}
\label{sec:densified}

\begin{figure*}[t!]
\centering
\includegraphics[width=0.95\textwidth]{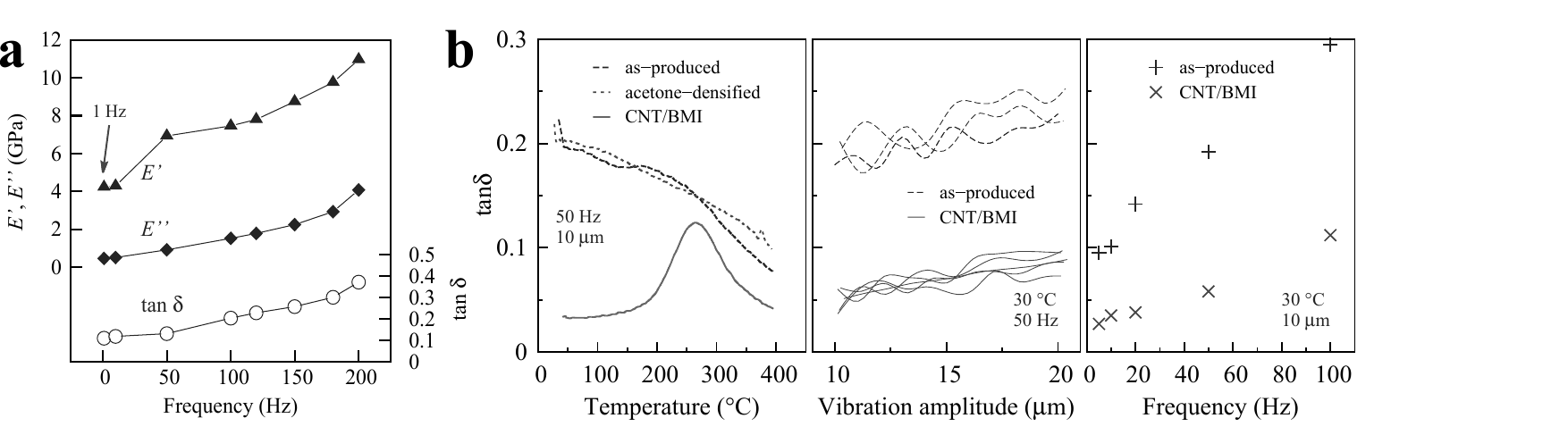}
\caption{(a) Dynamic mechanical properties ($E'$, $E''$, and $\tan\delta$) of iCVD CNT films as functions of frequency
\cite{liu.ql_2015}. (b) Loss tangent of various iCVD films as functions of temperature, vibration amplitude, and
frequency, from tension-mode DMA tests \cite{han.y_2015}.}
\label{fig:iCVDfilm}
\end{figure*}

CNT films are more 2D-like than CNT gels which are mostly 3D assemblies. (CNT sheets are another 2D assembly, with a
thickness less than tens to hundreds of nanometer, while CNT films usually refer to the assembly with a thickness over 1
\si{\um}.) The paper-like CNT films are also called buckypaper \cite{whitby.rld_2008, wang.d_2008}. Actually, it might
be difficult to distinguish CNT films from CNT gels when their thicknesses are in the micrometer ranges. However, CNT
films are often much more densified than CNT gels, and some films are just directly obtained by liquid densification on
CNT gels. Therefore, the volumetric mass density is a nice criterion. For example, densities of 0.15--0.21 \si{\gcc}
\cite{thevamaran.r_2017}, 0.26--0.42 \si{\gcc} \cite{whitby.rld_2008}, 0.47--0.51 \si{\gcc} \cite{liu.ql_2015},
0.54--0.62 \si{\gcc} \cite{wang.d_2008}, 0.50--0.85 \si{\gcc} \cite{wang.yj_2015}, 0.84--1.04 \si{\gcc}
\cite{zhang.lw_2015}, 1.21--1.35 \si{\gcc} \cite{wang.yj_2015, yu.xp_2016}, and 0.81--1.39 \si{\gcc}
\cite{zhang.l_2012}, have been reported recently. CNT films can also be used in sandwich beams to enhance the damping
performance \cite{ji.yg_2006}, similar to the application of CNT forests \cite{koratkar.n_2002}.

Although the CNTs are much more densified in these films, the entanglement feature of the CNTs is not different from the
undensified gels \cite{han.y_2015}. Therefore, the vibration reduction ability can be quite similar to the gels except
that the films also exhibit high modulus and strength. {\bf Figure \ref{fig:iCVDfilm}}a shows the dynamic properties of
the iCVD films at different vibration frequencies, from the tension-mode DMA tests \cite{liu.ql_2015}. At a low
frequency of 1 Hz, the loss tangent $\tan\delta$ was 0.11, smaller than the undensified gel ($\tan\delta\approx 0.3$ in
a wide frequency range of 0.1--100 Hz) \cite{xu.m_2010}, but much higher than many polymers whose Young's modulus is in
the similar range of several GPa, see Figure \ref{fig:lossmap}. Interestingly, the damping ratio increases monotonically
with frequency, nearly in a linear way, different from CNT gels. Furthermore, it is also surprising that the storage
modulus $E'$ could increase more than two-fold by increasing the frequency from $<$10 Hz to 200 Hz. Although the
zipping/unzipping of CNT connection nodes and the sliding friction between CNTs were found to play roles here, there
still should be some other issues to determine such frequency dependences in the damping ratio and storage modulus.

Besides the densification control, designing the intertube interactions is also efficient to tune the damping property.
For another group of iCVD films, the as-produced film (pre-densified by ethanol) exhibited $\tan\delta\approx 0.1$--0.3
in a frequency range of 5--100 Hz (Figure \ref{fig:iCVDfilm}b) \cite{han.y_2015}, higher than the previous report
\cite{liu.ql_2015}, despite of the introduction of a post densification by acetone. As compared to the previous group of
iCVD films, the CNTs here were less-walled (mainly double-walled) and the bundle size was also smaller (there was just
about 50 tubes in a CNT bundle). We suspect that the number of CNT connection nodes could determine the damping
performance at low frequencies, while the intertube interfaces (contact areas) inside CNT bundles could be triggered to
play roles with increasing the frequency.

Introducing polymers between CNTs is a widely used strategy to strengthen CNT assemblies \cite{cheng.qf_2009,
cheng.qf_2010, wang.x_20131, hu.dm_2017}. After the infiltration of BMI resins and post stretching and curing
treatments, high strength CNT/BMI composite films can be obtained \cite{han.y_2015}. Although the strength and modulus
were significantly increased for the composite film, the damping capacity was still non-negligible. For example, the
loss tangent was still larger than 0.05 at 50 Hz. This means that the CNT-BMI interfaces and CNT-CNT contacts inside
bundles still dissipate a certain vibration energy. Furthermore, the introduction of BMI caused the composite film to
show a clear temperature dependence due to the glass transition (Figure \ref{fig:iCVDfilm}b).

\begin{figure}[t!]
\centering
\includegraphics[width=0.48\textwidth]{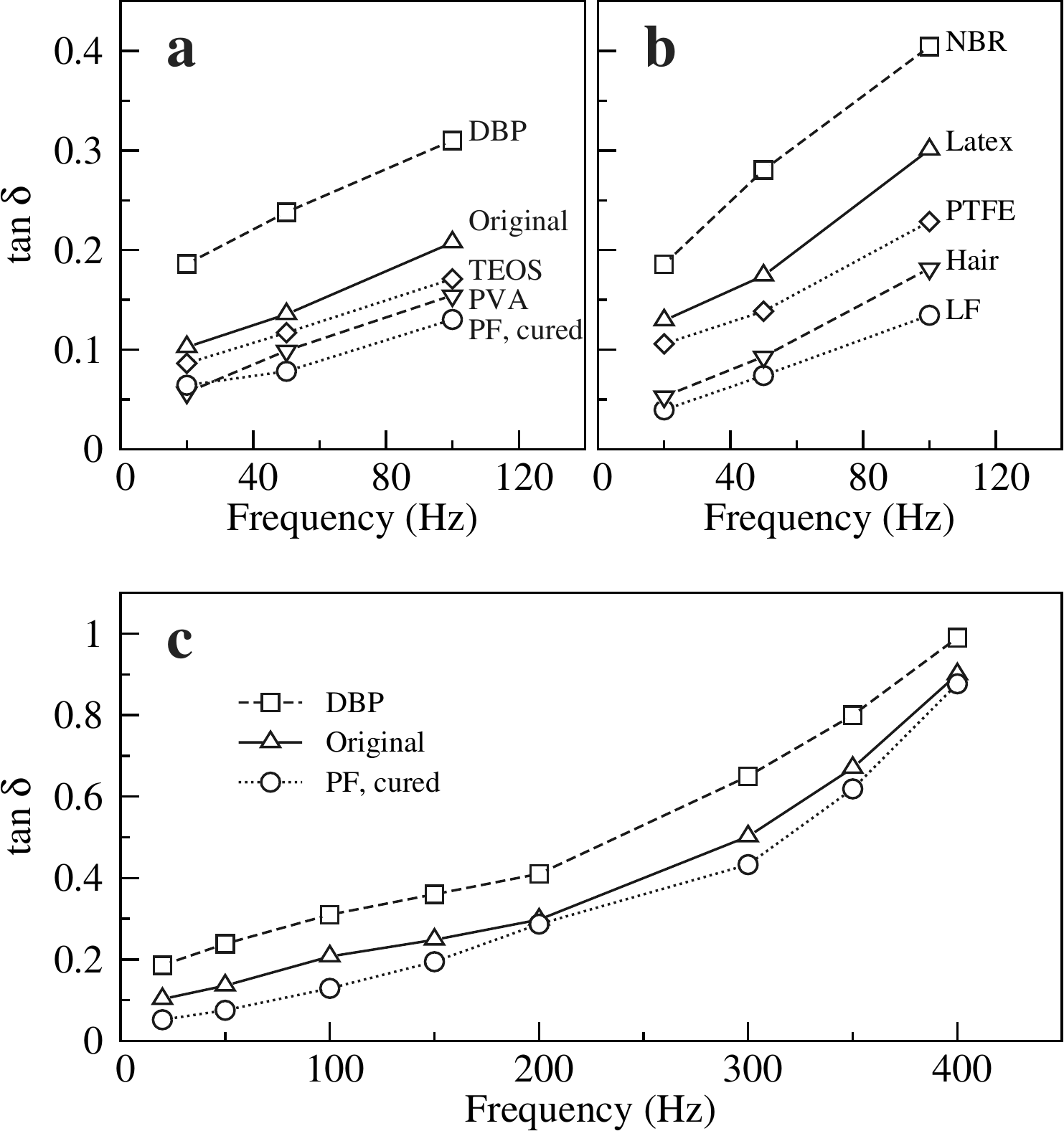}
\caption{Tuning the damping performance of CNT films by different organic compounds. (a) DBP infiltration can enhance
the loss tangent, while other compounds suspend the damping performance. (b) The damping performance of various
materials in our daily life. (c) A comparison between DPB-infiltrated, PF-infiltrated, and original CNT films in a wide
range of frequency.}
\label{fig:CNTfilms}
\end{figure}

{\bf Figure \ref{fig:CNTfilms}} provides a direct comparison in loss tangent for different treatments on densified CNT
networks, where dibutyl phthalate(DBP), tetraethyl orthosilicate (TEOS), poly(vinyl alcohol) (PVA), and phenol
formaldehyde (PF) were infiltrated into CNT network (see Supporting Information for the infiltration details). Due to
the different densification level, the original CNT film exhibited a loss tangent of 0.103, 0.136, and 0.208 at 20, 50,
and 100 Hz, respectively, slightly smaller than that shown in Figure \ref{fig:iCVDfilm}b. (Such difference could be
ascribed to the different wall thickness of the CNTs, see Supporting Information.) After the DBP infiltration, the loss
tangent was remarkably enhanced, up to 0.186, 0.238, and 0.310 at these frequencies. On the contrary, TEOS and PVA
slightly reduced the damping performance, and the cured PF resulted in the smallest loss tangent (0.064, 0.078, and
0.130). Clearly, the polymer infiltration is an efficient way to tune the damping capacity, in a wide range that can
reflect various materials in our daily life, such as nitrile butadiene rubber (NBR), latex film, polytetrafluoroethylene
(PTFE) film, human hair, and laminating film (LF), see Figure \ref{fig:CNTfilms}b. (All the film samples were cut into a
size of 15 mm $\times$ 0.5 mm, and tested in tensile mode DMA.)

The SEM characterization of the assembly structure provides evidences to understand the role of polymer infiltration
(Figure S1, Supporting Information). DBP molecules can not only infiltrate into CNT bundles but also adhere the
connection nodes. Such infiltration brings more viscous characteristics to CNT bundles and their connections, and thus
improves greatly the damping performance. On the contrary, the cured PF network is rather much more rigid and elastic,
and can store more energy as reflected by the reduced loss tangent.

Nevertheless, all these CNT films exhibited clear frequency dependence. By extending the vibration frequency up to 400
Hz, the loss tangent increased up to 0.900, 0.990, and 0.877 for the original, DBP-infiltrated, and PF-infiltrated films
(Figure \ref{fig:CNTfilms}c). Such frequency-dependent energy dissipation is strongly related to the intrinsic viscosity
of the constituents according to the Kelvin-Voigt model. This means that the simple zipping/unzipping behavior can not
fully capture the dissipation mechanism. For example, the wall thickness of CNT was found to play interesting roles in
the dynamic properties. Under the same iCVD process, the slight difference in growth condition could cause the tube's
wall number to change from 2--3 to around 10 \cite{han.y_2015, zou.jy_2016, zou.jy_2017, wang.h_2017}. By testing the
two different iCVD films, composed by the few-walled and multi-walled CNTs, a clear thickness dependence was observed.
For example, the few-walled films exhibited a loss tangent of 0.327 at 200 Hz, while such value was just 0.207 for the
multi-walled films (see Supporting Information for more details). We suspect that the wall thickness can affect the
intrinsic viscosity of the CNT bundle and that the higher flexibility and bendability of few-walled CNTs can also induce
energy dissipation during their structural deformation.

\subsection{CNT Fibers}

Vibration reduction of fiber-reinforced composites have been investigated for decades \cite{gibson.rf_1976,
reed.ke_1980, umadevi.l_1997, chandra.r_1999, kumar.ks_2014, carponcin.d_2015}, and the filler-matrix interface or
interphase region adjacent to fiber surface are found to effectively modify the viscoelastic nature of matrix, either
increasing or decreasing. Different from traditional fibers which are usually a solid structure without internal
interfaces, CNT fibers (indeed, not yarns) are a continuous length of interlocked nano-sized filaments of CNT bundles
\cite{zhao.jn_2015}. As results, the interfacial properties between CNT fiber and polymer matrix can be totally
different \cite{zu.m_2012, liu.yn_2013, lei.cs_2016}, and the fiber's vibrational mechanical property shows new features
as compared with the traditional fibers.

\begin{figure}[t!]
\centering
\includegraphics[width=0.48\textwidth]{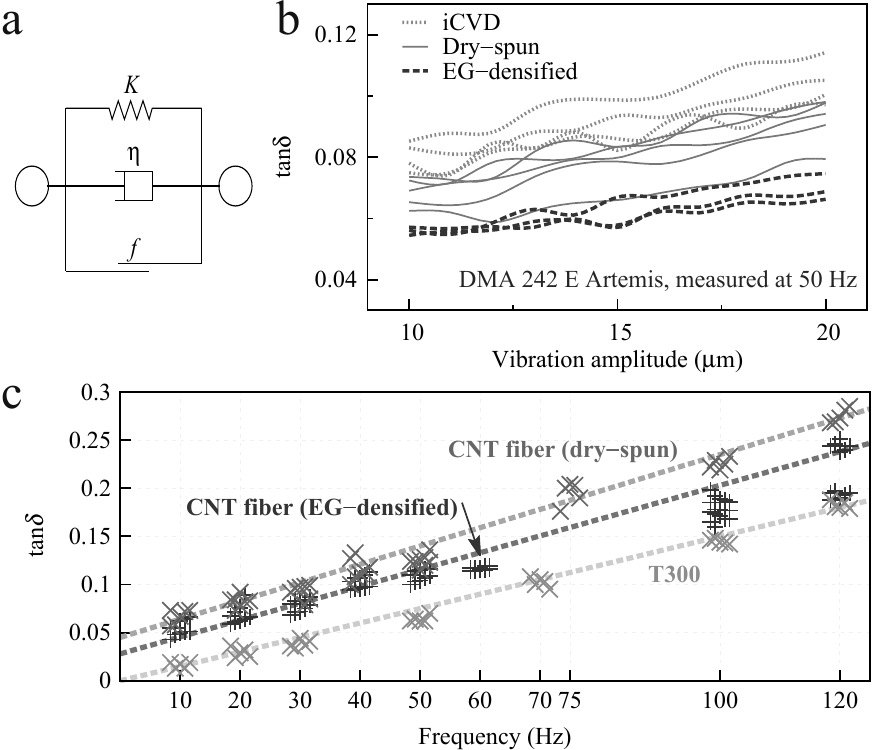}
\caption{Dynamic properties of CNT fibers \cite{zhao.jn_2015}. (a) Schematic of a modified Kelvin-Voigt model. (b)
$\tan\delta$ increase slightly with vibration amplitude, for different CNT fibers. (c) The frequency dependence of
$\tan\delta$ for CNT fibers, with a comparison to T300 carbon fiber.}
\label{fig:fiberDMA}
\end{figure}

\begin{figure*}[t!]
\centering
\includegraphics[width=0.80\textwidth]{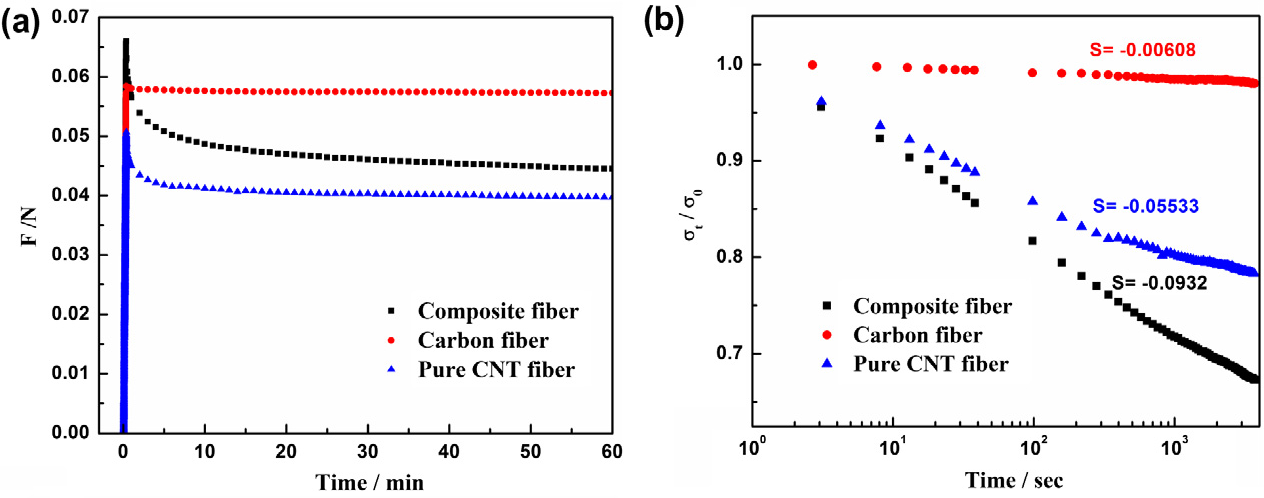}
\caption{Comparisons of relaxation behavior of carbon fiber, pure CNT fiber and CNT/epoxy composite fiber
\cite{zu.m_2013}. (a) Force as a function of time during loading and relaxation. (b) Stress ratio $\sigma_t/\sigma_0$ as
a function of time.}
\label{fig:fiberRelax}
\end{figure*}

Owing to the rich inter-filament contacts, the analysis of the dissipation mechanism for CNT fiber is quite reminiscent
to the case of staple yarns \cite{murayama.t_1979}. A modified Kelvin-Voigt model is used here, where a linear elastic
spring $K$ which stores energy (i.e., fiber's modulus $E$), an energy dissipation mechanism associated with the internal
viscosity $\eta$ of the filaments, and an energy dissipation mechanism of the coulomb form associated with the sum of
inter-bundle friction $f$ are connected in parallel ({\bf Figure \ref{fig:fiberDMA}}a). One should notice that due to
the hierarchical feature, $\eta$ is also related to the intertube friction within a CNT bundle. According to this model,
an effective damping coefficient $\eta* = \eta + 4 f / A \omega$ can be obtained, where $A$ and $\omega$ are the
vibration amplitude and frequency. Thus, the measured loss tangent contains frequency- and friction-dependent
components, by $\tan\delta = \eta\omega/K + 4f/AK$ \cite{zhao.jn_2015}.

Figure \ref{fig:fiberDMA}b compares the loss tangent for different CNT fibers. The iCVD fiber showed the highest
$\tan\delta$, even more than that of of the dry-spun fiber (spun from CNT forests). This means, the dissipation by the
CNT connection nodes might be missing in this modified Kelvin-Voigt model. Nevertheless, due to the high alignment of
the CNTs, such model is very useful to understand the effects of densification and interfacial locking. As one can find
from Figure \ref{fig:fiberDMA}b and c, the densified fibers by ethylene glycol (EG) exhibited the lower damping
capacity, and the loss tangent generally increased with vibration amplitude and frequency, in agreement with the model
analysis. Furthermore, the $\tan\delta$--$\omega$ plots have revealed the main difference between CNT fiber and carbon
fiber.

On one hand, the different $\tan\delta$-to-$\omega$ slopes captured the difference in elastic modulus. The T300 carbon
fiber had a modulus up to 204--238 GPa, while it was just 45--59 and 72--101 GPa for the dry-spun and EG-densified CNT
fibers, respectively. On the other hand, the nonzero extrapolated $\tan\delta$ at $\omega=0$ means that even without
external vibrations, there is always energy loss at the CNT interfaces. For the dry-spun and EG-densified fibers,
$\tan\delta|_{\omega = 0} = 0.045$ and 0.028, in agreement with the hysteresis test at 0.1 Hz where a dry-spun fiber
exhibited a loss ratio of 0.05 \cite{hehr.a_2014}. (Such ratio corresponded to the energy loss per stress-strain cycle,
for example, of around 30\% \cite{zhang.m_2004}.) This difference reflected their different frictional feature in the
assembly structures. After the densification, the improved packing density increased the inter-bundle contact area and
enhanced the interfacial frictional force. However, the tendency of CNT slippage was thus remarkably suppressed.
Therefore, $f$ in the model describes the level of overall frictional energy dissipation, reflecting not only the
magnitude of frictional force but also the total number of interfaces where sliding phenomenon might take place.

The introduction of cross-linking BMI network in CNT fibers can strongly suspend the sliding energy dissipation, as
reflected by the reduced loss tangent, which varied from 0.122 to 0.157 with an average of 0.136 at 100 Hz, in also
agreement with the study on CNT/BMI composite films ({Figure} \ref{fig:iCVDfilm}b).

According to such philosophy, the introduction of fiber-fiber contacts can increase the sliding dissipation. For
example, by multi-plying CNT fibers together, a real CNT yarn can be obtained. Due to the increased inter-fiber
contacts, a 400-ply CNT yarn showed a high loss tangent of $\tan\delta=0.185$ at 10 Hz, under a linear mass density of
65 tex. Such value is already higher than the cotton yarns ($\tan\delta = 0.13$--0.176 at 10 Hz).

For CNT fibers and multi-plied CNT yarns, the creep, creep recovery and stress relaxation behaviors are also of great
interests \cite{zhang.m_2004, zu.m_2013, misak.he_2013}. For a two-ply CNT yarn, the stress relaxed no more than 15\%
when it was held for 20 h at 6\% strain, and the majority of the stress relaxation occurred within the first 20 min
\cite{zhang.m_2004}. For a pure CNT fiber, there was significant load decay during the first 4 min of the relaxation
process, and the load dropped by $\sim$32\% after 18 h at constant strain \cite{zu.m_2013}. The CNT/epoxy composite
fiber showed the similar relaxation behavior, while the load of the carbon fiber was almost constant after 1 h and
decreased by only 5.4\% even after 18 h ({\bf Figure \ref{fig:fiberRelax}}a). This is because when a CNT fiber is held
at a constant strain, slippage among CNT bundles within the fiber can take place, resulting in a gradual load drop.
Figure \ref{fig:fiberRelax}b presents the stress relaxation plot of stress ratio, between the stress at a specific time
$\sigma_t$ and the maximum stress $\sigma_0$, as a function of time for the three specimens. The stress relaxation rate
of the composite fiber was higher than that of the pure fiber at the same initial strain level, even though the
composite fiber retained a higher load after 1 h. The possible mechanism could be the insufficient interaction between
CNT and epoxy; the polymer induces high efficient load transfer between CNTs and thus improves the modulus, but the
additional sliding at the CNT-epoxy interfaces could still occur during the relaxation process.

Differently, owing to the inter-fiber contacts, the multi-ply CNT yarns can show three distinct regions in the
stress-strain relationship when subjected to monotonic tensile load \cite{misak.he_2013}. Initially, the friction
between CNT fibers does not allow significant slippage of the yarns. In the intermediate stage, CNT fibers slip relative
to each other that results in the reduction of the helix angle up to a certain value. Thereafter in the third region,
stretching of yarns occurs causing further tightening.

\section{Discussion}
\label{sec:discuss}

The study on CNT-reinforced composites can cast lights on the analysis of CNT assemblies. In a recent study, to show the
effect of inherent damping of matrix and interfacial slip, with minimizing other energy dissipation mechanisms, such as
matrix tearing and plasticity, the CNTs were dominantly oriented parallel to the loading direction via hot-drawing
\cite{gardea.f_2015}. In such CNT/PS composites, a ``slip-stick'' mechanism is found as the source for energy
dissipation. At the CNT-polymer interface, debonding takes place upon a critical shear strain, and the initial bonds and
interactions are broken. Broken covalent bonds are irreversible, whereas the mechanical interlocking and vdW
interactions are reversible. Above the critical strain, when dynamic loading is applied, the interlocking and vdW
interactions can be repeatedly reformed and broken, corresponding to stick and slip occurring at the interface.

However, the mechanisms for energy dissipation differ greatly from sample to sample, and also vary when the same sample
is under different strains. For a CNT/epoxy composite containing $\sim$50\% volume fraction of CNTs \cite{suhr.j_2005},
there is a critical threshold of shear strain of about 5\%. When the strain is lower than the threshold, very few
tube-tube contacts reach the critical stress for interfacial slip, and the energy loss is small. At the critical strain,
intertube sliding is activated, particularly for CNT bundles closely aligned to the shear direction, and the damping
shows a sharp increase. Peak damping is achieved for about 10\% strain when most CNTs in the film will slip. The slight
decrease of loss modulus at larger strains ($>$10\%) could be related to dynamic friction between the CNT shells of an
MWCNT which is shown to be lower than static friction.

For CNT assemblies, the constituent nanotubes are built up by means of bundling, stacking, cross-linking, entangling,
and possible intertube covalent bonding. The interfacial engineering obviously determines the statical and dynamical
mechanical properties of the assembly materials \cite{zhang.xh_20171}. By learning from the CNT-reinforced composites,
we suggest the following issues for future consideration:

(1) Connection nodes of nano-assemblies. The connection here is restricted to the contact between non-entangled CNTs
(the entanglement is described below). The connection nodes are very essential to provide mechanical support for CNT
gels, sponges, and buckypapers. Zipping and unzipping at the nodes, usually the contacts of non-entangled CNTs
\cite{yu.xp_2016}, have been considered to contribute mostly the viscoelasticity. Welding these nodes, including
graphene coating \cite{kim.kh_2012}, deposition of amorphous carbon \cite{wang.h_2017}, and strapping with long-chain
polymers \cite{zou.jy_2016}, can suspend the zipping/unzipping and thus improve the elasticity or allow super-high
stretchability. For future development, strategies to enhance the node's damping capacity are still of great importance.
The design of CNT-polymer connection nodes is also helpful to introduce more damping ways.

(2) Entanglement of CNTs. Entanglement is another assembly parameter, and does also result in connection nodes in CNT
networks, but such nodes do not show zipping/unzipping but a ``stick-slip'' sliding motion upon stretching or
compressing \cite{bhushan.b_2008, bhushan.b_20081}. The entanglement also causes self-folding of two intersecting CNTs
and the intersecting point transfers load from one CNT to the other \cite{lu.wb_20111}. Due to such ``intersecting''
entanglement, CNT network shows also high viscoelasticity \cite{liu.ql_2015} and is highly stretchable \cite{han.y_2015,
zou.jy_2016}. Notice that, in many studies of the iCVD CNT assemblies \cite{liu.ql_2015, han.y_2015, zou.jy_2016,
wang.h_2017}, the ``intersecting'' entanglement and ``non-entangled'' contact co-exist, but unfortunately many previous
studies just payed attention to the latter one.

(3) Intertube covalent and non-covalent interactions. This is an interfacial point of view. The covalent bonding
benefits the load transfer between CNTs, especially for highly aligned tubes. The similar treatment is the introduction
of cross-linked polymer network inside the aligned CNTs. These treatments make as many as the CNTs response
simultaneously to strains \cite{kis.a_2004, ma.wj_2009}, resulting in significant improvement in material's modulus.
However, the tendency of sliding between CNTs is suspended, reducing the frictional energy dissipation
\cite{zhao.jn_2015}. Therefore, a certain level of non-bonded CNT contacts is still welcome towards high viscoelastic
assemblies.

(4) Inherent viscoelasticity of CNT and CNT bundles, the basic components of the assembly. Although CNT is a
super-elastic material with a modulus up to TPa scale, it can still cost energy by structural deformation such as
buckling and collapsing. Thus, the tube itself can exhibit inherent viscoelasticity in some cases like the compression
on CNT forests \cite{pathak.s_2009, bradford.pd_2011, cao.ch_2011, li.yp_2015}. Thus, there could be a strategy to make
CNT assemblies more viscoelastic by modifying the nanotube structure, e.g., introduction of defects that cause more
dissipation under interfacial sliding, changing straight tube structure into helical \cite{zhang.q_20101, zhao.mq_2012},
and so on.

\begin{figure}[t!]
\centering
\includegraphics[width=0.48\textwidth]{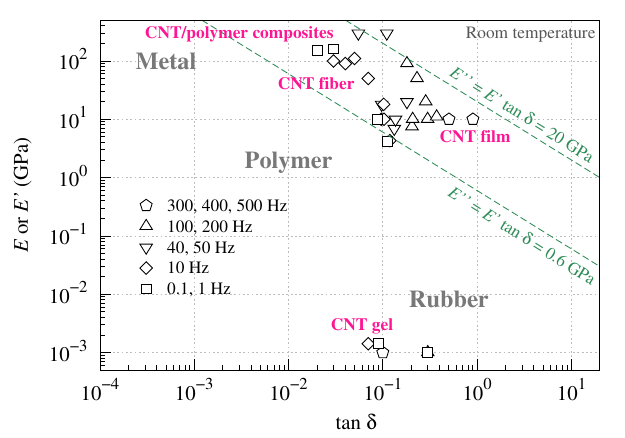}
\caption{Modulus-loss map for CNT assembly materials. The Young's modulus $E$ and storage modulus $E'$ are mixed as they
are collected from different literature and the present study. These CNT materials are CNT/BMI and CNT/polyimide
composite films \cite{han.y_2015, hu.dm_2017}, CNT films densified from CNT gels \cite{liu.ql_2015, han.y_2015} and
stacked-up using CNT sheets \cite{yu.xp_2016}, CNT fibers \cite{zhao.jn_2015}, CNT gels \cite{xu.m_2010, kim.kh_2012,
wang.h_2017}, and high-density CNT forests \cite{pathak.s_2009}. The two reference lines represent $E'' = E' \tan\delta
= 0.6$ GPa and 20 GPa, respectively.}
\label{fig:CNTlossmap}
\end{figure}

From the development, a new stiffness (modulus)-loss map for CNT assembly materials is provided in {\bf Figure
\ref{fig:CNTlossmap}}. So far, there are just countable studies on CNT assembly materials. However, due to the rich
interfacial areas between CNTs and between CNT and polymers, a high loss tangent can be realized at a high storage
modulus. For most materials, even practical polymer damping layers, $E' \tan \delta$ is less than 0.6 GPa (Figure
\ref{fig:lossmap}), while for CNT assemblies, the upper limit of $E' \tan \delta$ could be about 20 GPa. Clearly, such
performance is a big step towards the ``best  performance'', and such performance can only be observed in some extreme
viscoelastic metals and composites \cite{wang.yc_2004}.

Figure \ref{fig:CNTlossmap} also demonstrates an interesting phenomenon, that with densifying CNT gels into films, the
modulus increases by orders of magnitude while the loss tangent could be uninfluenced. As the CNT connection nodes,
intertube interaction, CNT alignment, and entanglement are totally different, the major contribution to the
viscoelasticity should also vary greatly. Therefore, we still should pay great efforts to CNT assembly materials.

\section{Conclusion}

We have introduced basic concepts of viscoelasticity and discussed recent progresses on the dynamic properties of CNT
assembly materials. Due to the rich intertube contacts and various ways to dissipate energy, CNT assemblies have shown
great advantages for developing novel high-damping materials, which can also show high strength or modulus for
engineering applications. Towards future development of such multifunctional assembly materials, the design and control
of interface structure, interactions, and connection ways between CNTs are still the key problems.

\begin{acknowledgements}
We thank financial supports from the National Natural Science Foundation of China (51561145008, 21503267, 11302241,
11404371, 21473238), Youth Innovation Promotion Association of the Chinese Academy of Sciences (2015256), and National
Key Research and Development Program of China (2016YFA0203301).
\end{acknowledgements}

\bibliography{ref}

\begin{thebibliography}{115}%
\makeatletter
\providecommand \@ifxundefined [1]{%
 \@ifx{#1\undefined}
}%
\providecommand \@ifnum [1]{%
 \ifnum #1\expandafter \@firstoftwo
 \else \expandafter \@secondoftwo
 \fi
}%
\providecommand \@ifx [1]{%
 \ifx #1\expandafter \@firstoftwo
 \else \expandafter \@secondoftwo
 \fi
}%
\providecommand \natexlab [1]{#1}%
\providecommand \enquote  [1]{``#1''}%
\providecommand \bibnamefont  [1]{#1}%
\providecommand \bibfnamefont [1]{#1}%
\providecommand \citenamefont [1]{#1}%
\providecommand \href@noop [0]{\@secondoftwo}%
\providecommand \href [0]{\begingroup \@sanitize@url \@href}%
\providecommand \@href[1]{\@@startlink{#1}\@@href}%
\providecommand \@@href[1]{\endgroup#1\@@endlink}%
\providecommand \@sanitize@url [0]{\catcode `\\12\catcode `\$12\catcode
  `\&12\catcode `\#12\catcode `\^12\catcode `\_12\catcode `\%12\relax}%
\providecommand \@@startlink[1]{}%
\providecommand \@@endlink[0]{}%
\providecommand \url  [0]{\begingroup\@sanitize@url \@url }%
\providecommand \@url [1]{\endgroup\@href {#1}{\urlprefix }}%
\providecommand \urlprefix  [0]{URL }%
\providecommand \Eprint [0]{\href }%
\providecommand \doibase [0]{http://dx.doi.org/}%
\providecommand \selectlanguage [0]{\@gobble}%
\providecommand \bibinfo  [0]{\@secondoftwo}%
\providecommand \bibfield  [0]{\@secondoftwo}%
\providecommand \translation [1]{[#1]}%
\providecommand \BibitemOpen [0]{}%
\providecommand \bibitemStop [0]{}%
\providecommand \bibitemNoStop [0]{.\EOS\space}%
\providecommand \EOS [0]{\spacefactor3000\relax}%
\providecommand \BibitemShut  [1]{\csname bibitem#1\endcsname}%
\let\auto@bib@innerbib\@empty
\bibitem [{\citenamefont {Chung}(2001)}]{chung.ddl_2001}%
  \BibitemOpen
  \bibfield  {author} {\bibinfo {author} {\bibfnamefont {D.~D.~L.}\
  \bibnamefont {Chung}},\ }\href {\doibase 10.1023/A:1012999616049} {\bibfield
  {journal} {\bibinfo  {journal} {J. Mater. Sci.}\ }\textbf {\bibinfo {volume}
  {36}},\ \bibinfo {pages} {5733} (\bibinfo {year} {2001})}\BibitemShut
  {NoStop}%
\bibitem [{\citenamefont {Lakes}(2002)}]{lakes.rs_2002}%
  \BibitemOpen
  \bibfield  {author} {\bibinfo {author} {\bibfnamefont {R.~S.}\ \bibnamefont
  {Lakes}},\ }\href {\doibase 10.1106/002199802023538} {\bibfield  {journal}
  {\bibinfo  {journal} {J. Compos. Mater.}\ }\textbf {\bibinfo {volume} {36}},\
  \bibinfo {pages} {287} (\bibinfo {year} {2002})}\BibitemShut {NoStop}%
\bibitem [{\citenamefont {Luo}\ \emph {et~al.}(2015)\citenamefont {Luo},
  \citenamefont {Duan}, \citenamefont {Xian}, \citenamefont {Li},\ and\
  \citenamefont {Zhao}}]{luo.jl_2015}%
  \BibitemOpen
  \bibfield  {author} {\bibinfo {author} {\bibfnamefont {J.}~\bibnamefont
  {Luo}}, \bibinfo {author} {\bibfnamefont {Z.}~\bibnamefont {Duan}}, \bibinfo
  {author} {\bibfnamefont {G.}~\bibnamefont {Xian}}, \bibinfo {author}
  {\bibfnamefont {Q.}~\bibnamefont {Li}}, \ and\ \bibinfo {author}
  {\bibfnamefont {T.}~\bibnamefont {Zhao}},\ }\href {\doibase
  10.1080/15376494.2012.736052} {\bibfield  {journal} {\bibinfo  {journal}
  {Mech. Adv. Mater. Struct.}\ }\textbf {\bibinfo {volume} {22}},\ \bibinfo
  {pages} {224} (\bibinfo {year} {2015})}\BibitemShut {NoStop}%
\bibitem [{\citenamefont {Zhou}\ \emph {et~al.}(2016)\citenamefont {Zhou},
  \citenamefont {Yu}, \citenamefont {Shao}, \citenamefont {Zhang},\ and\
  \citenamefont {Wang}}]{zhou.xq_2016}%
  \BibitemOpen
  \bibfield  {author} {\bibinfo {author} {\bibfnamefont {X.~Q.}\ \bibnamefont
  {Zhou}}, \bibinfo {author} {\bibfnamefont {D.~Y.}\ \bibnamefont {Yu}},
  \bibinfo {author} {\bibfnamefont {X.~Y.}\ \bibnamefont {Shao}}, \bibinfo
  {author} {\bibfnamefont {S.~Q.}\ \bibnamefont {Zhang}}, \ and\ \bibinfo
  {author} {\bibfnamefont {S.}~\bibnamefont {Wang}},\ }\href {\doibase
  10.1016/j.compstruct.2015.10.014} {\bibfield  {journal} {\bibinfo  {journal}
  {Compos. Struct.}\ }\textbf {\bibinfo {volume} {136}},\ \bibinfo {pages}
  {460} (\bibinfo {year} {2016})}\BibitemShut {NoStop}%
\bibitem [{\citenamefont {Chung}(2003)}]{chung.ddl_2003}%
  \BibitemOpen
  \bibfield  {author} {\bibinfo {author} {\bibfnamefont {D.~D.~L.}\
  \bibnamefont {Chung}},\ }\href {\doibase 10.1016/S0925-8388(03)00233-0}
  {\bibfield  {journal} {\bibinfo  {journal} {J. Alloys Compd.}\ }\textbf
  {\bibinfo {volume} {355}},\ \bibinfo {pages} {216} (\bibinfo {year}
  {2003})}\BibitemShut {NoStop}%
\bibitem [{\citenamefont {Chandra}, \citenamefont {Singh},\ and\ \citenamefont
  {Gupta}(1999)}]{chandra.r_1999}%
  \BibitemOpen
  \bibfield  {author} {\bibinfo {author} {\bibfnamefont {R.}~\bibnamefont
  {Chandra}}, \bibinfo {author} {\bibfnamefont {S.~P.}\ \bibnamefont {Singh}},
  \ and\ \bibinfo {author} {\bibfnamefont {K.}~\bibnamefont {Gupta}},\ }\href
  {\doibase 10.1016/S0263-8223(99)00041-0} {\bibfield  {journal} {\bibinfo
  {journal} {Compos. Struct.}\ }\textbf {\bibinfo {volume} {46}},\ \bibinfo
  {pages} {41} (\bibinfo {year} {1999})}\BibitemShut {NoStop}%
\bibitem [{\citenamefont {Khan}\ \emph {et~al.}(2011)\citenamefont {Khan},
  \citenamefont {Li}, \citenamefont {Siddiqui},\ and\ \citenamefont
  {Kim}}]{khan.su_2011}%
  \BibitemOpen
  \bibfield  {author} {\bibinfo {author} {\bibfnamefont {S.~U.}\ \bibnamefont
  {Khan}}, \bibinfo {author} {\bibfnamefont {C.~Y.}\ \bibnamefont {Li}},
  \bibinfo {author} {\bibfnamefont {N.~A.}\ \bibnamefont {Siddiqui}}, \ and\
  \bibinfo {author} {\bibfnamefont {J.-K.}\ \bibnamefont {Kim}},\ }\href
  {\doibase 10.1016/j.compscitech.2011.03.022} {\bibfield  {journal} {\bibinfo
  {journal} {Compos. Sci. Technol.}\ }\textbf {\bibinfo {volume} {71}},\
  \bibinfo {pages} {1486} (\bibinfo {year} {2011})}\BibitemShut {NoStop}%
\bibitem [{\citenamefont {Kumar}\ \emph {et~al.}(2014)\citenamefont {Kumar},
  \citenamefont {Siva}, \citenamefont {Jeyaraj}, \citenamefont {Jappes},
  \citenamefont {Amico},\ and\ \citenamefont {Rajini}}]{kumar.ks_2014}%
  \BibitemOpen
  \bibfield  {author} {\bibinfo {author} {\bibfnamefont {K.~S.}\ \bibnamefont
  {Kumar}}, \bibinfo {author} {\bibfnamefont {I.}~\bibnamefont {Siva}},
  \bibinfo {author} {\bibfnamefont {P.}~\bibnamefont {Jeyaraj}}, \bibinfo
  {author} {\bibfnamefont {J.~T.~W.}\ \bibnamefont {Jappes}}, \bibinfo {author}
  {\bibfnamefont {S.~C.}\ \bibnamefont {Amico}}, \ and\ \bibinfo {author}
  {\bibfnamefont {N.}~\bibnamefont {Rajini}},\ }\href {\doibase
  10.1016/j.matdes.2013.11.039} {\bibfield  {journal} {\bibinfo  {journal}
  {Mater. Des.}\ }\textbf {\bibinfo {volume} {56}},\ \bibinfo {pages} {379}
  (\bibinfo {year} {2014})}\BibitemShut {NoStop}%
\bibitem [{\citenamefont {Zhang}, \citenamefont {Heyne},\ and\ \citenamefont
  {To}(2015)}]{zhang.p_2015}%
  \BibitemOpen
  \bibfield  {author} {\bibinfo {author} {\bibfnamefont {P.}~\bibnamefont
  {Zhang}}, \bibinfo {author} {\bibfnamefont {M.~A.}\ \bibnamefont {Heyne}}, \
  and\ \bibinfo {author} {\bibfnamefont {A.~C.}\ \bibnamefont {To}},\ }\href
  {\doibase 10.1016/j.jmps.2015.06.015} {\bibfield  {journal} {\bibinfo
  {journal} {J. Mech. Phys. Solids}\ }\textbf {\bibinfo {volume} {83}},\
  \bibinfo {pages} {285} (\bibinfo {year} {2015})}\BibitemShut {NoStop}%
\bibitem [{\citenamefont {Liu}, \citenamefont {Ma},\ and\ \citenamefont
  {Zhang}(2011)}]{liu.lq_2011}%
  \BibitemOpen
  \bibfield  {author} {\bibinfo {author} {\bibfnamefont {L.}~\bibnamefont
  {Liu}}, \bibinfo {author} {\bibfnamefont {W.}~\bibnamefont {Ma}}, \ and\
  \bibinfo {author} {\bibfnamefont {Z.}~\bibnamefont {Zhang}},\ }\href
  {\doibase 10.1002/smll.201002198} {\bibfield  {journal} {\bibinfo  {journal}
  {Small}\ }\textbf {\bibinfo {volume} {7}},\ \bibinfo {pages} {1504} (\bibinfo
  {year} {2011})}\BibitemShut {NoStop}%
\bibitem [{\citenamefont {Lu}\ \emph {et~al.}(2012)\citenamefont {Lu},
  \citenamefont {Zu}, \citenamefont {Byun}, \citenamefont {Kim},\ and\
  \citenamefont {Chou}}]{lu.wb_2012}%
  \BibitemOpen
  \bibfield  {author} {\bibinfo {author} {\bibfnamefont {W.}~\bibnamefont
  {Lu}}, \bibinfo {author} {\bibfnamefont {M.}~\bibnamefont {Zu}}, \bibinfo
  {author} {\bibfnamefont {J.-H.}\ \bibnamefont {Byun}}, \bibinfo {author}
  {\bibfnamefont {B.-S.}\ \bibnamefont {Kim}}, \ and\ \bibinfo {author}
  {\bibfnamefont {T.-W.}\ \bibnamefont {Chou}},\ }\href {\doibase
  10.1002/adma.201104672} {\bibfield  {journal} {\bibinfo  {journal} {Adv.
  Mater.}\ }\textbf {\bibinfo {volume} {24}},\ \bibinfo {pages} {1805}
  (\bibinfo {year} {2012})}\BibitemShut {NoStop}%
\bibitem [{\citenamefont {Liu}\ and\ \citenamefont
  {Kumar}(2014)}]{liu.yd_2014}%
  \BibitemOpen
  \bibfield  {author} {\bibinfo {author} {\bibfnamefont {Y.}~\bibnamefont
  {Liu}}\ and\ \bibinfo {author} {\bibfnamefont {S.}~\bibnamefont {Kumar}},\
  }\href {\doibase 10.1021/am405136s} {\bibfield  {journal} {\bibinfo
  {journal} {ACS Appl. Mater. Interfaces}\ }\textbf {\bibinfo {volume} {6}},\
  \bibinfo {pages} {6069} (\bibinfo {year} {2014})}\BibitemShut {NoStop}%
\bibitem [{\citenamefont {Di}\ \emph {et~al.}(2016)\citenamefont {Di},
  \citenamefont {Zhang}, \citenamefont {Yong}, \citenamefont {Zhang},
  \citenamefont {Li}, \citenamefont {Li},\ and\ \citenamefont
  {Li}}]{di.jt_20161}%
  \BibitemOpen
  \bibfield  {author} {\bibinfo {author} {\bibfnamefont {J.}~\bibnamefont
  {Di}}, \bibinfo {author} {\bibfnamefont {X.}~\bibnamefont {Zhang}}, \bibinfo
  {author} {\bibfnamefont {Z.}~\bibnamefont {Yong}}, \bibinfo {author}
  {\bibfnamefont {Y.}~\bibnamefont {Zhang}}, \bibinfo {author} {\bibfnamefont
  {D.}~\bibnamefont {Li}}, \bibinfo {author} {\bibfnamefont {R.}~\bibnamefont
  {Li}}, \ and\ \bibinfo {author} {\bibfnamefont {Q.}~\bibnamefont {Li}},\
  }\href {\doibase 10.1002/adma.201601186} {\bibfield  {journal} {\bibinfo
  {journal} {Adv. Mater.}\ }\textbf {\bibinfo {volume} {28}},\ \bibinfo {pages}
  {10529} (\bibinfo {year} {2016})}\BibitemShut {NoStop}%
\bibitem [{\citenamefont {Zhang}\ \emph {et~al.}(2016)\citenamefont {Zhang},
  \citenamefont {Yu}, \citenamefont {Zhao},\ and\ \citenamefont
  {Li}}]{zhang.xh_2016}%
  \BibitemOpen
  \bibfield  {author} {\bibinfo {author} {\bibfnamefont {X.}~\bibnamefont
  {Zhang}}, \bibinfo {author} {\bibfnamefont {X.}~\bibnamefont {Yu}}, \bibinfo
  {author} {\bibfnamefont {J.}~\bibnamefont {Zhao}}, \ and\ \bibinfo {author}
  {\bibfnamefont {Q.}~\bibnamefont {Li}},\ }in\ \href@noop {} {\emph {\bibinfo
  {booktitle} {{Carbon Nanotubes - Current Progress of their Polymer
  Composites}}}},\ \bibinfo {editor} {edited by\ \bibinfo {editor}
  {\bibfnamefont {M.~R.}\ \bibnamefont {Berber}}\ and\ \bibinfo {editor}
  {\bibfnamefont {I.~H.}\ \bibnamefont {Hafez}}}\ (\bibinfo  {publisher}
  {InTech - Open Access},\ \bibinfo {address} {Rijeka, Croatia},\ \bibinfo
  {year} {2016})\ Chap.~\bibinfo {chapter} {3}, pp.\ \bibinfo {pages}
  {73--93}\BibitemShut {NoStop}%
\bibitem [{\citenamefont {Hashim}\ \emph {et~al.}(2012)\citenamefont {Hashim},
  \citenamefont {Narayanan}, \citenamefont {Romo-Herrera}, \citenamefont
  {Cullen}, \citenamefont {Hahm}, \citenamefont {Lezzi}, \citenamefont
  {Suttle}, \citenamefont {Kelkhoff}, \citenamefont {Mu\~{n}oz Sandoval},
  \citenamefont {Ganguli}, \citenamefont {Roy}, \citenamefont {Smith},
  \citenamefont {Vajtai}, \citenamefont {Sumpter}, \citenamefont {Meunier},
  \citenamefont {Terrones}, \citenamefont {Terrones},\ and\ \citenamefont
  {Ajayan}}]{hashim.dp_2012}%
  \BibitemOpen
  \bibfield  {author} {\bibinfo {author} {\bibfnamefont {D.~P.}\ \bibnamefont
  {Hashim}}, \bibinfo {author} {\bibfnamefont {N.~T.}\ \bibnamefont
  {Narayanan}}, \bibinfo {author} {\bibfnamefont {J.~M.}\ \bibnamefont
  {Romo-Herrera}}, \bibinfo {author} {\bibfnamefont {D.~A.}\ \bibnamefont
  {Cullen}}, \bibinfo {author} {\bibfnamefont {M.~G.}\ \bibnamefont {Hahm}},
  \bibinfo {author} {\bibfnamefont {P.}~\bibnamefont {Lezzi}}, \bibinfo
  {author} {\bibfnamefont {J.~R.}\ \bibnamefont {Suttle}}, \bibinfo {author}
  {\bibfnamefont {D.}~\bibnamefont {Kelkhoff}}, \bibinfo {author}
  {\bibfnamefont {E.}~\bibnamefont {Mu\~{n}oz Sandoval}}, \bibinfo {author}
  {\bibfnamefont {S.}~\bibnamefont {Ganguli}}, \bibinfo {author} {\bibfnamefont
  {A.~K.}\ \bibnamefont {Roy}}, \bibinfo {author} {\bibfnamefont {D.~J.}\
  \bibnamefont {Smith}}, \bibinfo {author} {\bibfnamefont {R.}~\bibnamefont
  {Vajtai}}, \bibinfo {author} {\bibfnamefont {B.~G.}\ \bibnamefont {Sumpter}},
  \bibinfo {author} {\bibfnamefont {V.}~\bibnamefont {Meunier}}, \bibinfo
  {author} {\bibfnamefont {H.}~\bibnamefont {Terrones}}, \bibinfo {author}
  {\bibfnamefont {M.}~\bibnamefont {Terrones}}, \ and\ \bibinfo {author}
  {\bibfnamefont {P.~M.}\ \bibnamefont {Ajayan}},\ }\href {\doibase
  10.1038/srep00363} {\bibfield  {journal} {\bibinfo  {journal} {Sci. Rep.}\
  }\textbf {\bibinfo {volume} {2}},\ \bibinfo {pages} {363} (\bibinfo {year}
  {2012})}\BibitemShut {NoStop}%
\bibitem [{\citenamefont {Lin}\ \emph {et~al.}(2016)\citenamefont {Lin},
  \citenamefont {Zeng}, \citenamefont {Gui}, \citenamefont {Tang},
  \citenamefont {Zou},\ and\ \citenamefont {Cao}}]{lin.zq_2016}%
  \BibitemOpen
  \bibfield  {author} {\bibinfo {author} {\bibfnamefont {Z.}~\bibnamefont
  {Lin}}, \bibinfo {author} {\bibfnamefont {Z.}~\bibnamefont {Zeng}}, \bibinfo
  {author} {\bibfnamefont {X.}~\bibnamefont {Gui}}, \bibinfo {author}
  {\bibfnamefont {Z.}~\bibnamefont {Tang}}, \bibinfo {author} {\bibfnamefont
  {M.}~\bibnamefont {Zou}}, \ and\ \bibinfo {author} {\bibfnamefont
  {A.}~\bibnamefont {Cao}},\ }\href {\doibase 10.1002/aenm.201600554}
  {\bibfield  {journal} {\bibinfo  {journal} {Adv. Energy Mater.}\ }\textbf
  {\bibinfo {volume} {6}},\ \bibinfo {pages} {1600554} (\bibinfo {year}
  {2016})}\BibitemShut {NoStop}%
\bibitem [{\citenamefont {Cao}\ \emph {et~al.}(2005)\citenamefont {Cao},
  \citenamefont {Dickrell}, \citenamefont {Sawyer}, \citenamefont
  {Ghasemi-Nejhad},\ and\ \citenamefont {Ajayan}}]{cao.ay_20051}%
  \BibitemOpen
  \bibfield  {author} {\bibinfo {author} {\bibfnamefont {A.}~\bibnamefont
  {Cao}}, \bibinfo {author} {\bibfnamefont {P.~L.}\ \bibnamefont {Dickrell}},
  \bibinfo {author} {\bibfnamefont {W.~G.}\ \bibnamefont {Sawyer}}, \bibinfo
  {author} {\bibfnamefont {M.~N.}\ \bibnamefont {Ghasemi-Nejhad}}, \ and\
  \bibinfo {author} {\bibfnamefont {P.~M.}\ \bibnamefont {Ajayan}},\ }\href
  {\doibase 10.1126/science.1118957} {\bibfield  {journal} {\bibinfo  {journal}
  {Science}\ }\textbf {\bibinfo {volume} {310}},\ \bibinfo {pages} {1307}
  (\bibinfo {year} {2005})}\BibitemShut {NoStop}%
\bibitem [{\citenamefont {Xu}\ \emph {et~al.}(2010)\citenamefont {Xu},
  \citenamefont {Futaba}, \citenamefont {Yamada}, \citenamefont {Yumura},\ and\
  \citenamefont {Hata}}]{xu.m_2010}%
  \BibitemOpen
  \bibfield  {author} {\bibinfo {author} {\bibfnamefont {M.}~\bibnamefont
  {Xu}}, \bibinfo {author} {\bibfnamefont {D.~N.}\ \bibnamefont {Futaba}},
  \bibinfo {author} {\bibfnamefont {T.}~\bibnamefont {Yamada}}, \bibinfo
  {author} {\bibfnamefont {M.}~\bibnamefont {Yumura}}, \ and\ \bibinfo {author}
  {\bibfnamefont {K.}~\bibnamefont {Hata}},\ }\href {\doibase
  10.1126/science.1194865} {\bibfield  {journal} {\bibinfo  {journal}
  {Science}\ }\textbf {\bibinfo {volume} {330}},\ \bibinfo {pages} {1364}
  (\bibinfo {year} {2010})}\BibitemShut {NoStop}%
\bibitem [{\citenamefont {Wang}\ \emph {et~al.}(2012)\citenamefont {Wang},
  \citenamefont {Xie}, \citenamefont {Liu},\ and\ \citenamefont
  {Xu}}]{wang.c_2012}%
  \BibitemOpen
  \bibfield  {author} {\bibinfo {author} {\bibfnamefont {C.}~\bibnamefont
  {Wang}}, \bibinfo {author} {\bibfnamefont {B.}~\bibnamefont {Xie}}, \bibinfo
  {author} {\bibfnamefont {Y.}~\bibnamefont {Liu}}, \ and\ \bibinfo {author}
  {\bibfnamefont {Z.}~\bibnamefont {Xu}},\ }\href {\doibase 10.1021/mz300422f}
  {\bibfield  {journal} {\bibinfo  {journal} {ACS Macro Lett.}\ }\textbf
  {\bibinfo {volume} {1}},\ \bibinfo {pages} {1176} (\bibinfo {year}
  {2012})}\BibitemShut {NoStop}%
\bibitem [{\citenamefont {Won}\ \emph {et~al.}(2013)\citenamefont {Won},
  \citenamefont {Gao}, \citenamefont {Panzer}, \citenamefont {Xiang},
  \citenamefont {Maruyama}, \citenamefont {Kenny}, \citenamefont {Cai},\ and\
  \citenamefont {Goodson}}]{won.y_2013}%
  \BibitemOpen
  \bibfield  {author} {\bibinfo {author} {\bibfnamefont {Y.}~\bibnamefont
  {Won}}, \bibinfo {author} {\bibfnamefont {Y.}~\bibnamefont {Gao}}, \bibinfo
  {author} {\bibfnamefont {M.~A.}\ \bibnamefont {Panzer}}, \bibinfo {author}
  {\bibfnamefont {R.}~\bibnamefont {Xiang}}, \bibinfo {author} {\bibfnamefont
  {S.}~\bibnamefont {Maruyama}}, \bibinfo {author} {\bibfnamefont {T.~W.}\
  \bibnamefont {Kenny}}, \bibinfo {author} {\bibfnamefont {W.}~\bibnamefont
  {Cai}}, \ and\ \bibinfo {author} {\bibfnamefont {K.~E.}\ \bibnamefont
  {Goodson}},\ }\href {\doibase 10.1073/pnas.1312253110} {\bibfield  {journal}
  {\bibinfo  {journal} {Proc. Natl. Acad. Sci.}\ }\textbf {\bibinfo {volume}
  {110}},\ \bibinfo {pages} {20426} (\bibinfo {year} {2013})}\BibitemShut
  {NoStop}%
\bibitem [{\citenamefont {Zhao}\ \emph {et~al.}(2015)\citenamefont {Zhao},
  \citenamefont {Zhang}, \citenamefont {Pan},\ and\ \citenamefont
  {Li}}]{zhao.jn_2015}%
  \BibitemOpen
  \bibfield  {author} {\bibinfo {author} {\bibfnamefont {J.}~\bibnamefont
  {Zhao}}, \bibinfo {author} {\bibfnamefont {X.}~\bibnamefont {Zhang}},
  \bibinfo {author} {\bibfnamefont {Z.}~\bibnamefont {Pan}}, \ and\ \bibinfo
  {author} {\bibfnamefont {Q.}~\bibnamefont {Li}},\ }\href {\doibase
  10.1002/admi.201500093} {\bibfield  {journal} {\bibinfo  {journal} {Adv.
  Mater. Interfaces}\ }\textbf {\bibinfo {volume} {2}},\ \bibinfo {pages}
  {1500093} (\bibinfo {year} {2015})}\BibitemShut {NoStop}%
\bibitem [{\citenamefont {Li}\ \emph {et~al.}(2015)\citenamefont {Li},
  \citenamefont {Kim}, \citenamefont {Wei}, \citenamefont {Kang}, \citenamefont
  {Choi}, \citenamefont {Nam},\ and\ \citenamefont {Suhr}}]{li.yp_2015}%
  \BibitemOpen
  \bibfield  {author} {\bibinfo {author} {\bibfnamefont {Y.}~\bibnamefont
  {Li}}, \bibinfo {author} {\bibfnamefont {H.-i.}\ \bibnamefont {Kim}},
  \bibinfo {author} {\bibfnamefont {B.}~\bibnamefont {Wei}}, \bibinfo {author}
  {\bibfnamefont {J.}~\bibnamefont {Kang}}, \bibinfo {author} {\bibfnamefont
  {J.-b.}\ \bibnamefont {Choi}}, \bibinfo {author} {\bibfnamefont {J.-D.}\
  \bibnamefont {Nam}}, \ and\ \bibinfo {author} {\bibfnamefont
  {J.}~\bibnamefont {Suhr}},\ }\href {\doibase 10.1039/c5nr03581c} {\bibfield
  {journal} {\bibinfo  {journal} {Nanoscale}\ }\textbf {\bibinfo {volume}
  {7}},\ \bibinfo {pages} {14299} (\bibinfo {year} {2015})}\BibitemShut
  {NoStop}%
\bibitem [{\citenamefont {Shen}\ \emph {et~al.}(2017)\citenamefont {Shen},
  \citenamefont {R\"{o}ding}, \citenamefont {Kr\"{o}ger},\ and\ \citenamefont
  {Li}}]{shen.zq_2017}%
  \BibitemOpen
  \bibfield  {author} {\bibinfo {author} {\bibfnamefont {Z.}~\bibnamefont
  {Shen}}, \bibinfo {author} {\bibfnamefont {M.}~\bibnamefont {R\"{o}ding}},
  \bibinfo {author} {\bibfnamefont {M.}~\bibnamefont {Kr\"{o}ger}}, \ and\
  \bibinfo {author} {\bibfnamefont {Y.}~\bibnamefont {Li}},\ }\href {\doibase
  10.3390/polym9040115} {\bibfield  {journal} {\bibinfo  {journal} {Polymers}\
  }\textbf {\bibinfo {volume} {9}},\ \bibinfo {pages} {115} (\bibinfo {year}
  {2017})}\BibitemShut {NoStop}%
\bibitem [{\citenamefont {Ferry}(1980)}]{ferry.jd_1980}%
  \BibitemOpen
  \bibfield  {author} {\bibinfo {author} {\bibfnamefont {J.~D.}\ \bibnamefont
  {Ferry}},\ }\href@noop {} {\emph {\bibinfo {title} {{Viscoelastic Properties
  of Polymers}}}},\ \bibinfo {edition} {3rd}\ ed.\ (\bibinfo  {publisher} {John
  Wiley \& Sons},\ \bibinfo {address} {New York},\ \bibinfo {year}
  {1980})\BibitemShut {NoStop}%
\bibitem [{\citenamefont {Meyers}\ and\ \citenamefont
  {Chawla}(2009)}]{meyers.ma_2009}%
  \BibitemOpen
  \bibfield  {author} {\bibinfo {author} {\bibfnamefont {M.~A.}\ \bibnamefont
  {Meyers}}\ and\ \bibinfo {author} {\bibfnamefont {K.~K.}\ \bibnamefont
  {Chawla}},\ }\href@noop {} {\emph {\bibinfo {title} {{Mechanical Behavior of
  Materials}}}},\ \bibinfo {edition} {2nd}\ ed.\ (\bibinfo  {publisher}
  {Cambridge University Press},\ \bibinfo {address} {New York},\ \bibinfo
  {year} {2009})\BibitemShut {NoStop}%
\bibitem [{\citenamefont {Lakes}(2009)}]{lakes.r_2009}%
  \BibitemOpen
  \bibfield  {author} {\bibinfo {author} {\bibfnamefont {R.}~\bibnamefont
  {Lakes}},\ }\href@noop {} {\emph {\bibinfo {title} {{Viscoelastic
  Materials}}}}\ (\bibinfo  {publisher} {Cambridge University Press},\ \bibinfo
  {address} {New York},\ \bibinfo {year} {2009})\BibitemShut {NoStop}%
\bibitem [{\citenamefont {Lemaitre}(2001)}]{lemaitre.j_2001}%
  \BibitemOpen
  \bibfield  {author} {\bibinfo {author} {\bibfnamefont {J.}~\bibnamefont
  {Lemaitre}},\ }in\ \href {\doibase 10.1016/B978-012443341-0/50006-5} {\emph
  {\bibinfo {booktitle} {Handbook of Materials Behavior Models}}},\ \bibinfo
  {editor} {edited by\ \bibinfo {editor} {\bibfnamefont {J.}~\bibnamefont
  {Lemaitre}}}\ (\bibinfo  {publisher} {Academic Press},\ \bibinfo {address}
  {San Diego, USA},\ \bibinfo {year} {2001})\ Chap.\ \bibinfo {chapter} {2.1},
  pp.\ \bibinfo {pages} {71--74}\BibitemShut {NoStop}%
\bibitem [{\citenamefont {Achorn}\ and\ \citenamefont
  {Ferrillo}(1994)}]{achorn.pj_1994}%
  \BibitemOpen
  \bibfield  {author} {\bibinfo {author} {\bibfnamefont {P.~J.}\ \bibnamefont
  {Achorn}}\ and\ \bibinfo {author} {\bibfnamefont {R.~G.}\ \bibnamefont
  {Ferrillo}},\ }\href {\doibase 10.1002/app.1994.070541305} {\bibfield
  {journal} {\bibinfo  {journal} {J. Appl. Polym. Sci.}\ }\textbf {\bibinfo
  {volume} {54}},\ \bibinfo {pages} {2033} (\bibinfo {year}
  {1994})}\BibitemShut {NoStop}%
\bibitem [{\citenamefont {Suhr}\ \emph {et~al.}(2006)\citenamefont {Suhr},
  \citenamefont {Zhang}, \citenamefont {Ajayan},\ and\ \citenamefont
  {Koratkar}}]{suhr.j_2006}%
  \BibitemOpen
  \bibfield  {author} {\bibinfo {author} {\bibfnamefont {J.}~\bibnamefont
  {Suhr}}, \bibinfo {author} {\bibfnamefont {W.}~\bibnamefont {Zhang}},
  \bibinfo {author} {\bibfnamefont {P.~M.}\ \bibnamefont {Ajayan}}, \ and\
  \bibinfo {author} {\bibfnamefont {N.~A.}\ \bibnamefont {Koratkar}},\ }\href
  {\doibase 10.1021/nl0521524} {\bibfield  {journal} {\bibinfo  {journal} {Nano
  Lett.}\ }\textbf {\bibinfo {volume} {6}},\ \bibinfo {pages} {219} (\bibinfo
  {year} {2006})}\BibitemShut {NoStop}%
\bibitem [{\citenamefont {Zhou}\ \emph {et~al.}(2004)\citenamefont {Zhou},
  \citenamefont {Shin}, \citenamefont {Wang},\ and\ \citenamefont
  {Bakis}}]{zhou.x_2004}%
  \BibitemOpen
  \bibfield  {author} {\bibinfo {author} {\bibfnamefont {X.}~\bibnamefont
  {Zhou}}, \bibinfo {author} {\bibfnamefont {E.}~\bibnamefont {Shin}}, \bibinfo
  {author} {\bibfnamefont {K.~W.}\ \bibnamefont {Wang}}, \ and\ \bibinfo
  {author} {\bibfnamefont {C.~E.}\ \bibnamefont {Bakis}},\ }\href {\doibase
  10.1016/j.compscitech.2004.06.001} {\bibfield  {journal} {\bibinfo  {journal}
  {Compos. Sci. Technol.}\ }\textbf {\bibinfo {volume} {64}},\ \bibinfo {pages}
  {2425} (\bibinfo {year} {2004})}\BibitemShut {NoStop}%
\bibitem [{\citenamefont {Suhr}\ \emph {et~al.}(2005)\citenamefont {Suhr},
  \citenamefont {Koratkar}, \citenamefont {Keblinski},\ and\ \citenamefont
  {Ajayan}}]{suhr.j_2005}%
  \BibitemOpen
  \bibfield  {author} {\bibinfo {author} {\bibfnamefont {J.}~\bibnamefont
  {Suhr}}, \bibinfo {author} {\bibfnamefont {N.}~\bibnamefont {Koratkar}},
  \bibinfo {author} {\bibfnamefont {P.}~\bibnamefont {Keblinski}}, \ and\
  \bibinfo {author} {\bibfnamefont {P.}~\bibnamefont {Ajayan}},\ }\href
  {\doibase 10.1038/nmat1293} {\bibfield  {journal} {\bibinfo  {journal} {Nat.
  Mater.}\ }\textbf {\bibinfo {volume} {4}},\ \bibinfo {pages} {134} (\bibinfo
  {year} {2005})}\BibitemShut {NoStop}%
\bibitem [{\citenamefont {Ajayan}, \citenamefont {Suhr},\ and\ \citenamefont
  {Koratkar}(2006)}]{ajayan.pm_2006}%
  \BibitemOpen
  \bibfield  {author} {\bibinfo {author} {\bibfnamefont {P.~M.}\ \bibnamefont
  {Ajayan}}, \bibinfo {author} {\bibfnamefont {J.}~\bibnamefont {Suhr}}, \ and\
  \bibinfo {author} {\bibfnamefont {N.}~\bibnamefont {Koratkar}},\ }\href
  {\doibase 10.1007/s10853-006-0693-4} {\bibfield  {journal} {\bibinfo
  {journal} {J. Mater. Sci.}\ }\textbf {\bibinfo {volume} {41}},\ \bibinfo
  {pages} {7824} (\bibinfo {year} {2006})}\BibitemShut {NoStop}%
\bibitem [{\citenamefont {Wang}\ \emph {et~al.}(2011)\citenamefont {Wang},
  \citenamefont {Whitby}, \citenamefont {Rousseau}, \citenamefont {Nevill},
  \citenamefont {Geaves},\ and\ \citenamefont {Mikhalovsky}}]{wang.zw_2011}%
  \BibitemOpen
  \bibfield  {author} {\bibinfo {author} {\bibfnamefont {Z.}~\bibnamefont
  {Wang}}, \bibinfo {author} {\bibfnamefont {R.~L.~D.}\ \bibnamefont {Whitby}},
  \bibinfo {author} {\bibfnamefont {M.}~\bibnamefont {Rousseau}}, \bibinfo
  {author} {\bibfnamefont {S.}~\bibnamefont {Nevill}}, \bibinfo {author}
  {\bibfnamefont {G.}~\bibnamefont {Geaves}}, \ and\ \bibinfo {author}
  {\bibfnamefont {S.~V.}\ \bibnamefont {Mikhalovsky}},\ }\href {\doibase
  10.1039/c0jm03335a} {\bibfield  {journal} {\bibinfo  {journal} {J. Mater.
  Chem.}\ }\textbf {\bibinfo {volume} {21}},\ \bibinfo {pages} {4150} (\bibinfo
  {year} {2011})}\BibitemShut {NoStop}%
\bibitem [{\citenamefont {Agrawal}\ \emph {et~al.}(2013)\citenamefont
  {Agrawal}, \citenamefont {Nieto}, \citenamefont {Chen}, \citenamefont
  {Mora},\ and\ \citenamefont {Agarwal}}]{agrawal.r_2013}%
  \BibitemOpen
  \bibfield  {author} {\bibinfo {author} {\bibfnamefont {R.}~\bibnamefont
  {Agrawal}}, \bibinfo {author} {\bibfnamefont {A.}~\bibnamefont {Nieto}},
  \bibinfo {author} {\bibfnamefont {H.}~\bibnamefont {Chen}}, \bibinfo {author}
  {\bibfnamefont {M.}~\bibnamefont {Mora}}, \ and\ \bibinfo {author}
  {\bibfnamefont {A.}~\bibnamefont {Agarwal}},\ }\href {\doibase
  10.1021/am4038678} {\bibfield  {journal} {\bibinfo  {journal} {ACS Appl.
  Mater. Interfaces}\ }\textbf {\bibinfo {volume} {5}},\ \bibinfo {pages}
  {12052} (\bibinfo {year} {2013})}\BibitemShut {NoStop}%
\bibitem [{\citenamefont {Carponcin}\ \emph {et~al.}(2015)\citenamefont
  {Carponcin}, \citenamefont {Dantras}, \citenamefont {Michon}, \citenamefont
  {Dandurand}, \citenamefont {Aridon}, \citenamefont {Levallois}, \citenamefont
  {Cadiergues},\ and\ \citenamefont {Lacabanne}}]{carponcin.d_2015}%
  \BibitemOpen
  \bibfield  {author} {\bibinfo {author} {\bibfnamefont {D.}~\bibnamefont
  {Carponcin}}, \bibinfo {author} {\bibfnamefont {E.}~\bibnamefont {Dantras}},
  \bibinfo {author} {\bibfnamefont {G.}~\bibnamefont {Michon}}, \bibinfo
  {author} {\bibfnamefont {J.}~\bibnamefont {Dandurand}}, \bibinfo {author}
  {\bibfnamefont {G.}~\bibnamefont {Aridon}}, \bibinfo {author} {\bibfnamefont
  {F.}~\bibnamefont {Levallois}}, \bibinfo {author} {\bibfnamefont
  {L.}~\bibnamefont {Cadiergues}}, \ and\ \bibinfo {author} {\bibfnamefont
  {C.}~\bibnamefont {Lacabanne}},\ }\href {\doibase
  10.1016/j.jnoncrysol.2014.11.008} {\bibfield  {journal} {\bibinfo  {journal}
  {J. Non-Cryst. Solids}\ }\textbf {\bibinfo {volume} {409}},\ \bibinfo {pages}
  {20} (\bibinfo {year} {2015})}\BibitemShut {NoStop}%
\bibitem [{\citenamefont {Chu}\ \emph {et~al.}(2016)\citenamefont {Chu},
  \citenamefont {Weng}, \citenamefont {Lin}, \citenamefont {Tsai},\ and\
  \citenamefont {Hsu}}]{chu.yc_2016}%
  \BibitemOpen
  \bibfield  {author} {\bibinfo {author} {\bibfnamefont {Y.-C.}\ \bibnamefont
  {Chu}}, \bibinfo {author} {\bibfnamefont {M.-H.}\ \bibnamefont {Weng}},
  \bibinfo {author} {\bibfnamefont {W.-Y.}\ \bibnamefont {Lin}}, \bibinfo
  {author} {\bibfnamefont {H.-J.}\ \bibnamefont {Tsai}}, \ and\ \bibinfo
  {author} {\bibfnamefont {W.-K.}\ \bibnamefont {Hsu}},\ }\href {\doibase
  10.1039/C6RA01239F} {\bibfield  {journal} {\bibinfo  {journal} {RSC Adv.}\
  }\textbf {\bibinfo {volume} {6}},\ \bibinfo {pages} {21271} (\bibinfo {year}
  {2016})}\BibitemShut {NoStop}%
\bibitem [{\citenamefont {Eftekhari}\ and\ \citenamefont
  {Fatemi}(2016)}]{eftekhari.m_2016}%
  \BibitemOpen
  \bibfield  {author} {\bibinfo {author} {\bibfnamefont {M.}~\bibnamefont
  {Eftekhari}}\ and\ \bibinfo {author} {\bibfnamefont {A.}~\bibnamefont
  {Fatemi}},\ }\href {\doibase 10.1016/j.ijfatigue.2016.01.014} {\bibfield
  {journal} {\bibinfo  {journal} {Int. J. Fatigue}\ }\textbf {\bibinfo {volume}
  {87}},\ \bibinfo {pages} {153} (\bibinfo {year} {2016})}\BibitemShut
  {NoStop}%
\bibitem [{\citenamefont {Wang}\ \emph {et~al.}(2015)\citenamefont {Wang},
  \citenamefont {Li}, \citenamefont {Gu}, \citenamefont {Zhang}, \citenamefont
  {Wang}, \citenamefont {Li},\ and\ \citenamefont {Zhang}}]{wang.yj_2015}%
  \BibitemOpen
  \bibfield  {author} {\bibinfo {author} {\bibfnamefont {Y.}~\bibnamefont
  {Wang}}, \bibinfo {author} {\bibfnamefont {M.}~\bibnamefont {Li}}, \bibinfo
  {author} {\bibfnamefont {Y.}~\bibnamefont {Gu}}, \bibinfo {author}
  {\bibfnamefont {X.}~\bibnamefont {Zhang}}, \bibinfo {author} {\bibfnamefont
  {S.}~\bibnamefont {Wang}}, \bibinfo {author} {\bibfnamefont {Q.}~\bibnamefont
  {Li}}, \ and\ \bibinfo {author} {\bibfnamefont {Z.}~\bibnamefont {Zhang}},\
  }\href {\doibase 10.1039/c4nr06401a} {\bibfield  {journal} {\bibinfo
  {journal} {Nanoscale}\ }\textbf {\bibinfo {volume} {7}},\ \bibinfo {pages}
  {3060} (\bibinfo {year} {2015})}\BibitemShut {NoStop}%
\bibitem [{\citenamefont {Zhang}\ \emph {et~al.}(2015)\citenamefont {Zhang},
  \citenamefont {Wang}, \citenamefont {Xu}, \citenamefont {Zhang},
  \citenamefont {Li}, \citenamefont {Bradford},\ and\ \citenamefont
  {Zhu}}]{zhang.lw_2015}%
  \BibitemOpen
  \bibfield  {author} {\bibinfo {author} {\bibfnamefont {L.}~\bibnamefont
  {Zhang}}, \bibinfo {author} {\bibfnamefont {X.}~\bibnamefont {Wang}},
  \bibinfo {author} {\bibfnamefont {W.}~\bibnamefont {Xu}}, \bibinfo {author}
  {\bibfnamefont {Y.}~\bibnamefont {Zhang}}, \bibinfo {author} {\bibfnamefont
  {Q.}~\bibnamefont {Li}}, \bibinfo {author} {\bibfnamefont {P.~D.}\
  \bibnamefont {Bradford}}, \ and\ \bibinfo {author} {\bibfnamefont
  {Y.}~\bibnamefont {Zhu}},\ }\href {\doibase 10.1002/smll.201500111}
  {\bibfield  {journal} {\bibinfo  {journal} {Small}\ }\textbf {\bibinfo
  {volume} {11}},\ \bibinfo {pages} {3830} (\bibinfo {year}
  {2015})}\BibitemShut {NoStop}%
\bibitem [{\citenamefont {Yu}\ \emph {et~al.}(2016)\citenamefont {Yu},
  \citenamefont {Zhang}, \citenamefont {Zou}, \citenamefont {Lan},
  \citenamefont {Jiang}, \citenamefont {Zhao}, \citenamefont {Zhang},
  \citenamefont {Miao},\ and\ \citenamefont {Li}}]{yu.xp_2016}%
  \BibitemOpen
  \bibfield  {author} {\bibinfo {author} {\bibfnamefont {X.}~\bibnamefont
  {Yu}}, \bibinfo {author} {\bibfnamefont {X.}~\bibnamefont {Zhang}}, \bibinfo
  {author} {\bibfnamefont {J.}~\bibnamefont {Zou}}, \bibinfo {author}
  {\bibfnamefont {Z.}~\bibnamefont {Lan}}, \bibinfo {author} {\bibfnamefont
  {C.}~\bibnamefont {Jiang}}, \bibinfo {author} {\bibfnamefont
  {J.}~\bibnamefont {Zhao}}, \bibinfo {author} {\bibfnamefont {D.}~\bibnamefont
  {Zhang}}, \bibinfo {author} {\bibfnamefont {M.}~\bibnamefont {Miao}}, \ and\
  \bibinfo {author} {\bibfnamefont {Q.}~\bibnamefont {Li}},\ }\href {\doibase
  10.1002/admi.201600352} {\bibfield  {journal} {\bibinfo  {journal} {Adv.
  Mater. Interfaces}\ }\textbf {\bibinfo {volume} {3}},\ \bibinfo {pages}
  {1600352} (\bibinfo {year} {2016})}\BibitemShut {NoStop}%
\bibitem [{\citenamefont {Cheng}\ \emph {et~al.}(2009)\citenamefont {Cheng},
  \citenamefont {Bao}, \citenamefont {Park}, \citenamefont {Liang},
  \citenamefont {Zhang},\ and\ \citenamefont {Wang}}]{cheng.qf_2009}%
  \BibitemOpen
  \bibfield  {author} {\bibinfo {author} {\bibfnamefont {Q.}~\bibnamefont
  {Cheng}}, \bibinfo {author} {\bibfnamefont {J.}~\bibnamefont {Bao}}, \bibinfo
  {author} {\bibfnamefont {J.}~\bibnamefont {Park}}, \bibinfo {author}
  {\bibfnamefont {Z.}~\bibnamefont {Liang}}, \bibinfo {author} {\bibfnamefont
  {C.}~\bibnamefont {Zhang}}, \ and\ \bibinfo {author} {\bibfnamefont
  {B.}~\bibnamefont {Wang}},\ }\href {\doibase 10.1002/adfm.200900663}
  {\bibfield  {journal} {\bibinfo  {journal} {Adv. Funct. Mater.}\ }\textbf
  {\bibinfo {volume} {19}},\ \bibinfo {pages} {3219} (\bibinfo {year}
  {2009})}\BibitemShut {NoStop}%
\bibitem [{\citenamefont {Cheng}\ \emph {et~al.}(2010)\citenamefont {Cheng},
  \citenamefont {Wang}, \citenamefont {Zhang},\ and\ \citenamefont
  {Liang}}]{cheng.qf_2010}%
  \BibitemOpen
  \bibfield  {author} {\bibinfo {author} {\bibfnamefont {Q.}~\bibnamefont
  {Cheng}}, \bibinfo {author} {\bibfnamefont {B.}~\bibnamefont {Wang}},
  \bibinfo {author} {\bibfnamefont {C.}~\bibnamefont {Zhang}}, \ and\ \bibinfo
  {author} {\bibfnamefont {Z.}~\bibnamefont {Liang}},\ }\href {\doibase
  10.1002/smll.200901957} {\bibfield  {journal} {\bibinfo  {journal} {Small}\
  }\textbf {\bibinfo {volume} {6}},\ \bibinfo {pages} {763} (\bibinfo {year}
  {2010})}\BibitemShut {NoStop}%
\bibitem [{\citenamefont {Wang}\ \emph {et~al.}(2013)\citenamefont {Wang},
  \citenamefont {Yong}, \citenamefont {Li}, \citenamefont {Bradford},
  \citenamefont {Liu}, \citenamefont {Tucker}, \citenamefont {Cai},
  \citenamefont {Wang}, \citenamefont {Yuan},\ and\ \citenamefont
  {Zhu}}]{wang.x_20131}%
  \BibitemOpen
  \bibfield  {author} {\bibinfo {author} {\bibfnamefont {X.}~\bibnamefont
  {Wang}}, \bibinfo {author} {\bibfnamefont {Z.~Z.}\ \bibnamefont {Yong}},
  \bibinfo {author} {\bibfnamefont {Q.~W.}\ \bibnamefont {Li}}, \bibinfo
  {author} {\bibfnamefont {P.~D.}\ \bibnamefont {Bradford}}, \bibinfo {author}
  {\bibfnamefont {W.}~\bibnamefont {Liu}}, \bibinfo {author} {\bibfnamefont
  {D.~S.}\ \bibnamefont {Tucker}}, \bibinfo {author} {\bibfnamefont
  {W.}~\bibnamefont {Cai}}, \bibinfo {author} {\bibfnamefont {H.}~\bibnamefont
  {Wang}}, \bibinfo {author} {\bibfnamefont {F.~G.}\ \bibnamefont {Yuan}}, \
  and\ \bibinfo {author} {\bibfnamefont {Y.~T.}\ \bibnamefont {Zhu}},\ }\href
  {\doibase 10.1080/21663831.2012.686586} {\bibfield  {journal} {\bibinfo
  {journal} {Mater. Res. Lett.}\ }\textbf {\bibinfo {volume} {1}},\ \bibinfo
  {pages} {19} (\bibinfo {year} {2013})}\BibitemShut {NoStop}%
\bibitem [{\citenamefont {Han}\ \emph {et~al.}(2015)\citenamefont {Han},
  \citenamefont {Zhang}, \citenamefont {Yu}, \citenamefont {Zhao},
  \citenamefont {Li}, \citenamefont {Liu}, \citenamefont {Gao}, \citenamefont
  {Zhang}, \citenamefont {Zhao},\ and\ \citenamefont {Li}}]{han.y_2015}%
  \BibitemOpen
  \bibfield  {author} {\bibinfo {author} {\bibfnamefont {Y.}~\bibnamefont
  {Han}}, \bibinfo {author} {\bibfnamefont {X.}~\bibnamefont {Zhang}}, \bibinfo
  {author} {\bibfnamefont {X.}~\bibnamefont {Yu}}, \bibinfo {author}
  {\bibfnamefont {J.}~\bibnamefont {Zhao}}, \bibinfo {author} {\bibfnamefont
  {S.}~\bibnamefont {Li}}, \bibinfo {author} {\bibfnamefont {F.}~\bibnamefont
  {Liu}}, \bibinfo {author} {\bibfnamefont {P.}~\bibnamefont {Gao}}, \bibinfo
  {author} {\bibfnamefont {Y.}~\bibnamefont {Zhang}}, \bibinfo {author}
  {\bibfnamefont {T.}~\bibnamefont {Zhao}}, \ and\ \bibinfo {author}
  {\bibfnamefont {Q.}~\bibnamefont {Li}},\ }\href {\doibase 10.1038/srep11533}
  {\bibfield  {journal} {\bibinfo  {journal} {Sci. Rep.}\ }\textbf {\bibinfo
  {volume} {5}},\ \bibinfo {pages} {11533} (\bibinfo {year}
  {2015})}\BibitemShut {NoStop}%
\bibitem [{\citenamefont {Koratkar}, \citenamefont {Wei},\ and\ \citenamefont
  {Ajayan}(2003)}]{koratkar.na_2003}%
  \BibitemOpen
  \bibfield  {author} {\bibinfo {author} {\bibfnamefont {N.~A.}\ \bibnamefont
  {Koratkar}}, \bibinfo {author} {\bibfnamefont {B.}~\bibnamefont {Wei}}, \
  and\ \bibinfo {author} {\bibfnamefont {P.~M.}\ \bibnamefont {Ajayan}},\
  }\href {\doibase 10.1016/S0266-3538(03)00065-4} {\bibfield  {journal}
  {\bibinfo  {journal} {Compos. Sci. Technol.}\ }\textbf {\bibinfo {volume}
  {63}},\ \bibinfo {pages} {1525} (\bibinfo {year} {2003})}\BibitemShut
  {NoStop}%
\bibitem [{\citenamefont {Pathak}\ \emph {et~al.}(2009)\citenamefont {Pathak},
  \citenamefont {Cambaz}, \citenamefont {Kalidindi}, \citenamefont {Swadener},\
  and\ \citenamefont {Gogotsi}}]{pathak.s_2009}%
  \BibitemOpen
  \bibfield  {author} {\bibinfo {author} {\bibfnamefont {S.}~\bibnamefont
  {Pathak}}, \bibinfo {author} {\bibfnamefont {Z.~G.}\ \bibnamefont {Cambaz}},
  \bibinfo {author} {\bibfnamefont {S.~R.}\ \bibnamefont {Kalidindi}}, \bibinfo
  {author} {\bibfnamefont {J.~G.}\ \bibnamefont {Swadener}}, \ and\ \bibinfo
  {author} {\bibfnamefont {Y.}~\bibnamefont {Gogotsi}},\ }\href {\doibase
  10.1016/j.carbon.2009.03.042} {\bibfield  {journal} {\bibinfo  {journal}
  {Carbon}\ }\textbf {\bibinfo {volume} {47}},\ \bibinfo {pages} {1969}
  (\bibinfo {year} {2009})}\BibitemShut {NoStop}%
\bibitem [{\citenamefont {Zhang}\ \emph
  {et~al.}(2010{\natexlab{a}})\citenamefont {Zhang}, \citenamefont {Lu},
  \citenamefont {Du}, \citenamefont {Dai}, \citenamefont {Baur},\ and\
  \citenamefont {Foster}}]{zhang.q_20102}%
  \BibitemOpen
  \bibfield  {author} {\bibinfo {author} {\bibfnamefont {Q.}~\bibnamefont
  {Zhang}}, \bibinfo {author} {\bibfnamefont {Y.~C.}\ \bibnamefont {Lu}},
  \bibinfo {author} {\bibfnamefont {F.}~\bibnamefont {Du}}, \bibinfo {author}
  {\bibfnamefont {L.}~\bibnamefont {Dai}}, \bibinfo {author} {\bibfnamefont
  {J.}~\bibnamefont {Baur}}, \ and\ \bibinfo {author} {\bibfnamefont {D.~C.}\
  \bibnamefont {Foster}},\ }\href {\doibase 10.1088/0022-3727/43/31/315401}
  {\bibfield  {journal} {\bibinfo  {journal} {J. Phys. D: Appl. Phys.}\
  }\textbf {\bibinfo {volume} {43}},\ \bibinfo {pages} {315401} (\bibinfo
  {year} {2010}{\natexlab{a}})}\BibitemShut {NoStop}%
\bibitem [{\citenamefont {Bradford}\ \emph {et~al.}(2011)\citenamefont
  {Bradford}, \citenamefont {Wang}, \citenamefont {Zhao},\ and\ \citenamefont
  {Zhu}}]{bradford.pd_2011}%
  \BibitemOpen
  \bibfield  {author} {\bibinfo {author} {\bibfnamefont {P.~D.}\ \bibnamefont
  {Bradford}}, \bibinfo {author} {\bibfnamefont {X.}~\bibnamefont {Wang}},
  \bibinfo {author} {\bibfnamefont {H.}~\bibnamefont {Zhao}}, \ and\ \bibinfo
  {author} {\bibfnamefont {Y.~T.}\ \bibnamefont {Zhu}},\ }\href {\doibase
  10.1016/j.carbon.2011.03.012} {\bibfield  {journal} {\bibinfo  {journal}
  {Carbon}\ }\textbf {\bibinfo {volume} {49}},\ \bibinfo {pages} {2834}
  (\bibinfo {year} {2011})}\BibitemShut {NoStop}%
\bibitem [{\citenamefont {Liu}\ \emph {et~al.}(2015)\citenamefont {Liu},
  \citenamefont {Li}, \citenamefont {Gu}, \citenamefont {Wang}, \citenamefont
  {Zhang}, \citenamefont {Li}, \citenamefont {Gao},\ and\ \citenamefont
  {Zhang}}]{liu.ql_2015}%
  \BibitemOpen
  \bibfield  {author} {\bibinfo {author} {\bibfnamefont {Q.}~\bibnamefont
  {Liu}}, \bibinfo {author} {\bibfnamefont {M.}~\bibnamefont {Li}}, \bibinfo
  {author} {\bibfnamefont {Y.}~\bibnamefont {Gu}}, \bibinfo {author}
  {\bibfnamefont {S.}~\bibnamefont {Wang}}, \bibinfo {author} {\bibfnamefont
  {Y.}~\bibnamefont {Zhang}}, \bibinfo {author} {\bibfnamefont
  {Q.}~\bibnamefont {Li}}, \bibinfo {author} {\bibfnamefont {L.}~\bibnamefont
  {Gao}}, \ and\ \bibinfo {author} {\bibfnamefont {Z.}~\bibnamefont {Zhang}},\
  }\href {\doibase 10.1016/j.carbon.2015.01.014} {\bibfield  {journal}
  {\bibinfo  {journal} {Carbon}\ }\textbf {\bibinfo {volume} {86}},\ \bibinfo
  {pages} {46} (\bibinfo {year} {2015})}\BibitemShut {NoStop}%
\bibitem [{\citenamefont {Li}\ and\ \citenamefont
  {Kr\"{o}ger}(2012{\natexlab{a}})}]{li.y_2012}%
  \BibitemOpen
  \bibfield  {author} {\bibinfo {author} {\bibfnamefont {Y.}~\bibnamefont
  {Li}}\ and\ \bibinfo {author} {\bibfnamefont {M.}~\bibnamefont
  {Kr\"{o}ger}},\ }\href {\doibase 10.1039/c2sm25561h} {\bibfield  {journal}
  {\bibinfo  {journal} {Soft Matter}\ }\textbf {\bibinfo {volume} {8}},\
  \bibinfo {pages} {7822} (\bibinfo {year} {2012}{\natexlab{a}})}\BibitemShut
  {NoStop}%
\bibitem [{\citenamefont {Koratkar}, \citenamefont {Wei},\ and\ \citenamefont
  {Ajayan}(2002)}]{koratkar.n_2002}%
  \BibitemOpen
  \bibfield  {author} {\bibinfo {author} {\bibfnamefont {N.}~\bibnamefont
  {Koratkar}}, \bibinfo {author} {\bibfnamefont {B.}~\bibnamefont {Wei}}, \
  and\ \bibinfo {author} {\bibfnamefont {P.~M.}\ \bibnamefont {Ajayan}},\
  }\href {\doibase
  10.1002/1521-4095(20020705)14:13/14<997::AID-ADMA997>3.0.CO;2-Y} {\bibfield
  {journal} {\bibinfo  {journal} {Adv. Mater.}\ }\textbf {\bibinfo {volume}
  {14}},\ \bibinfo {pages} {997} (\bibinfo {year} {2002})}\BibitemShut
  {NoStop}%
\bibitem [{\citenamefont {Fan}\ \emph {et~al.}(1999)\citenamefont {Fan},
  \citenamefont {Chapline}, \citenamefont {Franklin}, \citenamefont {Tombler},
  \citenamefont {Cassell},\ and\ \citenamefont {Dai}}]{fan.ss_1999}%
  \BibitemOpen
  \bibfield  {author} {\bibinfo {author} {\bibfnamefont {S.}~\bibnamefont
  {Fan}}, \bibinfo {author} {\bibfnamefont {M.~G.}\ \bibnamefont {Chapline}},
  \bibinfo {author} {\bibfnamefont {N.~R.}\ \bibnamefont {Franklin}}, \bibinfo
  {author} {\bibfnamefont {T.~W.}\ \bibnamefont {Tombler}}, \bibinfo {author}
  {\bibfnamefont {A.~M.}\ \bibnamefont {Cassell}}, \ and\ \bibinfo {author}
  {\bibfnamefont {H.}~\bibnamefont {Dai}},\ }\href {\doibase
  10.1126/science.283.5401.512} {\bibfield  {journal} {\bibinfo  {journal}
  {Science}\ }\textbf {\bibinfo {volume} {283}},\ \bibinfo {pages} {512}
  (\bibinfo {year} {1999})}\BibitemShut {NoStop}%
\bibitem [{\citenamefont {Zhang}, \citenamefont {Atkinson},\ and\ \citenamefont
  {Baughman}(2004)}]{zhang.m_2004}%
  \BibitemOpen
  \bibfield  {author} {\bibinfo {author} {\bibfnamefont {M.}~\bibnamefont
  {Zhang}}, \bibinfo {author} {\bibfnamefont {K.~R.}\ \bibnamefont {Atkinson}},
  \ and\ \bibinfo {author} {\bibfnamefont {R.~H.}\ \bibnamefont {Baughman}},\
  }\href {\doibase 10.1126/science.1104276} {\bibfield  {journal} {\bibinfo
  {journal} {Science}\ }\textbf {\bibinfo {volume} {306}},\ \bibinfo {pages}
  {1358} (\bibinfo {year} {2004})}\BibitemShut {NoStop}%
\bibitem [{\citenamefont {Wang}\ \emph {et~al.}(2008)\citenamefont {Wang},
  \citenamefont {Song}, \citenamefont {Liu}, \citenamefont {Wu},\ and\
  \citenamefont {Fan}}]{wang.d_2008}%
  \BibitemOpen
  \bibfield  {author} {\bibinfo {author} {\bibfnamefont {D.}~\bibnamefont
  {Wang}}, \bibinfo {author} {\bibfnamefont {P.}~\bibnamefont {Song}}, \bibinfo
  {author} {\bibfnamefont {C.}~\bibnamefont {Liu}}, \bibinfo {author}
  {\bibfnamefont {W.}~\bibnamefont {Wu}}, \ and\ \bibinfo {author}
  {\bibfnamefont {S.}~\bibnamefont {Fan}},\ }\href {\doibase
  10.1088/0957-4484/19/7/075609} {\bibfield  {journal} {\bibinfo  {journal}
  {Nanotechnology}\ }\textbf {\bibinfo {volume} {19}},\ \bibinfo {pages}
  {075609} (\bibinfo {year} {2008})}\BibitemShut {NoStop}%
\bibitem [{\citenamefont {Mizuno}\ \emph {et~al.}(2009)\citenamefont {Mizuno},
  \citenamefont {Ishii}, \citenamefont {Kishida}, \citenamefont {Hayamizu},
  \citenamefont {Yasuda}, \citenamefont {Futaba}, \citenamefont {Yumura},\ and\
  \citenamefont {Hata}}]{mizuno.k_2009}%
  \BibitemOpen
  \bibfield  {author} {\bibinfo {author} {\bibfnamefont {K.}~\bibnamefont
  {Mizuno}}, \bibinfo {author} {\bibfnamefont {J.}~\bibnamefont {Ishii}},
  \bibinfo {author} {\bibfnamefont {H.}~\bibnamefont {Kishida}}, \bibinfo
  {author} {\bibfnamefont {Y.}~\bibnamefont {Hayamizu}}, \bibinfo {author}
  {\bibfnamefont {S.}~\bibnamefont {Yasuda}}, \bibinfo {author} {\bibfnamefont
  {D.~N.}\ \bibnamefont {Futaba}}, \bibinfo {author} {\bibfnamefont
  {M.}~\bibnamefont {Yumura}}, \ and\ \bibinfo {author} {\bibfnamefont
  {K.}~\bibnamefont {Hata}},\ }\href {\doibase 10.1073/pnas.0900155106}
  {\bibfield  {journal} {\bibinfo  {journal} {Proc. Natl. Acad. Sci.}\ }\textbf
  {\bibinfo {volume} {106}},\ \bibinfo {pages} {6044} (\bibinfo {year}
  {2009})}\BibitemShut {NoStop}%
\bibitem [{\citenamefont {Fang}\ \emph {et~al.}(2010)\citenamefont {Fang},
  \citenamefont {Zhao}, \citenamefont {Jia}, \citenamefont {Zhang},
  \citenamefont {Zhang},\ and\ \citenamefont {Li}}]{fang.c_2010}%
  \BibitemOpen
  \bibfield  {author} {\bibinfo {author} {\bibfnamefont {C.}~\bibnamefont
  {Fang}}, \bibinfo {author} {\bibfnamefont {J.}~\bibnamefont {Zhao}}, \bibinfo
  {author} {\bibfnamefont {J.}~\bibnamefont {Jia}}, \bibinfo {author}
  {\bibfnamefont {Z.}~\bibnamefont {Zhang}}, \bibinfo {author} {\bibfnamefont
  {X.}~\bibnamefont {Zhang}}, \ and\ \bibinfo {author} {\bibfnamefont
  {Q.}~\bibnamefont {Li}},\ }\href {\doibase 10.1063/1.3511451} {\bibfield
  {journal} {\bibinfo  {journal} {Appl. Phys. Lett.}\ }\textbf {\bibinfo
  {volume} {97}},\ \bibinfo {pages} {181906} (\bibinfo {year}
  {2010})}\BibitemShut {NoStop}%
\bibitem [{\citenamefont {Jia}\ \emph {et~al.}(2011)\citenamefont {Jia},
  \citenamefont {Zhao}, \citenamefont {Xu}, \citenamefont {Di}, \citenamefont
  {Yong}, \citenamefont {Tao}, \citenamefont {Fang}, \citenamefont {Zhang},
  \citenamefont {Zhang}, \citenamefont {Zheng},\ and\ \citenamefont
  {Li}}]{jia.jj_2011}%
  \BibitemOpen
  \bibfield  {author} {\bibinfo {author} {\bibfnamefont {J.}~\bibnamefont
  {Jia}}, \bibinfo {author} {\bibfnamefont {J.}~\bibnamefont {Zhao}}, \bibinfo
  {author} {\bibfnamefont {G.}~\bibnamefont {Xu}}, \bibinfo {author}
  {\bibfnamefont {J.}~\bibnamefont {Di}}, \bibinfo {author} {\bibfnamefont
  {Z.}~\bibnamefont {Yong}}, \bibinfo {author} {\bibfnamefont {Y.}~\bibnamefont
  {Tao}}, \bibinfo {author} {\bibfnamefont {C.}~\bibnamefont {Fang}}, \bibinfo
  {author} {\bibfnamefont {Z.}~\bibnamefont {Zhang}}, \bibinfo {author}
  {\bibfnamefont {X.}~\bibnamefont {Zhang}}, \bibinfo {author} {\bibfnamefont
  {L.}~\bibnamefont {Zheng}}, \ and\ \bibinfo {author} {\bibfnamefont
  {Q.}~\bibnamefont {Li}},\ }\href {\doibase 10.1016/j.carbon.2010.11.054}
  {\bibfield  {journal} {\bibinfo  {journal} {Carbon}\ }\textbf {\bibinfo
  {volume} {49}},\ \bibinfo {pages} {1333} (\bibinfo {year}
  {2011})}\BibitemShut {NoStop}%
\bibitem [{\citenamefont {Polsen}, \citenamefont {Bedewy},\ and\ \citenamefont
  {Hart}(2013)}]{polsen.es_2013}%
  \BibitemOpen
  \bibfield  {author} {\bibinfo {author} {\bibfnamefont {E.~S.}\ \bibnamefont
  {Polsen}}, \bibinfo {author} {\bibfnamefont {M.}~\bibnamefont {Bedewy}}, \
  and\ \bibinfo {author} {\bibfnamefont {A.~J.}\ \bibnamefont {Hart}},\ }\href
  {\doibase 10.1002/smll.201202878} {\bibfield  {journal} {\bibinfo  {journal}
  {Small}\ }\textbf {\bibinfo {volume} {9}},\ \bibinfo {pages} {2564} (\bibinfo
  {year} {2013})}\BibitemShut {NoStop}%
\bibitem [{\citenamefont {Eom}\ \emph {et~al.}(2013)\citenamefont {Eom},
  \citenamefont {Nam}, \citenamefont {Jung}, \citenamefont {Kim}, \citenamefont
  {Strano}, \citenamefont {Han},\ and\ \citenamefont {Kwon}}]{eom.k_2013}%
  \BibitemOpen
  \bibfield  {author} {\bibinfo {author} {\bibfnamefont {K.}~\bibnamefont
  {Eom}}, \bibinfo {author} {\bibfnamefont {K.}~\bibnamefont {Nam}}, \bibinfo
  {author} {\bibfnamefont {H.}~\bibnamefont {Jung}}, \bibinfo {author}
  {\bibfnamefont {P.}~\bibnamefont {Kim}}, \bibinfo {author} {\bibfnamefont
  {M.~S.}\ \bibnamefont {Strano}}, \bibinfo {author} {\bibfnamefont {J.-H.}\
  \bibnamefont {Han}}, \ and\ \bibinfo {author} {\bibfnamefont
  {T.}~\bibnamefont {Kwon}},\ }\href {\doibase 10.1016/j.carbon.2013.08.030}
  {\bibfield  {journal} {\bibinfo  {journal} {Carbon}\ }\textbf {\bibinfo
  {volume} {65}},\ \bibinfo {pages} {305} (\bibinfo {year} {2013})}\BibitemShut
  {NoStop}%
\bibitem [{\citenamefont {Zeng}\ \emph {et~al.}(2010)\citenamefont {Zeng},
  \citenamefont {Ci}, \citenamefont {Carey}, \citenamefont {Vajtai},\ and\
  \citenamefont {Ajayan}}]{zeng.y_2010}%
  \BibitemOpen
  \bibfield  {author} {\bibinfo {author} {\bibfnamefont {Y.}~\bibnamefont
  {Zeng}}, \bibinfo {author} {\bibfnamefont {L.}~\bibnamefont {Ci}}, \bibinfo
  {author} {\bibfnamefont {B.~J.}\ \bibnamefont {Carey}}, \bibinfo {author}
  {\bibfnamefont {R.}~\bibnamefont {Vajtai}}, \ and\ \bibinfo {author}
  {\bibfnamefont {P.~M.}\ \bibnamefont {Ajayan}},\ }\href {\doibase
  10.1021/nn101650p} {\bibfield  {journal} {\bibinfo  {journal} {ACS Nano}\
  }\textbf {\bibinfo {volume} {4}},\ \bibinfo {pages} {6798} (\bibinfo {year}
  {2010})}\BibitemShut {NoStop}%
\bibitem [{\citenamefont {Mantena}\ \emph {et~al.}(2013)\citenamefont
  {Mantena}, \citenamefont {Tadepalli}, \citenamefont {Pramanik}, \citenamefont
  {Boddu}, \citenamefont {Brenner}, \citenamefont {Stephenson},\ and\
  \citenamefont {Kumar}}]{mantena.pr_2013}%
  \BibitemOpen
  \bibfield  {author} {\bibinfo {author} {\bibfnamefont {P.~R.}\ \bibnamefont
  {Mantena}}, \bibinfo {author} {\bibfnamefont {T.}~\bibnamefont {Tadepalli}},
  \bibinfo {author} {\bibfnamefont {B.}~\bibnamefont {Pramanik}}, \bibinfo
  {author} {\bibfnamefont {V.~M.}\ \bibnamefont {Boddu}}, \bibinfo {author}
  {\bibfnamefont {M.~W.}\ \bibnamefont {Brenner}}, \bibinfo {author}
  {\bibfnamefont {L.~D.}\ \bibnamefont {Stephenson}}, \ and\ \bibinfo {author}
  {\bibfnamefont {A.}~\bibnamefont {Kumar}},\ }\href {\doibase
  10.1155/2013/259458} {\bibfield  {journal} {\bibinfo  {journal} {J.
  Nanomater.}\ }\textbf {\bibinfo {volume} {2013}},\ \bibinfo {pages} {259458}
  (\bibinfo {year} {2013})}\BibitemShut {NoStop}%
\bibitem [{\citenamefont {Veedu}\ \emph {et~al.}(2006)\citenamefont {Veedu},
  \citenamefont {Cao}, \citenamefont {Li}, \citenamefont {Ma}, \citenamefont
  {Soldano}, \citenamefont {Kar}, \citenamefont {Ajayan},\ and\ \citenamefont
  {Ghasemi-Nejhad}}]{veedu.vp_2006}%
  \BibitemOpen
  \bibfield  {author} {\bibinfo {author} {\bibfnamefont {V.~P.}\ \bibnamefont
  {Veedu}}, \bibinfo {author} {\bibfnamefont {A.}~\bibnamefont {Cao}}, \bibinfo
  {author} {\bibfnamefont {X.}~\bibnamefont {Li}}, \bibinfo {author}
  {\bibfnamefont {K.}~\bibnamefont {Ma}}, \bibinfo {author} {\bibfnamefont
  {C.}~\bibnamefont {Soldano}}, \bibinfo {author} {\bibfnamefont
  {S.}~\bibnamefont {Kar}}, \bibinfo {author} {\bibfnamefont {P.~M.}\
  \bibnamefont {Ajayan}}, \ and\ \bibinfo {author} {\bibfnamefont {M.~N.}\
  \bibnamefont {Ghasemi-Nejhad}},\ }\href {\doibase 10.1038/nmat1650}
  {\bibfield  {journal} {\bibinfo  {journal} {Nat. Mater.}\ }\textbf {\bibinfo
  {volume} {5}},\ \bibinfo {pages} {457} (\bibinfo {year} {2006})}\BibitemShut
  {NoStop}%
\bibitem [{\citenamefont {Kim}\ \emph {et~al.}(2015)\citenamefont {Kim},
  \citenamefont {Mantena}, \citenamefont {Daryadel}, \citenamefont {Boddu},
  \citenamefont {Brenner},\ and\ \citenamefont {Patel}}]{kim.k_2015}%
  \BibitemOpen
  \bibfield  {author} {\bibinfo {author} {\bibfnamefont {K.}~\bibnamefont
  {Kim}}, \bibinfo {author} {\bibfnamefont {P.~R.}\ \bibnamefont {Mantena}},
  \bibinfo {author} {\bibfnamefont {S.~S.}\ \bibnamefont {Daryadel}}, \bibinfo
  {author} {\bibfnamefont {V.~M.}\ \bibnamefont {Boddu}}, \bibinfo {author}
  {\bibfnamefont {M.~W.}\ \bibnamefont {Brenner}}, \ and\ \bibinfo {author}
  {\bibfnamefont {J.~S.}\ \bibnamefont {Patel}},\ }\href {\doibase
  10.1155/2015/480549} {\bibfield  {journal} {\bibinfo  {journal} {J.
  Nanomater.}\ }\textbf {\bibinfo {volume} {2015}},\ \bibinfo {pages} {480549}
  (\bibinfo {year} {2015})}\BibitemShut {NoStop}%
\bibitem [{\citenamefont {Boddu}\ and\ \citenamefont
  {Brenner}(2016)}]{boddu.vm_2016}%
  \BibitemOpen
  \bibfield  {author} {\bibinfo {author} {\bibfnamefont {V.~M.}\ \bibnamefont
  {Boddu}}\ and\ \bibinfo {author} {\bibfnamefont {M.~W.}\ \bibnamefont
  {Brenner}},\ }\href {\doibase 10.1007/s00339-015-9571-8} {\bibfield
  {journal} {\bibinfo  {journal} {Appl. Phys. A}\ }\textbf {\bibinfo {volume}
  {122}},\ \bibinfo {pages} {88} (\bibinfo {year} {2016})}\BibitemShut
  {NoStop}%
\bibitem [{\citenamefont {Teo}\ \emph {et~al.}(2007)\citenamefont {Teo},
  \citenamefont {Yung}, \citenamefont {Chua},\ and\ \citenamefont
  {Tay}}]{teo.eht_2007}%
  \BibitemOpen
  \bibfield  {author} {\bibinfo {author} {\bibfnamefont {E.~H.~T.}\
  \bibnamefont {Teo}}, \bibinfo {author} {\bibfnamefont {W.~K.~P.}\
  \bibnamefont {Yung}}, \bibinfo {author} {\bibfnamefont {D.~H.~C.}\
  \bibnamefont {Chua}}, \ and\ \bibinfo {author} {\bibfnamefont {B.~K.}\
  \bibnamefont {Tay}},\ }\href {\doibase 10.1002/adma.200700351} {\bibfield
  {journal} {\bibinfo  {journal} {Adv. Mater.}\ }\textbf {\bibinfo {volume}
  {19}},\ \bibinfo {pages} {2941} (\bibinfo {year} {2007})}\BibitemShut
  {NoStop}%
\bibitem [{\citenamefont {Silva}\ \emph {et~al.}(2011)\citenamefont {Silva},
  \citenamefont {Rodrigues}, \citenamefont {Fantini}, \citenamefont {Borges},
  \citenamefont {Pimenta}, \citenamefont {Carey}, \citenamefont {Ci},\ and\
  \citenamefont {Ajayan}}]{silva.gg_2011}%
  \BibitemOpen
  \bibfield  {author} {\bibinfo {author} {\bibfnamefont {G.~G.}\ \bibnamefont
  {Silva}}, \bibinfo {author} {\bibfnamefont {M.-T.~F.}\ \bibnamefont
  {Rodrigues}}, \bibinfo {author} {\bibfnamefont {C.}~\bibnamefont {Fantini}},
  \bibinfo {author} {\bibfnamefont {R.~S.}\ \bibnamefont {Borges}}, \bibinfo
  {author} {\bibfnamefont {M.~A.}\ \bibnamefont {Pimenta}}, \bibinfo {author}
  {\bibfnamefont {B.~J.}\ \bibnamefont {Carey}}, \bibinfo {author}
  {\bibfnamefont {L.}~\bibnamefont {Ci}}, \ and\ \bibinfo {author}
  {\bibfnamefont {P.~M.}\ \bibnamefont {Ajayan}},\ }\href {\doibase
  10.1002/mame.201000276} {\bibfield  {journal} {\bibinfo  {journal} {Macromol.
  Mater. Eng.}\ }\textbf {\bibinfo {volume} {296}},\ \bibinfo {pages} {53}
  (\bibinfo {year} {2011})}\BibitemShut {NoStop}%
\bibitem [{\citenamefont {\"{U}rk}\ \emph {et~al.}(2016)\citenamefont
  {\"{U}rk}, \citenamefont {Demir}, \citenamefont {Bulut}, \citenamefont
  {\c{C}ak{\i}ro\u{g}lu}, \citenamefont {Cebeci}, \citenamefont
  {\"{O}ve\c{c}o\u{g}lu},\ and\ \citenamefont {Cebeci}}]{urk.d_2016}%
  \BibitemOpen
  \bibfield  {author} {\bibinfo {author} {\bibfnamefont {D.}~\bibnamefont
  {\"{U}rk}}, \bibinfo {author} {\bibfnamefont {E.}~\bibnamefont {Demir}},
  \bibinfo {author} {\bibfnamefont {O.}~\bibnamefont {Bulut}}, \bibinfo
  {author} {\bibfnamefont {D.}~\bibnamefont {\c{C}ak{\i}ro\u{g}lu}}, \bibinfo
  {author} {\bibfnamefont {F.~{\c{C}}.}\ \bibnamefont {Cebeci}}, \bibinfo
  {author} {\bibfnamefont {M.~L.}\ \bibnamefont {\"{O}ve\c{c}o\u{g}lu}}, \ and\
  \bibinfo {author} {\bibfnamefont {H.}~\bibnamefont {Cebeci}},\ }\href
  {\doibase 10.1016/j.compstruct.2016.05.087} {\bibfield  {journal} {\bibinfo
  {journal} {Compos. Struct.}\ }\textbf {\bibinfo {volume} {155}},\ \bibinfo
  {pages} {255} (\bibinfo {year} {2016})}\BibitemShut {NoStop}%
\bibitem [{\citenamefont {Xu}\ \emph {et~al.}(2011{\natexlab{a}})\citenamefont
  {Xu}, \citenamefont {Futaba}, \citenamefont {Yumura},\ and\ \citenamefont
  {Hata}}]{xu.m_20111}%
  \BibitemOpen
  \bibfield  {author} {\bibinfo {author} {\bibfnamefont {M.}~\bibnamefont
  {Xu}}, \bibinfo {author} {\bibfnamefont {D.~N.}\ \bibnamefont {Futaba}},
  \bibinfo {author} {\bibfnamefont {M.}~\bibnamefont {Yumura}}, \ and\ \bibinfo
  {author} {\bibfnamefont {K.}~\bibnamefont {Hata}},\ }\href {\doibase
  10.1002/adma.201101412} {\bibfield  {journal} {\bibinfo  {journal} {Adv.
  Mater.}\ }\textbf {\bibinfo {volume} {23}},\ \bibinfo {pages} {3686}
  (\bibinfo {year} {2011}{\natexlab{a}})}\BibitemShut {NoStop}%
\bibitem [{\citenamefont {Xu}\ \emph {et~al.}(2011{\natexlab{b}})\citenamefont
  {Xu}, \citenamefont {Futaba}, \citenamefont {Yumura},\ and\ \citenamefont
  {Hata}}]{xu.m_2011}%
  \BibitemOpen
  \bibfield  {author} {\bibinfo {author} {\bibfnamefont {M.}~\bibnamefont
  {Xu}}, \bibinfo {author} {\bibfnamefont {D.~N.}\ \bibnamefont {Futaba}},
  \bibinfo {author} {\bibfnamefont {M.}~\bibnamefont {Yumura}}, \ and\ \bibinfo
  {author} {\bibfnamefont {K.}~\bibnamefont {Hata}},\ }\href {\doibase
  10.1021/nl201632m} {\bibfield  {journal} {\bibinfo  {journal} {Nano Lett.}\
  }\textbf {\bibinfo {volume} {11}},\ \bibinfo {pages} {3279} (\bibinfo {year}
  {2011}{\natexlab{b}})}\BibitemShut {NoStop}%
\bibitem [{\citenamefont {Gui}\ \emph {et~al.}(2010)\citenamefont {Gui},
  \citenamefont {Wei}, \citenamefont {Wang}, \citenamefont {Cao}, \citenamefont
  {Zhu}, \citenamefont {Jia}, \citenamefont {Shu},\ and\ \citenamefont
  {Wu}}]{gui.xc_2010}%
  \BibitemOpen
  \bibfield  {author} {\bibinfo {author} {\bibfnamefont {X.}~\bibnamefont
  {Gui}}, \bibinfo {author} {\bibfnamefont {J.}~\bibnamefont {Wei}}, \bibinfo
  {author} {\bibfnamefont {K.}~\bibnamefont {Wang}}, \bibinfo {author}
  {\bibfnamefont {A.}~\bibnamefont {Cao}}, \bibinfo {author} {\bibfnamefont
  {H.}~\bibnamefont {Zhu}}, \bibinfo {author} {\bibfnamefont {Y.}~\bibnamefont
  {Jia}}, \bibinfo {author} {\bibfnamefont {Q.}~\bibnamefont {Shu}}, \ and\
  \bibinfo {author} {\bibfnamefont {D.}~\bibnamefont {Wu}},\ }\href {\doibase
  10.1002/adma.200902986} {\bibfield  {journal} {\bibinfo  {journal} {Adv.
  Mater.}\ }\textbf {\bibinfo {volume} {22}},\ \bibinfo {pages} {617} (\bibinfo
  {year} {2010})}\BibitemShut {NoStop}%
\bibitem [{\citenamefont {Kim}, \citenamefont {Oh},\ and\ \citenamefont
  {Islam}(2012)}]{kim.kh_2012}%
  \BibitemOpen
  \bibfield  {author} {\bibinfo {author} {\bibfnamefont {K.~H.}\ \bibnamefont
  {Kim}}, \bibinfo {author} {\bibfnamefont {Y.}~\bibnamefont {Oh}}, \ and\
  \bibinfo {author} {\bibfnamefont {M.~F.}\ \bibnamefont {Islam}},\ }\href
  {\doibase 10.1038/nnano.2012.118} {\bibfield  {journal} {\bibinfo  {journal}
  {Nat. Nanotechnol.}\ }\textbf {\bibinfo {volume} {7}},\ \bibinfo {pages}
  {562} (\bibinfo {year} {2012})}\BibitemShut {NoStop}%
\bibitem [{\citenamefont {Dai}\ \emph {et~al.}(2016)\citenamefont {Dai},
  \citenamefont {Liu}, \citenamefont {Qi}, \citenamefont {Kuang}, \citenamefont
  {Wei}, \citenamefont {Zhu},\ and\ \citenamefont {Zhang}}]{dai.zh_2016}%
  \BibitemOpen
  \bibfield  {author} {\bibinfo {author} {\bibfnamefont {Z.}~\bibnamefont
  {Dai}}, \bibinfo {author} {\bibfnamefont {L.}~\bibnamefont {Liu}}, \bibinfo
  {author} {\bibfnamefont {X.}~\bibnamefont {Qi}}, \bibinfo {author}
  {\bibfnamefont {J.}~\bibnamefont {Kuang}}, \bibinfo {author} {\bibfnamefont
  {Y.}~\bibnamefont {Wei}}, \bibinfo {author} {\bibfnamefont {H.}~\bibnamefont
  {Zhu}}, \ and\ \bibinfo {author} {\bibfnamefont {Z.}~\bibnamefont {Zhang}},\
  }\href {\doibase 10.1038/srep18930} {\bibfield  {journal} {\bibinfo
  {journal} {Sci. Rep.}\ }\textbf {\bibinfo {volume} {6}},\ \bibinfo {pages}
  {18930} (\bibinfo {year} {2016})}\BibitemShut {NoStop}%
\bibitem [{\citenamefont {Wang}\ \emph {et~al.}(2017)\citenamefont {Wang},
  \citenamefont {Lu}, \citenamefont {Di}, \citenamefont {Li}, \citenamefont
  {Zhang}, \citenamefont {Li}, \citenamefont {Zhang}, \citenamefont {Zheng},\
  and\ \citenamefont {Li}}]{wang.h_2017}%
  \BibitemOpen
  \bibfield  {author} {\bibinfo {author} {\bibfnamefont {H.}~\bibnamefont
  {Wang}}, \bibinfo {author} {\bibfnamefont {W.}~\bibnamefont {Lu}}, \bibinfo
  {author} {\bibfnamefont {J.}~\bibnamefont {Di}}, \bibinfo {author}
  {\bibfnamefont {D.}~\bibnamefont {Li}}, \bibinfo {author} {\bibfnamefont
  {X.}~\bibnamefont {Zhang}}, \bibinfo {author} {\bibfnamefont
  {M.}~\bibnamefont {Li}}, \bibinfo {author} {\bibfnamefont {Z.}~\bibnamefont
  {Zhang}}, \bibinfo {author} {\bibfnamefont {L.}~\bibnamefont {Zheng}}, \ and\
  \bibinfo {author} {\bibfnamefont {Q.}~\bibnamefont {Li}},\ }\href {\doibase
  10.1002/adfm.201606220} {\bibfield  {journal} {\bibinfo  {journal} {Adv.
  Funct. Mater.}\ }\textbf {\bibinfo {volume} {27}},\ \bibinfo {pages}
  {1606220} (\bibinfo {year} {2017})}\BibitemShut {NoStop}%
\bibitem [{\citenamefont {Hough}\ \emph {et~al.}(2006)\citenamefont {Hough},
  \citenamefont {Islam}, \citenamefont {Hammouda}, \citenamefont {Yodh},\ and\
  \citenamefont {Heiney}}]{hough.la_2006}%
  \BibitemOpen
  \bibfield  {author} {\bibinfo {author} {\bibfnamefont {L.~A.}\ \bibnamefont
  {Hough}}, \bibinfo {author} {\bibfnamefont {M.~F.}\ \bibnamefont {Islam}},
  \bibinfo {author} {\bibfnamefont {B.}~\bibnamefont {Hammouda}}, \bibinfo
  {author} {\bibfnamefont {A.~G.}\ \bibnamefont {Yodh}}, \ and\ \bibinfo
  {author} {\bibfnamefont {P.~A.}\ \bibnamefont {Heiney}},\ }\href {\doibase
  10.1021/nl051871f} {\bibfield  {journal} {\bibinfo  {journal} {Nano Lett.}\
  }\textbf {\bibinfo {volume} {6}},\ \bibinfo {pages} {313} (\bibinfo {year}
  {2006})}\BibitemShut {NoStop}%
\bibitem [{\citenamefont {Bryning}\ \emph {et~al.}(2007)\citenamefont
  {Bryning}, \citenamefont {Milkie}, \citenamefont {Islam}, \citenamefont
  {Hough}, \citenamefont {Kikkawa},\ and\ \citenamefont
  {Yodh}}]{bryning.mb_2007}%
  \BibitemOpen
  \bibfield  {author} {\bibinfo {author} {\bibfnamefont {M.~B.}\ \bibnamefont
  {Bryning}}, \bibinfo {author} {\bibfnamefont {D.~E.}\ \bibnamefont {Milkie}},
  \bibinfo {author} {\bibfnamefont {M.~F.}\ \bibnamefont {Islam}}, \bibinfo
  {author} {\bibfnamefont {L.~A.}\ \bibnamefont {Hough}}, \bibinfo {author}
  {\bibfnamefont {J.~M.}\ \bibnamefont {Kikkawa}}, \ and\ \bibinfo {author}
  {\bibfnamefont {A.~G.}\ \bibnamefont {Yodh}},\ }\href {\doibase
  10.1002/adma.200601748} {\bibfield  {journal} {\bibinfo  {journal} {Adv.
  Mater.}\ }\textbf {\bibinfo {volume} {19}},\ \bibinfo {pages} {661} (\bibinfo
  {year} {2007})}\BibitemShut {NoStop}%
\bibitem [{\citenamefont {Zou}\ \emph {et~al.}(2010)\citenamefont {Zou},
  \citenamefont {Liu}, \citenamefont {Karakoti}, \citenamefont {Kumar},
  \citenamefont {Joung}, \citenamefont {Li}, \citenamefont {Khondaker},
  \citenamefont {Seal},\ and\ \citenamefont {Zhai}}]{zou.jh_2010}%
  \BibitemOpen
  \bibfield  {author} {\bibinfo {author} {\bibfnamefont {J.}~\bibnamefont
  {Zou}}, \bibinfo {author} {\bibfnamefont {J.}~\bibnamefont {Liu}}, \bibinfo
  {author} {\bibfnamefont {A.~S.}\ \bibnamefont {Karakoti}}, \bibinfo {author}
  {\bibfnamefont {A.}~\bibnamefont {Kumar}}, \bibinfo {author} {\bibfnamefont
  {D.}~\bibnamefont {Joung}}, \bibinfo {author} {\bibfnamefont
  {Q.}~\bibnamefont {Li}}, \bibinfo {author} {\bibfnamefont {S.~I.}\
  \bibnamefont {Khondaker}}, \bibinfo {author} {\bibfnamefont {S.}~\bibnamefont
  {Seal}}, \ and\ \bibinfo {author} {\bibfnamefont {L.}~\bibnamefont {Zhai}},\
  }\href {\doibase 10.1021/nn102246a} {\bibfield  {journal} {\bibinfo
  {journal} {ACS Nano}\ }\textbf {\bibinfo {volume} {4}},\ \bibinfo {pages}
  {7293} (\bibinfo {year} {2010})}\BibitemShut {NoStop}%
\bibitem [{\citenamefont {Kim}, \citenamefont {Vural},\ and\ \citenamefont
  {Islam}(2011)}]{kim.kh_2011}%
  \BibitemOpen
  \bibfield  {author} {\bibinfo {author} {\bibfnamefont {K.~H.}\ \bibnamefont
  {Kim}}, \bibinfo {author} {\bibfnamefont {M.}~\bibnamefont {Vural}}, \ and\
  \bibinfo {author} {\bibfnamefont {M.~F.}\ \bibnamefont {Islam}},\ }\href
  {\doibase 10.1002/adma.201100310} {\bibfield  {journal} {\bibinfo  {journal}
  {Adv. Mater.}\ }\textbf {\bibinfo {volume} {23}},\ \bibinfo {pages} {2865}
  (\bibinfo {year} {2011})}\BibitemShut {NoStop}%
\bibitem [{\citenamefont {Faraji}\ \emph {et~al.}(2015)\citenamefont {Faraji},
  \citenamefont {Stano}, \citenamefont {Yildiz}, \citenamefont {Li},
  \citenamefont {Zhu},\ and\ \citenamefont {Bradford}}]{faraji.s_2015}%
  \BibitemOpen
  \bibfield  {author} {\bibinfo {author} {\bibfnamefont {S.}~\bibnamefont
  {Faraji}}, \bibinfo {author} {\bibfnamefont {K.~L.}\ \bibnamefont {Stano}},
  \bibinfo {author} {\bibfnamefont {O.}~\bibnamefont {Yildiz}}, \bibinfo
  {author} {\bibfnamefont {A.}~\bibnamefont {Li}}, \bibinfo {author}
  {\bibfnamefont {Y.}~\bibnamefont {Zhu}}, \ and\ \bibinfo {author}
  {\bibfnamefont {P.~D.}\ \bibnamefont {Bradford}},\ }\href {\doibase
  10.1039/c5nr03899e} {\bibfield  {journal} {\bibinfo  {journal} {Nanoscale}\
  }\textbf {\bibinfo {volume} {7}},\ \bibinfo {pages} {17038} (\bibinfo {year}
  {2015})}\BibitemShut {NoStop}%
\bibitem [{\citenamefont {Whitby}\ \emph {et~al.}(2008)\citenamefont {Whitby},
  \citenamefont {Fukuda}, \citenamefont {Maekawa}, \citenamefont {James},\ and\
  \citenamefont {Mikhalovsky}}]{whitby.rld_2008}%
  \BibitemOpen
  \bibfield  {author} {\bibinfo {author} {\bibfnamefont {R.~L.~D.}\
  \bibnamefont {Whitby}}, \bibinfo {author} {\bibfnamefont {T.}~\bibnamefont
  {Fukuda}}, \bibinfo {author} {\bibfnamefont {T.}~\bibnamefont {Maekawa}},
  \bibinfo {author} {\bibfnamefont {S.~L.}\ \bibnamefont {James}}, \ and\
  \bibinfo {author} {\bibfnamefont {S.~V.}\ \bibnamefont {Mikhalovsky}},\
  }\href {\doibase 10.1016/j.carbon.2008.02.028} {\bibfield  {journal}
  {\bibinfo  {journal} {Carbon}\ }\textbf {\bibinfo {volume} {46}},\ \bibinfo
  {pages} {949} (\bibinfo {year} {2008})}\BibitemShut {NoStop}%
\bibitem [{\citenamefont {Whitby}\ \emph {et~al.}(2010)\citenamefont {Whitby},
  \citenamefont {Fukuda}, \citenamefont {Maekawa}, \citenamefont
  {Mikhalovsky},\ and\ \citenamefont {Cundy}}]{whitby.rld_2010}%
  \BibitemOpen
  \bibfield  {author} {\bibinfo {author} {\bibfnamefont {R.~L.~D.}\
  \bibnamefont {Whitby}}, \bibinfo {author} {\bibfnamefont {T.}~\bibnamefont
  {Fukuda}}, \bibinfo {author} {\bibfnamefont {T.}~\bibnamefont {Maekawa}},
  \bibinfo {author} {\bibfnamefont {S.~V.}\ \bibnamefont {Mikhalovsky}}, \ and\
  \bibinfo {author} {\bibfnamefont {A.~B.}\ \bibnamefont {Cundy}},\ }\href
  {\doibase 10.1088/0957-4484/21/7/075707} {\bibfield  {journal} {\bibinfo
  {journal} {Nanotechnology}\ }\textbf {\bibinfo {volume} {21}},\ \bibinfo
  {pages} {075707} (\bibinfo {year} {2010})}\BibitemShut {NoStop}%
\bibitem [{\citenamefont {Kim}, \citenamefont {Tsui},\ and\ \citenamefont
  {Islam}(2017)}]{kim.kh_2017}%
  \BibitemOpen
  \bibfield  {author} {\bibinfo {author} {\bibfnamefont {K.~H.}\ \bibnamefont
  {Kim}}, \bibinfo {author} {\bibfnamefont {M.~N.}\ \bibnamefont {Tsui}}, \
  and\ \bibinfo {author} {\bibfnamefont {M.~F.}\ \bibnamefont {Islam}},\ }\href
  {\doibase 10.1021/acs.chemmater.6b04460} {\bibfield  {journal} {\bibinfo
  {journal} {Chem. Mater.}\ }\textbf {\bibinfo {volume} {29}},\ \bibinfo
  {pages} {2748} (\bibinfo {year} {2017})}\BibitemShut {NoStop}%
\bibitem [{\citenamefont {Li}, \citenamefont {Kinloch},\ and\ \citenamefont
  {Windle}(2004)}]{li.yl_2004}%
  \BibitemOpen
  \bibfield  {author} {\bibinfo {author} {\bibfnamefont {Y.-L.}\ \bibnamefont
  {Li}}, \bibinfo {author} {\bibfnamefont {I.~A.}\ \bibnamefont {Kinloch}}, \
  and\ \bibinfo {author} {\bibfnamefont {A.~H.}\ \bibnamefont {Windle}},\
  }\href {\doibase 10.1126/science.1094982} {\bibfield  {journal} {\bibinfo
  {journal} {Science}\ }\textbf {\bibinfo {volume} {304}},\ \bibinfo {pages}
  {276} (\bibinfo {year} {2004})}\BibitemShut {NoStop}%
\bibitem [{\citenamefont {Motta}\ \emph {et~al.}(2005)\citenamefont {Motta},
  \citenamefont {Li}, \citenamefont {Kinloch},\ and\ \citenamefont
  {Windle}}]{motta.m_2005}%
  \BibitemOpen
  \bibfield  {author} {\bibinfo {author} {\bibfnamefont {M.}~\bibnamefont
  {Motta}}, \bibinfo {author} {\bibfnamefont {Y.-L.}\ \bibnamefont {Li}},
  \bibinfo {author} {\bibfnamefont {I.}~\bibnamefont {Kinloch}}, \ and\
  \bibinfo {author} {\bibfnamefont {A.}~\bibnamefont {Windle}},\ }\href@noop {}
  {\bibfield  {journal} {\bibinfo  {journal} {Nano Lett.}\ }\textbf {\bibinfo
  {volume} {5}},\ \bibinfo {pages} {1529} (\bibinfo {year} {2005})}\BibitemShut
  {NoStop}%
\bibitem [{\citenamefont {Zou}\ \emph {et~al.}(2016)\citenamefont {Zou},
  \citenamefont {Zhang}, \citenamefont {Zhao}, \citenamefont {Lei},
  \citenamefont {Zhao}, \citenamefont {Zhu},\ and\ \citenamefont
  {Li}}]{zou.jy_2016}%
  \BibitemOpen
  \bibfield  {author} {\bibinfo {author} {\bibfnamefont {J.}~\bibnamefont
  {Zou}}, \bibinfo {author} {\bibfnamefont {X.}~\bibnamefont {Zhang}}, \bibinfo
  {author} {\bibfnamefont {J.}~\bibnamefont {Zhao}}, \bibinfo {author}
  {\bibfnamefont {C.}~\bibnamefont {Lei}}, \bibinfo {author} {\bibfnamefont
  {Y.}~\bibnamefont {Zhao}}, \bibinfo {author} {\bibfnamefont {Y.}~\bibnamefont
  {Zhu}}, \ and\ \bibinfo {author} {\bibfnamefont {Q.}~\bibnamefont {Li}},\
  }\href {\doibase 10.1016/j.compscitech.2016.09.019} {\bibfield  {journal}
  {\bibinfo  {journal} {Compos. Sci. Technol.}\ }\textbf {\bibinfo {volume}
  {135}},\ \bibinfo {pages} {123} (\bibinfo {year} {2016})}\BibitemShut
  {NoStop}%
\bibitem [{\citenamefont {Zou}\ \emph {et~al.}(2017)\citenamefont {Zou},
  \citenamefont {Zhang}, \citenamefont {Xu}, \citenamefont {Zhao},
  \citenamefont {Zhu},\ and\ \citenamefont {Li}}]{zou.jy_2017}%
  \BibitemOpen
  \bibfield  {author} {\bibinfo {author} {\bibfnamefont {J.}~\bibnamefont
  {Zou}}, \bibinfo {author} {\bibfnamefont {X.}~\bibnamefont {Zhang}}, \bibinfo
  {author} {\bibfnamefont {C.}~\bibnamefont {Xu}}, \bibinfo {author}
  {\bibfnamefont {J.}~\bibnamefont {Zhao}}, \bibinfo {author} {\bibfnamefont
  {Y.~T.}\ \bibnamefont {Zhu}}, \ and\ \bibinfo {author} {\bibfnamefont
  {Q.}~\bibnamefont {Li}},\ }\href {\doibase 10.1016/j.carbon.2017.05.091}
  {\bibfield  {journal} {\bibinfo  {journal} {Carbon}\ }\textbf {\bibinfo
  {volume} {121}},\ \bibinfo {pages} {242} (\bibinfo {year}
  {2017})}\BibitemShut {NoStop}%
\bibitem [{\citenamefont {Cranford}\ and\ \citenamefont
  {Buehler}(2010)}]{cranford.sw_2010}%
  \BibitemOpen
  \bibfield  {author} {\bibinfo {author} {\bibfnamefont {S.~W.}\ \bibnamefont
  {Cranford}}\ and\ \bibinfo {author} {\bibfnamefont {M.~J.}\ \bibnamefont
  {Buehler}},\ }\href {\doibase 10.1088/0957-4484/21/26/265706} {\bibfield
  {journal} {\bibinfo  {journal} {Nanotechnology}\ }\textbf {\bibinfo {volume}
  {21}},\ \bibinfo {pages} {265706} (\bibinfo {year} {2010})}\BibitemShut
  {NoStop}%
\bibitem [{\citenamefont {Xie}\ \emph {et~al.}(2011)\citenamefont {Xie},
  \citenamefont {Liu}, \citenamefont {Ding}, \citenamefont {Zheng},\ and\
  \citenamefont {Xu}}]{xie.b_2011}%
  \BibitemOpen
  \bibfield  {author} {\bibinfo {author} {\bibfnamefont {B.}~\bibnamefont
  {Xie}}, \bibinfo {author} {\bibfnamefont {Y.}~\bibnamefont {Liu}}, \bibinfo
  {author} {\bibfnamefont {Y.}~\bibnamefont {Ding}}, \bibinfo {author}
  {\bibfnamefont {Q.}~\bibnamefont {Zheng}}, \ and\ \bibinfo {author}
  {\bibfnamefont {Z.}~\bibnamefont {Xu}},\ }\href {\doibase 10.1039/c1sm06034a}
  {\bibfield  {journal} {\bibinfo  {journal} {Soft Matter}\ }\textbf {\bibinfo
  {volume} {7}},\ \bibinfo {pages} {10039} (\bibinfo {year}
  {2011})}\BibitemShut {NoStop}%
\bibitem [{\citenamefont {Yang}, \citenamefont {He},\ and\ \citenamefont
  {Gao}(2011)}]{yang.xd_2011}%
  \BibitemOpen
  \bibfield  {author} {\bibinfo {author} {\bibfnamefont {X.}~\bibnamefont
  {Yang}}, \bibinfo {author} {\bibfnamefont {P.}~\bibnamefont {He}}, \ and\
  \bibinfo {author} {\bibfnamefont {H.}~\bibnamefont {Gao}},\ }\href {\doibase
  10.1007/s12274-011-0169-y} {\bibfield  {journal} {\bibinfo  {journal} {Nano
  Res.}\ }\textbf {\bibinfo {volume} {4}},\ \bibinfo {pages} {1191} (\bibinfo
  {year} {2011})}\BibitemShut {NoStop}%
\bibitem [{\citenamefont {Li}\ and\ \citenamefont
  {Kr\"{o}ger}(2012{\natexlab{b}})}]{li.y_20121}%
  \BibitemOpen
  \bibfield  {author} {\bibinfo {author} {\bibfnamefont {Y.}~\bibnamefont
  {Li}}\ and\ \bibinfo {author} {\bibfnamefont {M.}~\bibnamefont
  {Kr\"{o}ger}},\ }\href {\doibase 10.1016/j.carbon.2011.12.027} {\bibfield
  {journal} {\bibinfo  {journal} {Carbon}\ }\textbf {\bibinfo {volume} {50}},\
  \bibinfo {pages} {1793} (\bibinfo {year} {2012}{\natexlab{b}})}\BibitemShut
  {NoStop}%
\bibitem [{\citenamefont {Chen}\ \emph {et~al.}(2016)\citenamefont {Chen},
  \citenamefont {Zhang}, \citenamefont {Chen}, \citenamefont {Becton},
  \citenamefont {Wang},\ and\ \citenamefont {Nie}}]{chen.h_2016}%
  \BibitemOpen
  \bibfield  {author} {\bibinfo {author} {\bibfnamefont {H.}~\bibnamefont
  {Chen}}, \bibinfo {author} {\bibfnamefont {L.}~\bibnamefont {Zhang}},
  \bibinfo {author} {\bibfnamefont {J.}~\bibnamefont {Chen}}, \bibinfo {author}
  {\bibfnamefont {M.}~\bibnamefont {Becton}}, \bibinfo {author} {\bibfnamefont
  {X.}~\bibnamefont {Wang}}, \ and\ \bibinfo {author} {\bibfnamefont
  {H.}~\bibnamefont {Nie}},\ }\href {\doibase 10.1016/j.carbon.2016.07.055}
  {\bibfield  {journal} {\bibinfo  {journal} {Carbon}\ }\textbf {\bibinfo
  {volume} {109}},\ \bibinfo {pages} {19} (\bibinfo {year} {2016})}\BibitemShut
  {NoStop}%
\bibitem [{\citenamefont {Thevamaran}\ \emph {et~al.}(2017)\citenamefont
  {Thevamaran}, \citenamefont {Saini}, \citenamefont {Karakaya}, \citenamefont
  {Zhu}, \citenamefont {Podila}, \citenamefont {Rao},\ and\ \citenamefont
  {Daraio}}]{thevamaran.r_2017}%
  \BibitemOpen
  \bibfield  {author} {\bibinfo {author} {\bibfnamefont {R.}~\bibnamefont
  {Thevamaran}}, \bibinfo {author} {\bibfnamefont {D.}~\bibnamefont {Saini}},
  \bibinfo {author} {\bibfnamefont {M.}~\bibnamefont {Karakaya}}, \bibinfo
  {author} {\bibfnamefont {J.}~\bibnamefont {Zhu}}, \bibinfo {author}
  {\bibfnamefont {R.}~\bibnamefont {Podila}}, \bibinfo {author} {\bibfnamefont
  {A.~M.}\ \bibnamefont {Rao}}, \ and\ \bibinfo {author} {\bibfnamefont
  {C.}~\bibnamefont {Daraio}},\ }\href {\doibase 10.1088/1361-6528/aa6904}
  {\bibfield  {journal} {\bibinfo  {journal} {Nanotechnology}\ }\textbf
  {\bibinfo {volume} {28}},\ \bibinfo {pages} {184002} (\bibinfo {year}
  {2017})}\BibitemShut {NoStop}%
\bibitem [{\citenamefont {Zhang}\ \emph {et~al.}(2012)\citenamefont {Zhang},
  \citenamefont {Zhang}, \citenamefont {Liu},\ and\ \citenamefont
  {Fan}}]{zhang.l_2012}%
  \BibitemOpen
  \bibfield  {author} {\bibinfo {author} {\bibfnamefont {L.}~\bibnamefont
  {Zhang}}, \bibinfo {author} {\bibfnamefont {G.}~\bibnamefont {Zhang}},
  \bibinfo {author} {\bibfnamefont {C.}~\bibnamefont {Liu}}, \ and\ \bibinfo
  {author} {\bibfnamefont {S.}~\bibnamefont {Fan}},\ }\href {\doibase
  10.1021/nl3023274} {\bibfield  {journal} {\bibinfo  {journal} {Nano Lett.}\
  }\textbf {\bibinfo {volume} {12}},\ \bibinfo {pages} {4848} (\bibinfo {year}
  {2012})}\BibitemShut {NoStop}%
\bibitem [{\citenamefont {Ji}, \citenamefont {Lin},\ and\ \citenamefont
  {Wong}(2006)}]{ji.yg_2006}%
  \BibitemOpen
  \bibfield  {author} {\bibinfo {author} {\bibfnamefont {Y.}~\bibnamefont
  {Ji}}, \bibinfo {author} {\bibfnamefont {Y.-J.}\ \bibnamefont {Lin}}, \ and\
  \bibinfo {author} {\bibfnamefont {J.~S.~C.}\ \bibnamefont {Wong}},\ }in\
  \href {\doibase 10.1109/nems.2006.334882} {\emph {\bibinfo {booktitle} {2006
  1st IEEE International Conference on Nano/Micro Engineered and Molecular
  Systems}}}\ (\bibinfo {address} {Zhuhai, China},\ \bibinfo {year} {2006})\
  pp.\ \bibinfo {pages} {725--729}\BibitemShut {NoStop}%
\bibitem [{\citenamefont {Hu}\ \emph {et~al.}(2017)\citenamefont {Hu},
  \citenamefont {Xing}, \citenamefont {Chen}, \citenamefont {Gu}, \citenamefont
  {Sun},\ and\ \citenamefont {Li}}]{hu.dm_2017}%
  \BibitemOpen
  \bibfield  {author} {\bibinfo {author} {\bibfnamefont {D.}~\bibnamefont
  {Hu}}, \bibinfo {author} {\bibfnamefont {Y.}~\bibnamefont {Xing}}, \bibinfo
  {author} {\bibfnamefont {M.}~\bibnamefont {Chen}}, \bibinfo {author}
  {\bibfnamefont {B.}~\bibnamefont {Gu}}, \bibinfo {author} {\bibfnamefont
  {B.}~\bibnamefont {Sun}}, \ and\ \bibinfo {author} {\bibfnamefont
  {Q.}~\bibnamefont {Li}},\ }\href {\doibase 10.1016/j.compscitech.2017.01.019}
  {\bibfield  {journal} {\bibinfo  {journal} {Compos. Sci. Technol.}\ }\textbf
  {\bibinfo {volume} {141}},\ \bibinfo {pages} {137} (\bibinfo {year}
  {2017})}\BibitemShut {NoStop}%
\bibitem [{\citenamefont {Gibson}\ and\ \citenamefont
  {Plunkett}(1976)}]{gibson.rf_1976}%
  \BibitemOpen
  \bibfield  {author} {\bibinfo {author} {\bibfnamefont {R.~F.}\ \bibnamefont
  {Gibson}}\ and\ \bibinfo {author} {\bibfnamefont {R.}~\bibnamefont
  {Plunkett}},\ }\href {\doibase 10.1177/002199837601000406} {\bibfield
  {journal} {\bibinfo  {journal} {J. Compos. Mater.}\ }\textbf {\bibinfo
  {volume} {10}},\ \bibinfo {pages} {325} (\bibinfo {year} {1976})}\BibitemShut
  {NoStop}%
\bibitem [{\citenamefont {Reed}(1980)}]{reed.ke_1980}%
  \BibitemOpen
  \bibfield  {author} {\bibinfo {author} {\bibfnamefont {K.~E.}\ \bibnamefont
  {Reed}},\ }\href {\doibase 10.1002/pc.750010109} {\bibfield  {journal}
  {\bibinfo  {journal} {Polym. Compos.}\ }\textbf {\bibinfo {volume} {1}},\
  \bibinfo {pages} {44} (\bibinfo {year} {1980})}\BibitemShut {NoStop}%
\bibitem [{\citenamefont {Uma~Devi}, \citenamefont {Bhagawan},\ and\
  \citenamefont {Thomas}(1997)}]{umadevi.l_1997}%
  \BibitemOpen
  \bibfield  {author} {\bibinfo {author} {\bibfnamefont {L.}~\bibnamefont
  {Uma~Devi}}, \bibinfo {author} {\bibfnamefont {S.~S.}\ \bibnamefont
  {Bhagawan}}, \ and\ \bibinfo {author} {\bibfnamefont {S.}~\bibnamefont
  {Thomas}},\ }\href {\doibase
  10.1002/(SICI)1097-4628(19970531)64:9<1739::AID-APP10>3.0.CO;2-T} {\bibfield
  {journal} {\bibinfo  {journal} {J. Appl. Polym. Sci.}\ }\textbf {\bibinfo
  {volume} {64}},\ \bibinfo {pages} {1739} (\bibinfo {year}
  {1997})}\BibitemShut {NoStop}%
\bibitem [{\citenamefont {Zu}\ \emph {et~al.}(2012)\citenamefont {Zu},
  \citenamefont {Li}, \citenamefont {Zhu}, \citenamefont {Dey}, \citenamefont
  {Wang}, \citenamefont {Lu}, \citenamefont {Deitzel}, \citenamefont
  {Gillespie}, \citenamefont {Byun},\ and\ \citenamefont {Chou}}]{zu.m_2012}%
  \BibitemOpen
  \bibfield  {author} {\bibinfo {author} {\bibfnamefont {M.}~\bibnamefont
  {Zu}}, \bibinfo {author} {\bibfnamefont {Q.}~\bibnamefont {Li}}, \bibinfo
  {author} {\bibfnamefont {Y.}~\bibnamefont {Zhu}}, \bibinfo {author}
  {\bibfnamefont {M.}~\bibnamefont {Dey}}, \bibinfo {author} {\bibfnamefont
  {G.}~\bibnamefont {Wang}}, \bibinfo {author} {\bibfnamefont {W.}~\bibnamefont
  {Lu}}, \bibinfo {author} {\bibfnamefont {J.~M.}\ \bibnamefont {Deitzel}},
  \bibinfo {author} {\bibfnamefont {J.~W.}\ \bibnamefont {Gillespie}}, \bibinfo
  {author} {\bibfnamefont {J.-H.}\ \bibnamefont {Byun}}, \ and\ \bibinfo
  {author} {\bibfnamefont {T.-W.}\ \bibnamefont {Chou}},\ }\href {\doibase
  10.1016/j.carbon.2011.10.047} {\bibfield  {journal} {\bibinfo  {journal}
  {Carbon}\ }\textbf {\bibinfo {volume} {50}},\ \bibinfo {pages} {1271}
  (\bibinfo {year} {2012})}\BibitemShut {NoStop}%
\bibitem [{\citenamefont {Liu}\ \emph {et~al.}(2013)\citenamefont {Liu},
  \citenamefont {Li}, \citenamefont {Gu}, \citenamefont {Zhang}, \citenamefont
  {Zhao}, \citenamefont {Li},\ and\ \citenamefont {Zhang}}]{liu.yn_2013}%
  \BibitemOpen
  \bibfield  {author} {\bibinfo {author} {\bibfnamefont {Y.-N.}\ \bibnamefont
  {Liu}}, \bibinfo {author} {\bibfnamefont {M.}~\bibnamefont {Li}}, \bibinfo
  {author} {\bibfnamefont {Y.}~\bibnamefont {Gu}}, \bibinfo {author}
  {\bibfnamefont {X.}~\bibnamefont {Zhang}}, \bibinfo {author} {\bibfnamefont
  {J.}~\bibnamefont {Zhao}}, \bibinfo {author} {\bibfnamefont {Q.}~\bibnamefont
  {Li}}, \ and\ \bibinfo {author} {\bibfnamefont {Z.}~\bibnamefont {Zhang}},\
  }\href {\doibase 10.1016/j.carbon.2012.10.011} {\bibfield  {journal}
  {\bibinfo  {journal} {Carbon}\ }\textbf {\bibinfo {volume} {52}},\ \bibinfo
  {pages} {550} (\bibinfo {year} {2013})}\BibitemShut {NoStop}%
\bibitem [{\citenamefont {Lei}\ \emph {et~al.}(2016)\citenamefont {Lei},
  \citenamefont {Zhao}, \citenamefont {Zou}, \citenamefont {Jiang},
  \citenamefont {Li}, \citenamefont {Zhang}, \citenamefont {Zhang},\ and\
  \citenamefont {Li}}]{lei.cs_2016}%
  \BibitemOpen
  \bibfield  {author} {\bibinfo {author} {\bibfnamefont {C.}~\bibnamefont
  {Lei}}, \bibinfo {author} {\bibfnamefont {J.}~\bibnamefont {Zhao}}, \bibinfo
  {author} {\bibfnamefont {J.}~\bibnamefont {Zou}}, \bibinfo {author}
  {\bibfnamefont {C.}~\bibnamefont {Jiang}}, \bibinfo {author} {\bibfnamefont
  {M.}~\bibnamefont {Li}}, \bibinfo {author} {\bibfnamefont {X.}~\bibnamefont
  {Zhang}}, \bibinfo {author} {\bibfnamefont {Z.}~\bibnamefont {Zhang}}, \ and\
  \bibinfo {author} {\bibfnamefont {Q.}~\bibnamefont {Li}},\ }\href {\doibase
  10.1002/adem.201500379} {\bibfield  {journal} {\bibinfo  {journal} {Adv. Eng.
  Mater.}\ }\textbf {\bibinfo {volume} {18}},\ \bibinfo {pages} {839} (\bibinfo
  {year} {2016})}\BibitemShut {NoStop}%
\bibitem [{\citenamefont {Zu}\ \emph {et~al.}(2013)\citenamefont {Zu},
  \citenamefont {Li}, \citenamefont {Zhu}, \citenamefont {Zhu}, \citenamefont
  {Wang}, \citenamefont {Byun},\ and\ \citenamefont {Chou}}]{zu.m_2013}%
  \BibitemOpen
  \bibfield  {author} {\bibinfo {author} {\bibfnamefont {M.}~\bibnamefont
  {Zu}}, \bibinfo {author} {\bibfnamefont {Q.}~\bibnamefont {Li}}, \bibinfo
  {author} {\bibfnamefont {Y.}~\bibnamefont {Zhu}}, \bibinfo {author}
  {\bibfnamefont {Y.}~\bibnamefont {Zhu}}, \bibinfo {author} {\bibfnamefont
  {G.}~\bibnamefont {Wang}}, \bibinfo {author} {\bibfnamefont {J.-H.}\
  \bibnamefont {Byun}}, \ and\ \bibinfo {author} {\bibfnamefont {T.-W.}\
  \bibnamefont {Chou}},\ }\href {\doibase 10.1016/j.carbon.2012.09.036}
  {\bibfield  {journal} {\bibinfo  {journal} {Carbon}\ }\textbf {\bibinfo
  {volume} {52}},\ \bibinfo {pages} {347} (\bibinfo {year} {2013})}\BibitemShut
  {NoStop}%
\bibitem [{\citenamefont {Murayama}(1979)}]{murayama.t_1979}%
  \BibitemOpen
  \bibfield  {author} {\bibinfo {author} {\bibfnamefont {T.}~\bibnamefont
  {Murayama}},\ }\href {\doibase 10.1002/app.1979.070240602} {\bibfield
  {journal} {\bibinfo  {journal} {J. Appl. Polym. Sci.}\ }\textbf {\bibinfo
  {volume} {24}},\ \bibinfo {pages} {1413} (\bibinfo {year}
  {1979})}\BibitemShut {NoStop}%
\bibitem [{\citenamefont {Hehr}\ \emph {et~al.}(2014)\citenamefont {Hehr},
  \citenamefont {Schulz}, \citenamefont {Shanov},\ and\ \citenamefont
  {Song}}]{hehr.a_2014}%
  \BibitemOpen
  \bibfield  {author} {\bibinfo {author} {\bibfnamefont {A.}~\bibnamefont
  {Hehr}}, \bibinfo {author} {\bibfnamefont {M.}~\bibnamefont {Schulz}},
  \bibinfo {author} {\bibfnamefont {V.}~\bibnamefont {Shanov}}, \ and\ \bibinfo
  {author} {\bibfnamefont {A.}~\bibnamefont {Song}},\ }\href {\doibase
  10.1177/1045389X13500578} {\bibfield  {journal} {\bibinfo  {journal} {J.
  Intell. Mater. Syst. Struct.}\ }\textbf {\bibinfo {volume} {25}},\ \bibinfo
  {pages} {713} (\bibinfo {year} {2014})}\BibitemShut {NoStop}%
\bibitem [{\citenamefont {Misak}\ \emph {et~al.}(2013)\citenamefont {Misak},
  \citenamefont {Sabelkin}, \citenamefont {Miller}, \citenamefont {Asmatulu},\
  and\ \citenamefont {Mall}}]{misak.he_2013}%
  \BibitemOpen
  \bibfield  {author} {\bibinfo {author} {\bibfnamefont {H.~E.}\ \bibnamefont
  {Misak}}, \bibinfo {author} {\bibfnamefont {V.}~\bibnamefont {Sabelkin}},
  \bibinfo {author} {\bibfnamefont {L.}~\bibnamefont {Miller}}, \bibinfo
  {author} {\bibfnamefont {R.}~\bibnamefont {Asmatulu}}, \ and\ \bibinfo
  {author} {\bibfnamefont {S.}~\bibnamefont {Mall}},\ }\href {\doibase
  10.1166/jnn.2013.8654} {\bibfield  {journal} {\bibinfo  {journal} {J.
  Nanosci. Nanotechnol.}\ }\textbf {\bibinfo {volume} {13}},\ \bibinfo {pages}
  {8331} (\bibinfo {year} {2013})}\BibitemShut {NoStop}%
\bibitem [{\citenamefont {Gardea}\ \emph {et~al.}(2015)\citenamefont {Gardea},
  \citenamefont {Glaz}, \citenamefont {Riddick}, \citenamefont {Lagoudas},\
  and\ \citenamefont {Naraghi}}]{gardea.f_2015}%
  \BibitemOpen
  \bibfield  {author} {\bibinfo {author} {\bibfnamefont {F.}~\bibnamefont
  {Gardea}}, \bibinfo {author} {\bibfnamefont {B.}~\bibnamefont {Glaz}},
  \bibinfo {author} {\bibfnamefont {J.}~\bibnamefont {Riddick}}, \bibinfo
  {author} {\bibfnamefont {D.~C.}\ \bibnamefont {Lagoudas}}, \ and\ \bibinfo
  {author} {\bibfnamefont {M.}~\bibnamefont {Naraghi}},\ }\href {\doibase
  10.1021/acsami.5b01459} {\bibfield  {journal} {\bibinfo  {journal} {ACS Appl.
  Mater. Interfaces}\ }\textbf {\bibinfo {volume} {7}},\ \bibinfo {pages}
  {9725} (\bibinfo {year} {2015})}\BibitemShut {NoStop}%
\bibitem [{\citenamefont {Zhang}(2017)}]{zhang.xh_20171}%
  \BibitemOpen
  \bibfield  {author} {\bibinfo {author} {\bibfnamefont {X.}~\bibnamefont
  {Zhang}},\ }in\ \href@noop {} {\emph {\bibinfo {booktitle}
  {{Nanomechanics}}}},\ \bibinfo {editor} {edited by\ \bibinfo {editor}
  {\bibfnamefont {A.}~\bibnamefont {Vakhrushev}}}\ (\bibinfo  {publisher}
  {InTech - Open Access},\ \bibinfo {address} {Rijeka, Croatia},\ \bibinfo
  {year} {2017})\ Chap.~\bibinfo {chapter} {1}, pp.\ \bibinfo {pages}
  {1--29}\BibitemShut {NoStop}%
\bibitem [{\citenamefont {Bhushan}\ \emph {et~al.}(2008)\citenamefont
  {Bhushan}, \citenamefont {Ling}, \citenamefont {Jungen},\ and\ \citenamefont
  {Hierold}}]{bhushan.b_2008}%
  \BibitemOpen
  \bibfield  {author} {\bibinfo {author} {\bibfnamefont {B.}~\bibnamefont
  {Bhushan}}, \bibinfo {author} {\bibfnamefont {X.}~\bibnamefont {Ling}},
  \bibinfo {author} {\bibfnamefont {A.}~\bibnamefont {Jungen}}, \ and\ \bibinfo
  {author} {\bibfnamefont {C.}~\bibnamefont {Hierold}},\ }\href {\doibase
  10.1103/PhysRevB.77.165428} {\bibfield  {journal} {\bibinfo  {journal} {Phys.
  Rev. B}\ }\textbf {\bibinfo {volume} {77}},\ \bibinfo {pages} {165428}
  (\bibinfo {year} {2008})}\BibitemShut {NoStop}%
\bibitem [{\citenamefont {Bhushan}\ and\ \citenamefont
  {Ling}(2008)}]{bhushan.b_20081}%
  \BibitemOpen
  \bibfield  {author} {\bibinfo {author} {\bibfnamefont {B.}~\bibnamefont
  {Bhushan}}\ and\ \bibinfo {author} {\bibfnamefont {X.}~\bibnamefont {Ling}},\
  }\href {\doibase 10.1103/PhysRevB.78.045429} {\bibfield  {journal} {\bibinfo
  {journal} {Phys. Rev. B}\ }\textbf {\bibinfo {volume} {78}},\ \bibinfo
  {pages} {045429} (\bibinfo {year} {2008})}\BibitemShut {NoStop}%
\bibitem [{\citenamefont {Lu}\ and\ \citenamefont {Chou}(2011)}]{lu.wb_20111}%
  \BibitemOpen
  \bibfield  {author} {\bibinfo {author} {\bibfnamefont {W.}~\bibnamefont
  {Lu}}\ and\ \bibinfo {author} {\bibfnamefont {T.-W.}\ \bibnamefont {Chou}},\
  }\href {\doibase 10.1016/j.jmps.2011.01.004} {\bibfield  {journal} {\bibinfo
  {journal} {J. Mech. Phys. Solids}\ }\textbf {\bibinfo {volume} {59}},\
  \bibinfo {pages} {511} (\bibinfo {year} {2011})}\BibitemShut {NoStop}%
\bibitem [{\citenamefont {Kis}\ \emph {et~al.}(2004)\citenamefont {Kis},
  \citenamefont {Cs\'{a}nyi}, \citenamefont {Salvetat}, \citenamefont {Lee},
  \citenamefont {Couteau}, \citenamefont {Kulik}, \citenamefont {Benoit},
  \citenamefont {Brugger},\ and\ \citenamefont {Forr\'{o}}}]{kis.a_2004}%
  \BibitemOpen
  \bibfield  {author} {\bibinfo {author} {\bibfnamefont {A.}~\bibnamefont
  {Kis}}, \bibinfo {author} {\bibfnamefont {G.}~\bibnamefont {Cs\'{a}nyi}},
  \bibinfo {author} {\bibfnamefont {J.-P.}\ \bibnamefont {Salvetat}}, \bibinfo
  {author} {\bibfnamefont {T.-N.}\ \bibnamefont {Lee}}, \bibinfo {author}
  {\bibfnamefont {E.}~\bibnamefont {Couteau}}, \bibinfo {author} {\bibfnamefont
  {A.~J.}\ \bibnamefont {Kulik}}, \bibinfo {author} {\bibfnamefont
  {W.}~\bibnamefont {Benoit}}, \bibinfo {author} {\bibfnamefont
  {J.}~\bibnamefont {Brugger}}, \ and\ \bibinfo {author} {\bibfnamefont
  {L.}~\bibnamefont {Forr\'{o}}},\ }\href {\doibase 10.1038/nmat1076}
  {\bibfield  {journal} {\bibinfo  {journal} {Nat. Mater.}\ }\textbf {\bibinfo
  {volume} {3}},\ \bibinfo {pages} {153} (\bibinfo {year} {2004})}\BibitemShut
  {NoStop}%
\bibitem [{\citenamefont {Ma}\ \emph {et~al.}(2009)\citenamefont {Ma},
  \citenamefont {Liu}, \citenamefont {Zhang}, \citenamefont {Yang},
  \citenamefont {Liu}, \citenamefont {Zhang}, \citenamefont {An}, \citenamefont
  {Yi}, \citenamefont {Ren}, \citenamefont {Niu}, \citenamefont {Li},
  \citenamefont {Dong}, \citenamefont {Zhou}, \citenamefont {Ajayan},\ and\
  \citenamefont {Xie}}]{ma.wj_2009}%
  \BibitemOpen
  \bibfield  {author} {\bibinfo {author} {\bibfnamefont {W.}~\bibnamefont
  {Ma}}, \bibinfo {author} {\bibfnamefont {L.}~\bibnamefont {Liu}}, \bibinfo
  {author} {\bibfnamefont {Z.}~\bibnamefont {Zhang}}, \bibinfo {author}
  {\bibfnamefont {R.}~\bibnamefont {Yang}}, \bibinfo {author} {\bibfnamefont
  {G.}~\bibnamefont {Liu}}, \bibinfo {author} {\bibfnamefont {T.}~\bibnamefont
  {Zhang}}, \bibinfo {author} {\bibfnamefont {X.}~\bibnamefont {An}}, \bibinfo
  {author} {\bibfnamefont {X.}~\bibnamefont {Yi}}, \bibinfo {author}
  {\bibfnamefont {Y.}~\bibnamefont {Ren}}, \bibinfo {author} {\bibfnamefont
  {Z.}~\bibnamefont {Niu}}, \bibinfo {author} {\bibfnamefont {J.}~\bibnamefont
  {Li}}, \bibinfo {author} {\bibfnamefont {H.}~\bibnamefont {Dong}}, \bibinfo
  {author} {\bibfnamefont {W.}~\bibnamefont {Zhou}}, \bibinfo {author}
  {\bibfnamefont {P.~M.}\ \bibnamefont {Ajayan}}, \ and\ \bibinfo {author}
  {\bibfnamefont {S.}~\bibnamefont {Xie}},\ }\href {\doibase 10.1021/nl901035v}
  {\bibfield  {journal} {\bibinfo  {journal} {Nano Lett.}\ }\textbf {\bibinfo
  {volume} {9}},\ \bibinfo {pages} {2855} (\bibinfo {year} {2009})}\BibitemShut
  {NoStop}%
\bibitem [{\citenamefont {Cao}\ \emph {et~al.}(2011)\citenamefont {Cao},
  \citenamefont {Reiner}, \citenamefont {Chung}, \citenamefont {Chang},
  \citenamefont {Kao}, \citenamefont {Kukta},\ and\ \citenamefont
  {Korach}}]{cao.ch_2011}%
  \BibitemOpen
  \bibfield  {author} {\bibinfo {author} {\bibfnamefont {C.}~\bibnamefont
  {Cao}}, \bibinfo {author} {\bibfnamefont {A.}~\bibnamefont {Reiner}},
  \bibinfo {author} {\bibfnamefont {C.}~\bibnamefont {Chung}}, \bibinfo
  {author} {\bibfnamefont {S.-H.}\ \bibnamefont {Chang}}, \bibinfo {author}
  {\bibfnamefont {I.}~\bibnamefont {Kao}}, \bibinfo {author} {\bibfnamefont
  {R.~V.}\ \bibnamefont {Kukta}}, \ and\ \bibinfo {author} {\bibfnamefont
  {C.~S.}\ \bibnamefont {Korach}},\ }\href {\doibase
  10.1016/j.carbon.2011.03.043} {\bibfield  {journal} {\bibinfo  {journal}
  {Carbon}\ }\textbf {\bibinfo {volume} {49}},\ \bibinfo {pages} {3190}
  (\bibinfo {year} {2011})}\BibitemShut {NoStop}%
\bibitem [{\citenamefont {Zhang}\ \emph
  {et~al.}(2010{\natexlab{b}})\citenamefont {Zhang}, \citenamefont {Zhao},
  \citenamefont {Tang}, \citenamefont {Li}, \citenamefont {Huang},
  \citenamefont {Liu}, \citenamefont {Zhu}, \citenamefont {Zhang},\ and\
  \citenamefont {Wei}}]{zhang.q_20101}%
  \BibitemOpen
  \bibfield  {author} {\bibinfo {author} {\bibfnamefont {Q.}~\bibnamefont
  {Zhang}}, \bibinfo {author} {\bibfnamefont {M.-Q.}\ \bibnamefont {Zhao}},
  \bibinfo {author} {\bibfnamefont {D.-M.}\ \bibnamefont {Tang}}, \bibinfo
  {author} {\bibfnamefont {F.}~\bibnamefont {Li}}, \bibinfo {author}
  {\bibfnamefont {J.-Q.}\ \bibnamefont {Huang}}, \bibinfo {author}
  {\bibfnamefont {B.}~\bibnamefont {Liu}}, \bibinfo {author} {\bibfnamefont
  {W.-C.}\ \bibnamefont {Zhu}}, \bibinfo {author} {\bibfnamefont {Y.-H.}\
  \bibnamefont {Zhang}}, \ and\ \bibinfo {author} {\bibfnamefont
  {F.}~\bibnamefont {Wei}},\ }\href {\doibase 10.1002/anie.200907130}
  {\bibfield  {journal} {\bibinfo  {journal} {Angew. Chem. Int. Ed.}\ }\textbf
  {\bibinfo {volume} {49}},\ \bibinfo {pages} {3642} (\bibinfo {year}
  {2010}{\natexlab{b}})}\BibitemShut {NoStop}%
\bibitem [{\citenamefont {Zhao}\ \emph {et~al.}(2012)\citenamefont {Zhao},
  \citenamefont {Zhang}, \citenamefont {Tian}, \citenamefont {Huang},\ and\
  \citenamefont {Wei}}]{zhao.mq_2012}%
  \BibitemOpen
  \bibfield  {author} {\bibinfo {author} {\bibfnamefont {M.-Q.}\ \bibnamefont
  {Zhao}}, \bibinfo {author} {\bibfnamefont {Q.}~\bibnamefont {Zhang}},
  \bibinfo {author} {\bibfnamefont {G.-L.}\ \bibnamefont {Tian}}, \bibinfo
  {author} {\bibfnamefont {J.-Q.}\ \bibnamefont {Huang}}, \ and\ \bibinfo
  {author} {\bibfnamefont {F.}~\bibnamefont {Wei}},\ }\href {\doibase
  10.1021/nn301421x} {\bibfield  {journal} {\bibinfo  {journal} {ACS Nano}\
  }\textbf {\bibinfo {volume} {6}},\ \bibinfo {pages} {4520} (\bibinfo {year}
  {2012})}\BibitemShut {NoStop}%
\bibitem [{\citenamefont {Wang}, \citenamefont {Ludwigson},\ and\ \citenamefont
  {Lakes}(2004)}]{wang.yc_2004}%
  \BibitemOpen
  \bibfield  {author} {\bibinfo {author} {\bibfnamefont {Y.~C.}\ \bibnamefont
  {Wang}}, \bibinfo {author} {\bibfnamefont {M.}~\bibnamefont {Ludwigson}}, \
  and\ \bibinfo {author} {\bibfnamefont {R.~S.}\ \bibnamefont {Lakes}},\ }\href
  {\doibase 10.1016/j.msea.2003.08.071} {\bibfield  {journal} {\bibinfo
  {journal} {Mater. Sci. Eng. A}\ }\textbf {\bibinfo {volume} {370}},\ \bibinfo
  {pages} {41} (\bibinfo {year} {2004})}\BibitemShut {NoStop}%
\end{thebibliography}%

\clearpage

\setcounter{figure}{0}
\renewcommand{\thefigure}{S\arabic{figure}}
\setcounter{table}{0}
\renewcommand{\thetable}{S\arabic{table}}
\setcounter{page}{1}
\renewcommand{\thepage}{S\arabic{page}}

\noindent
{\Large\bf Supporting Information}

\bigskip

In this supporting information, the synthesis details and characterization results are provided to support the
discussion in Sec. \ref{sec:densified}. These results were for the CNT/DBP, CNT/TEOS, CNT/PVA, and CNT/PF composite
films, as compared to the untreated iCVD films. For the iCVD films, under the similar assembly structure, there was also
evidences to show the effect of tube structure on the film's damping performance.

The reference numbers here are the same as in the main text.

{\bf Fabrication of iCVD films:} For the CNT growth, a tube furnace was heated up to 1300 \si{\celsius} with a heating
rate of 10 \si{\celsius\per\min} in Ar atmosphere. Then, a solution of ethyl alcohol with 1.2 vol\% ferrocene and 0.4
vol\% thiophene was injected at a rate of 20 \si{\mL\per\hour} into the furnace, carried by an Ar/\ce{H2} gas mixture
\cite{wang.h_2017}. Such group of growth parameters is different from our previous attempt where ethanol was the carbon
source and the ferrocene and thiophene contents were 2 wt\% and 1 vol\% \cite{han.y_2015, zou.jy_2016, zou.jy_2017}. As
a result, the present CNTs were mainly multi-walled \cite{wang.h_2017}, while the previous result showed a large amount
of double- and triple-walled CNTs were observed \cite{han.y_2015, zou.jy_2016, zou.jy_2017}.

The as-grown CNTs, highly entangled due to the gas flow, were collected layer-by-layer on a rotating mandrel (diameter
400 mm). A gel structure was obtained by the direct winding, and a simultaneous spraying of liquid (mostly ethanol)
during the winding could produce the densified CNT films.

{\bf Effect of tube thickness on damping performance:} As we have provided in the main text, the wall thickness has
shown clear influence on the loss tangent. {\bf Table \ref{tab:tan}} shows a comparison of loss tangent between two
different iCVD films. At the similar bundle size of $\sim$50 nm, the wall thickness can greatly affect the intrinsic
viscosity of the CNT bundle. First, the less wall number, the smaller tube diameter, and thus the larger contact area
between CNTs in the bundle. Second, the few-walled CNTs are generally much softer than the multi-walled tubes, in
flexibility and bendability. These features can easily induce intertube sliding and structural deformation, resulting in
the increase in loss tangent.

\begin{table*}[!t]
\centering
\caption{A comparison of loss tangent between two different iCVD films, which were composed by few-walled and
multi-walled CNTs, respectively.}
\label{tab:tan}
\begin{tabular}{ccccccccc}
\hline\hline
Wall        & Bundle &           \multicolumn{6}{c}{Frequency (Hz)}             & \\ \cline{3-8}
number & size (nm) & 1 & 10 & 20 & 50 & 100 & 200 & Reference\\
\hline
2--3          & $\sim$50 & & {\bf 0.101} & {\bf 0.142} & {\bf 0.192} & {\bf 0.295} &             &Sci. Rep. 2015 \cite{han.y_2015} \\
2--3          & $\sim$50 & & {\bf 0.119} &                   & 0.132 & 0.204 &  {\bf 0.372} & Carbon 2015 \cite{liu.ql_2015} \\
$\sim$10 &                  & 0.090 &            &           &             &           &           & AFM 2017 \cite{wang.h_2017} \\
$\sim$10 & $\sim$50 & 0.088 & {\bf 0.096} & {\bf 0.103} & {\bf 0.136} & {\bf 0.207}  & {\bf 0.298} & this study \\
\hline\hline
\end{tabular}
\end{table*}

{\bf Preparation of CNT composite films:} The as-produced ethanol-densified iCVD films were post treated in the
following ways. (1) Infiltration of DBP or TEOS: After being soaked in DBP or TEOS liquid for 8 h, the CNT/DBP or
CNT/TEOS composite film was compressed with a compression roller at 150 \si{\celsius}, below the boiling point of the
organics. (2) Infiltration of PVA: The 8-h soaking was performed in a 0.5 wt\% PVA/water solution. After the soaking,
the film was dried at 80 \si{\celsius} for 3 h, and then compressed. (3) Infiltration and curing of PF: The solution was
a 13.7 wt\% PF resin/acetone solution. After the 8-h soaking, the film was washed by acetone and dried naturally. The
curing was performed at 10 MPa and 150 \si{\celsius} for 5 min. (4) The control sample of iCVD film was also compressed
with a compression roller, in oder to obtain the similar CNT packing density for all samples.

\begin{figure}[t!]
\centering
\includegraphics[width=0.48\textwidth]{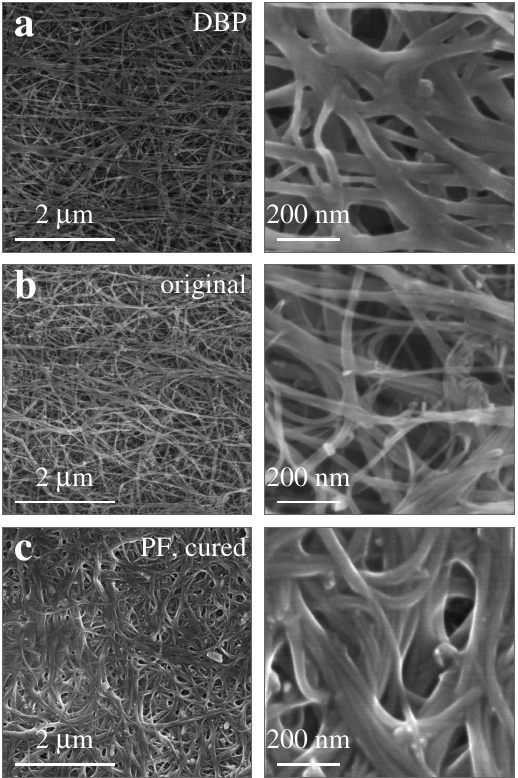}
\caption{Assembly morphologies of the CNT/DBP, pure CNT (iCVD), and CNT/PF films.}
\label{fig:assmorph}
\end{figure}

{\bf Polymer-tuned intertube contacts:} {\bf Figure \ref{fig:assmorph}} compares the assembly morphologies of the
CNT/DBP, pure CNT (iCVD), and CNT/PF films. The lower magnification results (left panels) show clearly that the CNTs
were packed together under a similar density. However, the amount of infiltrated organic compounds were different due to
the different molecular weights. There was about 11--14 wt\% infiltration of DBP according to thermogravimetry (divided
by the total mass of the composite film), and the infiltration was much larger for PF, about 28--33 wt\%.

The higher magnification results (right panels in Figure \ref{fig:assmorph}) show totally different intertube contacts
in these films. The infiltrated DBP molecules mostly locate inside the CNT bundles and/or around the connection nodes of
the bundles. However, the overall CNT network seems to be as similar as to the original sample. Due to the viscosity of
DBP, both the intrinsic viscosity of CNT bundle and the zipping/unzipping behavior at the connection nodes could induce
more energy dissipation, as reflected in the increase in loss tangent. For the CNT/PF composite film, although the
polymer fraction was much higher, the cured polymer network is indeed an elastic material. This phenomenon can not only
hinder the zipping/unzipping behavior, but also increase the elasticity of CNT bundle. Therefore, the CNT/PF composite
film exhibited the lowest loss tangent.

{\bf Testing mechods:} A T150 Universal Testing Machine (Keysight Technologies, Inc., Santa Rosa, USA) and a DMA 242 E
Artemis (NETZSCH-Ger\"{a}tebau GmbH, Selb, Germany) were used to characterize the dynamic properties. Frequencies
ranging from 10 to 400 Hz were used in the T150 system, and 1 to 100 Hz in the DMA 242 E Artemis. The two testers
provided the same results for the testings at the frequencies \cite{zhao.jn_2015}. The vibration amplitude was
controlled as load force variation and displacement variation in the two tester, respectively. For the T150 system, the
default value of $A_f = 4.5$ mN for the force variation corresponded to a displacement amplitude ($A$) of several μm for
a given gauge length $L$, modulus $E$, and cross sectional area $S$, by $A_f / S E =  A / L$ \cite{zhao.jn_2015}. For
the tests with DMA 242 E Artemis, 10 \si{\um} was used for the displacement amplitude.

\end{document}